\documentclass[10pt]{iopart}
\usepackage{stmaryrd}
\usepackage[normalem]{ulem}
\usepackage{comment}
\usepackage{color} 
\usepackage{bm}
\usepackage{graphicx}
\definecolor{dred}{rgb}{0.6, 0.0, 0.0}
\definecolor{dgreen}{rgb}{0.0, 0.5, 0.0}

\usepackage{iopams}  
\begin{document}
%\review[Hermitian and non-Hermitian topology in active matter]{Hermitian and non-Hermitian topology in active matter}
\title[Hermitian and non-Hermitian topology in active matter]{Hermitian and non-Hermitian topology in active matter}

\author{Kazuki Sone$^{1,2\ast}$, Kazuki Yokomizo$^3$, Kyogo Kawaguchi$^{3,4,5,6}$, and Yuto Ashida$^{3,5}$}
\address{$^1$Department of Physics, University of Tsukuba, Tsukuba, Ibaraki 305-8571, Japan\\
$^2$Department of Applied Physics, The University of Tokyo, 7-3-1 Hongo, Bunkyo-ku, Tokyo 113-8656, Japan\\
$^3$Department of Physics, The University of Tokyo, 7-3-1 Hongo, Bunkyo-ku, Tokyo 113-0033, Japan\\
$^4$Nonequilibrium Physics of living Matter laboratory, RIKEN Pioneering Research institute, 2-2-3 Minatojima-minamimachi, chuo-ku, Kobe, Hyogo 650-0047, Japan\\
%$^4$Nonequilibrium Physics of Living Matter RIKEN Hakubi Research Team, RIKEN Center for Biosystems Dynamics Research, 2-2-3 Minatojima-minamimachi, Chuo-ku, Kobe 650-0047, Japan\\
%$^5$RIKEN Cluster for Pioneering Research, 2-2-3 Minatojima-minamimachi, Chuo-ku, Kobe 650-0047, Japan\\
$^5$Institute for Physics of Intelligence, The University of Tokyo, 7-3-1 Hongo, Tokyo 113-0033, Japan\\
$^6$Universal Biology Institute, The University of Tokyo, Bunkyo-ku, Tokyo 113-0033, Japan
}

\ead{sone@rhodia.ph.tsukuba.ac.jp}
\vspace{10pt}
\begin{indented}
\item[]%July 2022
\end{indented}

\begin{abstract}
Self-propulsion is a quintessential aspect of biological systems, which can induce nonequilibrium phenomena that have no counterparts in passive systems. Motivated by biophysical interest together with recent advances in experimental techniques, active matter has been a rapidly developing field in physics. Meanwhile, over the past few decades, topology has played a crucial role to understand certain robust properties appearing in condensed matter systems. For instance, the nontrivial topology of band structures leads to the notion of topological insulators, where one can find robust gapless edge modes protected by the bulk band topology. We here review recent progress in an interdisciplinary area of research at the intersection of these two fields. Specifically, we give brief introductions to active matter and band topology in Hermitian systems, and then explain how the notion of band topology can be extended to nonequilibrium (and thus non-Hermitian) systems including active matter. We review recent studies that have demonstrated the intimate connections between active matter and topological materials, where exotic topological phenomena that are unfeasible in passive systems have been found. A possible extension of the band topology to nonlinear systems is also briefly discussed. Active matter can thus provide an ideal playground to explore topological phenomena in qualitatively new realms beyond conservative linear systems. 
\end{abstract}

%
% Uncomment for keywords
%\vspace{2pc}
%\noindent{\it Keywords}: XXXXXX, YYYYYYYY, ZZZZZZZZZ
%
% Uncomment for Submitted to journal title message
%\submitto{\JPA}
%
% Uncomment if a separate title page is required
%\maketitle
% 
% For two-column output uncomment the next line and choose [10pt] rather than [12pt] in the \documentclass declaration
\ioptwocol
\setcounter{tocdepth}{3}
\tableofcontents

\section{Introduction}\label{sec:1}
The notion of active matter, a collection of self-propelled particles, has provided simplified frameworks to model complex systems, such as bird flocking and cell dynamics. 
Since numerical studies of the models of flocking implied the presence of phase transitions inherent to active systems~\cite{Vicsek1995,Vicsek2012}, active matter has attracted interest in theoretical statistical physics. More recently, dynamics of active matter have been observed in certain biological systems \cite{Dunkel2013,Trepat2009} and artificial colloids \cite{Jiang2010,Bricard2013} in highly controllable settings. Thus, a number of theoretically predicted nonequilibrium phases or phenomena, many of which were once considered to be of purely academic interest, are now within experimental reach. Given that physical systems falling into a class of active matter are fairly abundant in biological systems \cite{Vicsek2012,Marchetti2013}, understanding active matter is also important from the viewpoint of biophysics.

Activity is also of practical interest in, e.g., metamaterial science. In particular, nonequilibrium features of active matter can lead to nonreciprocity, which breaks the time-reversal symmetry and thus enables spontaneous functionalities of metamaterials. The emergence of such time-reversal-odd forces is theoretically predicted from microscopic models of active particles \cite{Banerjee2017}. Furthermore, experimental studies have realized spontaneously moving metamaterials by using artificial electrical devices \cite{Chen2021}. Motivated by these findings, the frontier of studies on active matter has also been expanded into various fields of physics including nonbiological materials.

On another front, over the past few decades, the notion of topology has been widely applied to condensed matter physics \cite{Berezinskii1971,Kosterlitz1972,Orlandini2007}. Topology can tell us physical properties that are preserved under continuous deformations of parameters, such as the strength of interactions and geometry of the system. A representative example of topological properties is a genus of a manifold, which cannot be unchanged until we cut the manifold or stick its surfaces. Such a topological invariant distinguishes manifolds that can be continuously deformed into each other; for example, a donut can be deformed into a mug, while it cannot be deformed into a ball. Similar topology has been found in real-space configurations of spins and liquid crystals \cite{Muhlbauer2009,Yu2010}, where topological charges are defined around their defective structures.

In the field of condensed matter physics, a different kind of topology has also been discussed in the wavenumber space, i.e., the band structures of electrons. Such nontrivial topology in band structures brings the notion of topological materials, which is exemplified by topological insulators. The first discovery of the topological band structure is the quantum Hall effect \cite{Klitzing1980}, where the Hall conductance is quantized. While the conventional quantum Hall effect is observed under the existence of the external magnetic field, the quantized Hall conductance is proportional to the topological number (Chern number) of occupied bands \cite{Thouless1982}, which can be nonzero even without external magnetic fields by instead using internal magnetic structures. Furthermore, symmetries enrich the topological phases of matter. In fact, by using the spin-orbit interaction, the quantum spin Hall effect, the time-reversal symmetric version of the quantum Hall effect, was proposed \cite{Kane2005a,Kane2005b}, which stimulates the subsequent studies on topological materials.

Topology of band structures also affects the surface properties of the materials; the nontrivial band topology predicts the emergence of localized boundary modes with gapless dispersions \cite{Hatsugai1993b}. While the conventional studies of band topology have focused on electronic band structures, recent studies \cite{Haldane2008,Prodan2009,Rechtsman2013,Khanikaev2013,Huber2016} have also aimed to construct classical analogs of topological insulators. In such classical topological systems, one can directly observe the topological boundary modes as localized waves, which can enrich the application of band topology. In addition, classical systems can often realize controlled dissipation and/or complex nonlinear interactions \cite{Ashida2020,Smirnova2020,Ota2020}, which allow one to explore interesting non-Hermitian or nonlinear effects that are difficult to realize in electronic systems. Classical systems can thus provide ideal platforms to study band topology beyond conservative linear systems.

Among various attempts to realize classical topological systems, topological active matter \cite{Souslov2017,Shankar2022} is one of the most exciting possibilities; since the dissipation of energy is inherent to active matter owing to its genuine nonequilibrium nature, one can expect that various non-Hermitian topological phenomena should be observed in active systems. In addition, topological boundary modes may be of interest in biophysics to elucidate the mechanism of localization \cite{Leduc2012,Mayer2010} seen in various biological systems. The aim of this review is to introduce the basic concepts in active matter and topological materials and to review recent attempts to bridge these two different fields.  

The remaining part of this review is organized as follows. In Sec.~\ref{sec:2}, we introduce experimental setups of active matter. We also explain the hydrodynamic equation of active matter in Sec.~\ref{sec:2c}, which is the starting point to discuss the active-matter counterparts of topological insulators. In Sec.~\ref{sec:3}, we introduce basic concepts of band topology in condensed matter physics. In Sec.~\ref{sec:4}, we introduce topological phenomena unique to non-Hermitian systems, which is an emerging field dealing with the interplay between band topology and out-of-equilibrium physics. We also provide theoretical tools to analyze non-Hermitian topology called the non-Bloch band theory. In Sec.~\ref{sec:5}, we review the studies on band topology in active matter. We also explain how to combine classical physics and topological physics explained in the above sections via the linearization of active hydrodynamics. Finally, we discuss future perspectives in Sec.~\ref{sec:6}.

\begin{figure}[]
\centering
\includegraphics[width=8cm]{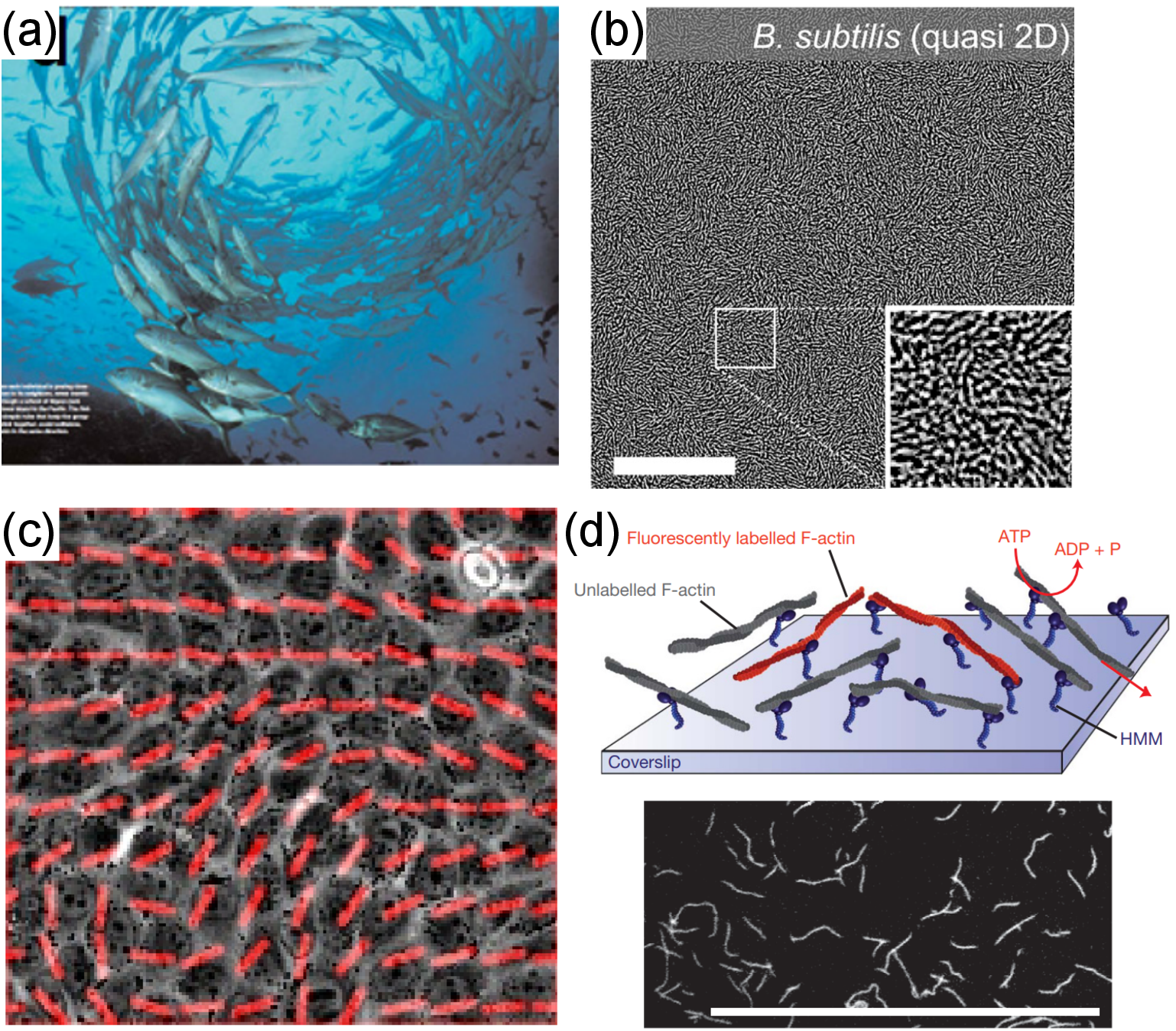}
\caption{\label{example_active_matter_fig} Examples of biological active matter. (a) School of fish is a typical example of active matter. Reprinted from Rep. Phys., 517, T. Vicsek and A. Zafeiris, ``Collective motion,'' 71--140, Copyright \copyright\  (2012) \cite{Vicsek2012}, with permission from Elsevier. (b) Bacteria exhibit active turbulence even in a low-Reynolds-number regime.  This figure is adapted from H. H. Wensink et al., ``Meso-scale turbulence in living fluids'' Proc. Natl. Acad. Sci. USA 109, 14308--14313 (2012) \cite{Wensink2012}. (c) The collective dynamics of cells is governed by their topological defects. This figure is adapted from T. B. Saw et al., ``Topological defects in epithelia govern cell death and extrusion'' Nature 544, 212--216 (2017) \cite{Saw2017}, Springer Nature. Copyright \copyright\  2017. (d) A schematic and a snapshot of actin filaments are shown. This figure is adapted from V. Schaller et al., ``Polar patterns of driven filaments'' Nature 467, 73--77 (2010) \cite{Schaller2010}, Springer Nature. Copyright \copyright\  2010. }
\end{figure}

\section{Active matter}\label{sec:2}
\subsection{Self-propelled particles}\label{sec:2a}
\subsubsection{Flocking}\label{sec:2a1}
Self-propelled particles are of prevalent interest in the field of statistical physics owing to their ability to collectively move in the same direction without a leader or direct aligning interactions. Animals, such as birds and fish, serve as typical examples of these particles \cite{Vicsek2012}, which can form flocks and schools at a collective scale (cf.~Fig.~\ref{example_active_matter_fig}(a)). One early attempt to describe the collective dynamics of birds is the numerical simulation study of boids \cite{Reynolds1987}, which refers to bird-like (bird-oid) particles. Boids possess attractive, exclusive, and aligning interactions, allowing them to imitate the flocks of birds despite the symmetric role of the particles.

An important study regarding the mechanism of flocking appeared in 1995 \cite{Vicsek1995}, which analyzed a simple model using only an aligning interaction. The Vicsek model is described by three rules: (1) Each particle moves at a constant and homogeneous speed. (2) Alignment interactions are applied to neighbor particles. (3) The direction of motion of each particle fluctuates. These rules are expressed by the following time evolution:
\begin{eqnarray}
 \mathbf{r}_i(t+\Delta t) &=&  \mathbf{r}_i(t) + v_0 \mathbf{e}_{\theta_i(t)}, \label{Vicsek1}\\
 \theta_i(t+\Delta t) &=& \arg \left( \sum_{j \in U_i} e^{i\theta_j(t)} \right) + \xi_i(t), \label{Vicsek2}
\end{eqnarray}
where $\mathbf{r}_i(t)$ and $\theta_i(t)$ are the position and angle of the $i$th particle at time $t$. $\mathbf{e}_{\theta_i(t)}$ is a unit vector whose angle is $\theta_i(t)$. $U_i$ represents the vicinity of the $i$th particle, $U_i = \{ \mathbf{r}' | ||\mathbf{r}' - \mathbf{r}_i|| < r_0 \}$ with $r_0$ being a range of the aligning interaction. $\xi_i(t)$ is a random variable that introduces the fluctuation of the direction of each particle. With large density and small fluctuation, the Vicsek model exhibits a flocking phase where the directions of motions of active particles are aligned in one direction. More interestingly, the flocking phase also exhibits a similar feature to a flock of animals in its density distribution, i.e., giant density fluctuation \cite{Toner1998,Chate2008}, where the size dependence of density fluctuation follows $\delta \rho \sim L^{-\alpha}$, $\alpha<2$ and thus is larger than that found in random distributions. Due to its simplicity, the Vicsek model and its variants \cite{Shimoyama1996,Gregoire2004} have been widely used to analyze aspects of active systems as detailed later. It is also noteworthy that particle models like the Vicsek model are mutually related to continuum models introduced later in Sec.~\ref{sec:2c}, based on which the connection between active matter and topological insulators is discussed.

While the simple rules in the Vicsek model reproduce collective motion, the real interaction can be more complex. For example, observations in the bird flock indicate that the birds interact on average with a fixed number of neighbors~\cite{Ballerini2008,Ginelli2010} rather than with all the neighbors within a fixed distance as assumed in the original Vicsek model. Also, it has been proposed that specific rules of interaction might be responsible for realizing certain structures of the flocks~\cite{Cattivelli2011,Corcoran2019}.

Within the collective dynamics of living animals, a number of studies have been conducted on the natural swarm of midges~\cite{Attanasi2014,Attanasi2014b,Cavagna2017}. 
The dynamical exponent, i.e., an exponent that relates the smarm size to the timescale of autocorrelation, has been explained by the dynamic renormalization procedure that incorporates inertia and friction in the latest model by Cavagna {\it et al.}~\cite{Cavagna2023}.
Other interesting examples of collective dynamics include the analysis of the dynamics of sheep~\cite{Gomez2022}, fish~\cite{Giannini2020}, and insects~\cite{Sinhuber2017,Tennenbaum2016}, as well as human swarms~\cite{Silverberg2013}, although their relevance to simple models of the Vicsek type is often unclear. 
The situation is presumably even more complex in social animals including ants and humans, which are also affected by social or psychological forces \cite{Helbing2000}.

\subsubsection{Cells and bacteria}\label{sec:2a2}
Bacteria and cultured cells are widely used to observe the dynamics of active matter due to their availability and controllability. Bacteria, such as \textit{E. coli}, can swim in the medium by utilizing their flagella, which are the whiplike organs that rotate due to the rotor machinery placed at their base. The flagella can also create flow around the bacteria, which leads to the hydrodynamic interaction between bacteria. The direction of the created flow depends on the type of flagella motion, coined pusher and puller, which correspond to propeller-type and crawling-type swimming, respectively~\cite{Lauga2009,Lauga2016}. Theoretically, the flow and the induced hydrodynamic interaction are analyzed by expanding the multipole of the flow called the Stokeslet~\cite{Lauga2009}.

The motion of individual bacteria is characterized by a run-and-tumble feature \cite{Tailleur2008,Aranson2022,Kurzthaler2024}; the bacteria can move straight and randomly rotate. 
During the straight motion, bacteria synchronously move their flagella, while in a tumbling phase, the flagella often extend to different directions, and the average force created by flagella just rotates the bacteria. 
To effectively seek nutrients, bacteria also control the rate of alternation between runs and tumbles \cite{Segall1986}. The run-and-tumble motion is also widely used in theoretical and numerical studies because of its simplicity \cite{Angelani2014,Malakar2018}.

The interaction between flagella and the substrate or boundaries can also affect the motion of bacteria. In particular, collision and hydrodynamic interaction between rotating flagella and wall can induce chirality of active fluid, i.e., bacteria near the wall show a curved motion \cite{DiLuzio2005,Li2008}. Hydrodynamic interaction between pusher bacteria like \textit{E. coli} and a wall stabilizes swimming in a parallel direction to the wall \cite{Berke2008}. Then, the rotating body and flagella further induce the hydrodynamic interaction that leads to the chiral bacterial motion. At the same time, fluctuating dynamics of bacteria induce collisions with the wall, which is also a source of chiral motions of the bacteria. Experimental studies (such as Ref.~\cite{Drescher2011}) imply that the collision may be a major contribution to the chiral motion rather than the hydrodynamic interaction.

The hydrodynamic interactions between bacteria can lead to the dynamical instability of the polar or nematic alignment, which can lead to chaotic bending flows observed in experiments of the collective motion of bacteria (cf.~Fig.~\ref{example_active_matter_fig}(b)) \cite{Dunkel2013}. This bacterial turbulence, although an interesting phenomenon on its own, hinders the application of simple active fluid dynamics to predict the behavior of the collective dynamics of bacteria. Recent studies have proposed methods to control the active turbulence by using pillars or a liquid-crystal medium \cite{Nishiguchi2018, Reinken2020}. Periodically located pillars can rectify the flow of active matter and the clockwise and anti-clockwise vortices can alternately appear in the squared regions surrounded by pillars. If one uses other periodic alignments of pillars that have sublattice structures, the rectified flow can break symmetries and enable the realization of an active-matter counterpart of topological insulators discussed in Sec.~\ref{sec:5b2}. Another method uses rod-like passive particles, which will allow bacteria to align parallel or anti-parallel to the passive particles \cite{Zhou2014,Peng2016}.

Cultured cells, usually referring to mammalian-derived cells that are harvested in a dish, have been considered as model systems for active matter experiments, although 
the properties of the migration of cells and the cell-to-cell interactions are diverse.
Cells within real tissues and development have been observed to undergo collective cell dynamics~\cite{Trepat2009}, which has been explained by the combination of the cytoskeletal dynamics as well as the adhesive properties between particular cell types.

One of the early observations reported for the alignment pattern of cultured cells involved melanocytes, fibroblasts, osteoblasts, and adipocytes from humans~\cite{Kemkemer1998}.
Under high cell density in the dish, all of these bipolar-shaped cells exhibit a liquid-crystal-like nematic pattern, where the alignment interaction does not discriminate the head direction from the tail if there is any such distinction within the cells.
Similar nematic patterns have been investigated further in mouse fibroblasts~\cite{Duclos2014}, mouse myoblasts~\cite{Duclos2017,Kawaguchi2017}, as well as in mouse neural progenitor cells~\cite{Kawaguchi2017}.

The nematic pattern can be easily characterized by the topological defects observed in the cell culture. Topological defects are defective structures around which the degree of the direction of active particles changes $\pm \pi$ or more, and thus at the center, one cannot determine the direction. Since the rotation of the direction of active particles must be quantized by $\pi$, we cannot continuously remove this defect (i.e., no rotations of directions of nematics), which leads to the long-term stability of the defect with a topological origin.
Typical defects called the +1/2 and -1/2 defects, where around $n/2$ defects, the direction of active particles changes $n\pi$, can only be observed when the cell interactions are nematic.
The same topological defects have been observed in rod-like shaped cultured cells~\cite{Kemkemer1998,Duclos2017,Kawaguchi2017} as well as in epithelial cells~\cite{Saw2017} (cf.~Fig.~\ref{example_active_matter_fig}(c)), which are more isotropic and less bipolar at the single-cell level, yet can exhibit coordinated patterns that allow for defining the nematic order~\cite{Mueller2019}.
Defects in the crystal order of epithelial sheets in real animals have also been studied in Ref.~\cite{Roshal2020}.

Cell motility and proliferation in the presence of such liquid-crystal-like orientation order and topological defects are also observed in bacterial systems, and the role of topological defects in colony growth and their behavior in three-dimensional accumulation and biofilm dynamics have been reported  \cite{Copenhagen2021,Meacock2021,Shimaya2022,Prasad2023}.
Bacteria have also been used in Ref.~\cite{Nishiguchi2017}, which compares experimental results to models such as the Vicsek model with polar interactions modified to nematic interactions, and studies extracting the time evolution equation of active nematic systems from the dynamics of bacterial populations have been reported~\cite{Li2019,Colen2021}.

Although the effective nematic interaction of the cells is easy to deduce, the detailed mechanism behind the nematic alignment is difficult to probe. It has been shown by experiments and numerical simulation that self-propelled bipolar-shaped particles can exhibit large regions of nematic pattern simply through the alignment induced by collision~\cite{Narayan2007, Shi2013}.
For the nematic cell alignment, however, it has been proposed that non-local alignment interaction mediated by the extracellular matrix on the substrate is required to establish long-range ordering~\cite{Li2017}.
The detail of the contact interaction between the cells can also depend on the cell type, as they can express distinct patterns of adhesion molecules.
It has been shown, for example, that the knockdown of one of the key components of cell adhesion can switch the direction of the collective motion of the topological defects in epithelial cells~\cite{Balasubramaniam2021}.

From the theoretical point of view, such nematic interactions can enrich the dynamical phase, which is not restricted to a ferromagnetic phase but can exhibit nematic collective motion \cite{Abkenar2013,Doostmohammadi2018}. Of course, the other interactions, such as hydrodynamic and repulsive ones, can have nonnegligible effects, which may lead to effective polar interactions. Furthermore, one can control the strength of the interactions and activity by tuning the circumstances, e.g., the concentration of oxygens \cite{Sokolov2012}. Thus, cells and bacteria are highly controllable active matter, which may imitate various aspects of active systems.

\subsubsection{Molecular motors}\label{sec:2a3}
Active matter has been also studied at the molecular scale, which consists of components that are orders of magnitude smaller than cells and bacteria. Molecular motors \cite{Mogilner2002} are typical examples of such microscopic active matter, which are abundant in cells. Specifically, two types of molecular motors and associated biological filaments are frequently used to study active matter dynamics; myosins and actin filaments \cite{Mizuno2007}, and kinesins and microtubules \cite{Sanchez2012}. Myosins create a power of muscles by pulling filaments in opposite directions. On the other hand, kinesins walk on the microtubules and carry cargo that is attached to their heads. The kinesin's walk can be regarded as an active particle on a one-dimensional chain and analyzed in some particle-based models, such as the one similar to the asymmetric exclusion process \cite{Leduc2012}, which can be utilized as a typical setup to study non-Hermitian topology discussed in Sec.~\ref{sec:3}. 

While molecular motors are prominent examples of active particles, filaments attached to these motors are also typically used to study the collective motion as active matter \cite{Schaller2010}. The motility assay is a typical setup of such experiments of active systems where molecular motors are fixed on a substrate and microtubules or actin filaments are free to move (cf.~Fig.~\ref{example_active_matter_fig}(d)). This setup allows for the observation of the motor proteins' ability to transport the filaments by moving their heads.  From the coarse-grained view, the collective motion of such carried filaments is active rods. It is noteworthy that the motion of the filaments often becomes chiral, i.e., circular motions corresponding to the chirality of the molecules \cite{Schaller2010,Sumino2012}.

The collective dynamics and pattern formation of cytoskeletons have been long studied in the context of biophysics and nonequilibrium physics \cite{Ndlec1997,Surrey2001}.
Two-dimensional patterns of purified cytoskeletal and molecular motors prepared at high concentrations were subsequently developed~\cite{Sanchez2011}, where the motion of topological defects in active liquid crystal by microtubules and kinesin was analyzed for a 2D situation~\cite{Sanchez2012} and also within vesicles~\cite{Keber2014}.

\subsubsection{Other systems}\label{sec:2a4}
To study the dynamics of active matter, many studies have also utilized physical setups using non-biological active particles or agents. One straightforward way to realize active matter is using robots \cite{Bayindir2016,Fruchart2021}. Although the dynamics and interactions can be directly programmed in robots, there is often the problem of preparing them in macroscopic numbers since sophisticated robots tend to be expensive. However, they are still useful to examine topological physics as discussed in Sec.~\ref{sec:5c1}, because topological phenomena tend to be robust against the finite-size effect and seen in a small system.

Another common approach to realizing active matter involves the use of artificial, non-biological particles driven by external fields. This approach offers a controlled and scalable platform for studying large-scale active matter behavior, making it a valuable tool for investigating the fundamental principles governing these systems. These particles can be broadly categorized into those that are driven by electric fields, heat, mechanical vibrations, as well as chemical reactions.

An example of particles driven by electric fields is the quincke rollers, which are insulating particles subjected to electric fields above a critical strength. Quincke rollers undergo sustained movement due to the creation of a tilted electric dipole around the particle, breaking front-back symmetry and interacting with the field \cite{Bricard2013}. These dipoles also lead to polar interactions between the particles, favoring polar alignment. Another example of particles driven by external fields involves rod-shaped grains subjected to external vibrations \cite{Narayan2007,Blair2003,Aranson2003,Kumar2014}. These rods exhibit active fluctuations and nematic interactions, mimicking self-propelled rods and exhibiting a similar self-correlation.

Janus particles, named after the two-faced Roman god, possess distinct surface properties on their opposing hemispheres. This difference in reactivity allows them to interact with the surrounding medium or external fields in an asymmetric manner, leading to self-propulsion. For example, Janus particles with different heat-absorbing properties can be propelled by laser irradiation, creating a temperature gradient that drives thermo-osmotic flow \cite{Jiang2010,Buttinoni2012}. Similarly, applying AC electric fields to Janus particles can induce charge electrophoresis, leading to their self-propulsion \cite{Nishiguchi2018,Gangwal2008,Nishiguchi2015,Iwasawa2021}.

\begin{figure}[]
\centering
\includegraphics[width=7cm]{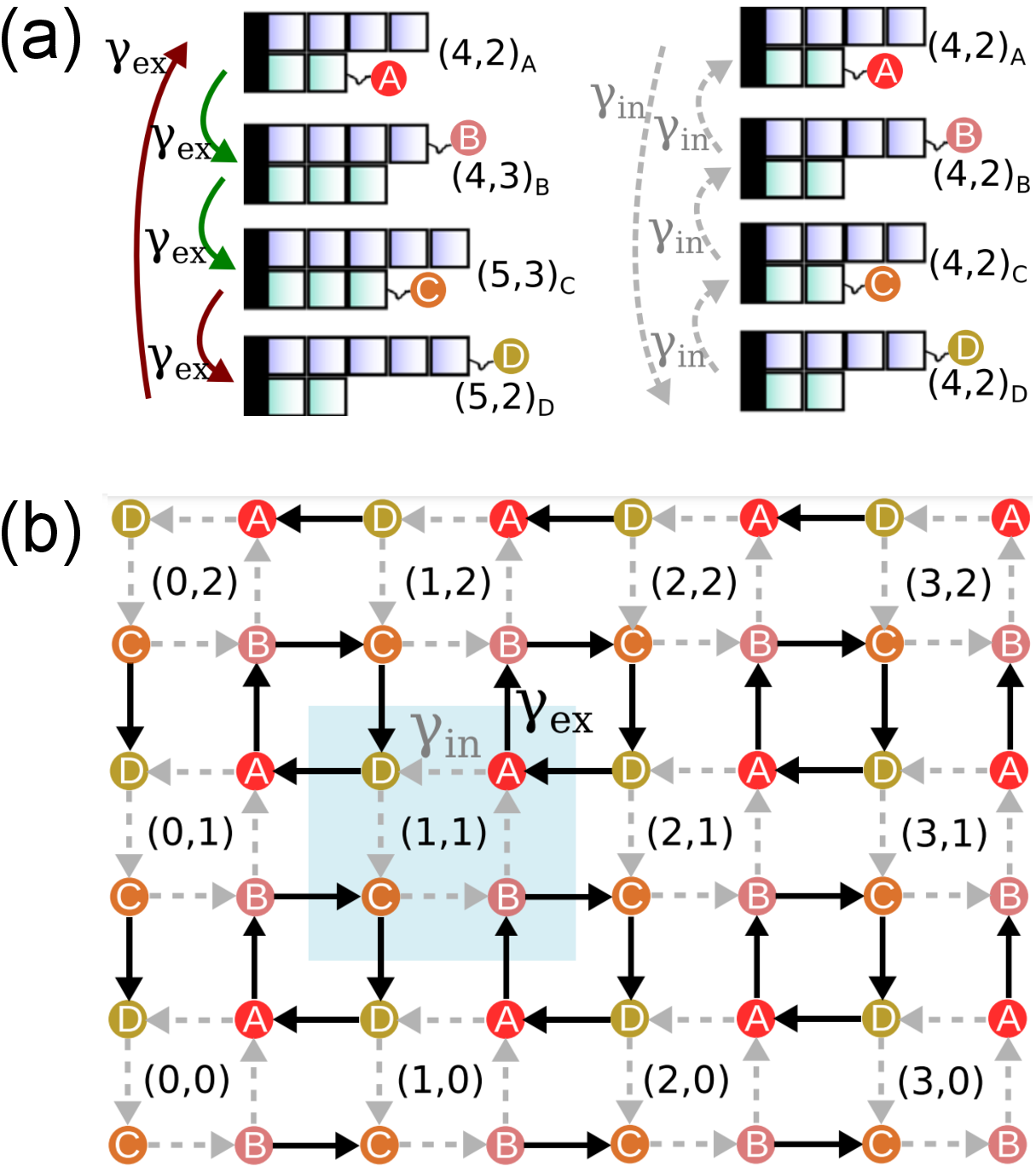}
\caption{\label{chemical_EvelynTang_fig} (a) Four-state model of a chemical compound. The compound is composed of two types of monomers and takes four internal states labeled A, B, C, and D. If these reactions can be regarded as a single molecule reaction, the rates of reactions are proportional to the concentration of reactants, whose coefficients are $\gamma_{\rm in}$ or $\gamma_{\rm ex}$. (b) Schematic of the reaction network of the four-state model. The blue square shows the unit cell of this model and the numbers show those of monomers attached to the chemical compound at each state.  These figures are adapted from E. Tang, J. Agudo-Canalejo, and R. Golestanian, ``Topology Protects Chiral Edge Currents in Stochastic Systems'' Phys. Rev. X 11, 031015 (2021) \cite{Tang2021}, licensed under a Creative Commons Attribution 4.0 International License (http://creativecommons.org/licenses/by/4.0/).}
\end{figure}

\subsection{Biochemical network}\label{sec:2b}
Activity of biological systems is supported by various biochemical reactions, which can be regarded as active systems themselves. To realize the nontrivial functions of cells, biological systems combine several chemical reactions and utilize the dynamical or statistical features of the network of such chemical reactions. 
For example, in kinetic proofreading \cite{Hopfield1974,Murugan2012}, biological molecules (cf.~proteins and tRNAs) are chemically synthesized or modified in multiple steps. The length of the synthesized polymer or the level of chemical modification can be used as an effective position, and the chemical reaction network becomes a one-dimensional ladder. Such a network is reminiscent of lattice models in condensed matter physics, and thus we can apply the notion of topological insulators to chemical reaction networks.

More specifically, the dynamics of (bio)chemical reactions is described by the rate equations, which are nonlinear equations of the concentrations of chemical species. In general, the rate equations are denoted by
\begin{equation}
 \frac{d \mathbf{n} }{dt} = S \mathbf{f} (\mathbf{n}), \label{chemical_rate_equation}
\end{equation}
where each component of $\mathbf{n} = (n_1,n_2,\cdots,n_N)^T$ represents the concentration of a chemical species, and $\mathbf{f} (\mathbf{n}) = (f_1(\mathbf{n}),f_2(\mathbf{n}),\cdots,f_M(\mathbf{n}))$ is the nonlinear vector function whose $j$th components are determined by the reaction rate of the $j$th chemical reaction. $S$ is a stoichiometric matrix whose $j$th row represents the numbers of molecules changed under the $j$th chemical reaction. In most cases, $f_j(\mathbf{n})$ is a nonlinear function of $\mathbf{n}$, while if the chemical reaction can be regarded as an effectively unimolecular reaction, $f_j(\mathbf{n})$ is proportional to the concentration of one chemical species $f_j(\mathbf{n}) \propto n_k$. Therefore, assuming all the reactions can be regarded as unimolecular reactions, one can obtain a linear rate equation and construct an analogy between chemical reactions and quantum mechanics \cite{Tang2021} (Fig.~\ref{chemical_EvelynTang_fig}). In fact, a previous study \cite{Murugan2017} has discussed the topological origin of localization in chemical reactions as we will review in Sec.~\ref{sec:5d}.

One can also obtain linear dynamics by considering the stochastic dynamics of chemical reactions, which takes the fluctuation of chemical reactions into account under the small numbers of molecules. By focusing on the probabilistic distribution of the set of the numbers of molecules, one can obtain chemical master equations \cite{McQuarrie1967,Gillespie1992}. The chemical master equations are linear equations of the distribution functions as in conventional master equations of stochastic dynamics. Because of the linearity of the chemical master equations, it should be possible to make an analogy to quantum dynamics but so far topological analysis of chemical reactions via the chemical master equations is lacking.

\subsection{Active hydrodynamics}\label{sec:2c}
\subsubsection{Toner-Tu model}\label{sec:2c1}
As mentioned in the previous section theoretical studies of active matter have been motivated by particle-based numerical simulations based, for example, on the Vicsek model \cite{Vicsek1995}. The numerical studies have revealed the unique nonequilibrium feature of active matter, while analytical results are relatively scarce. Nevertheless, field-theoretic approaches can be developed to theoretically study active matter as we briefly describe here. 

Hydrodynamic equations are widely used to analyze active-matter dynamics \cite{Marchetti2013,Bar2020,Chate2020}. Soon after the pioneering work by Vicsek et al., Toner and Tu \cite{Toner1995} derived the hydrodynamic equations of the Vicsek-type active particles by maintaining all the relevant terms that are permitted under the symmetry constraint. The equations read
\begin{eqnarray}
&{}& \partial_t \rho + \nabla \cdot (\rho \mathbf{v}) = 0, \label{toner-tu-eq1}\\
&{}& \partial_t \mathbf{v} + \lambda (\mathbf{v} \cdot \nabla) \mathbf{v} + \lambda_2 (\nabla \cdot \mathbf{v}) \mathbf{v} + \lambda_3  \nabla |\mathbf{v}|^2 \nonumber\\
&=& (\alpha-\beta|\mathbf{v}|^2)\mathbf{v}-\nabla P \nonumber\\
&&+D_B \nabla (\nabla \cdot \mathbf{v}) + D_T \nabla^2 \mathbf{v} + D_2(\mathbf{v} \cdot \nabla)^2 \mathbf{v} + \mathbf{f},
\label{toner-tu-eq2}
\end{eqnarray}
where $\rho$ and $\mathbf{v}$ represent the density and mean velocity, respectively. These equations include similar terms to those in conventional hydrodynamics, i.e., the Navier-Stokes equations, while the effect of activity and symmetry breaking lead to additional terms. Especially, the term $(\alpha-\beta|\mathbf{v}|^2)\mathbf{v}$ represents the tendency to a constant speed $|\mathbf{v}| = \sqrt{\alpha/\beta}$ that has been imposed in the Vicsek model.

Toner and Tu \cite{Toner1995} also performed the renormalization analysis of the hydrodynamic equations. They first assumed a homogeneous ordered state $\rho=\rho_0$, $\mathbf{v}=v_0 \mathbf{e}_x$, where $\mathbf{e}_x$ is the unit vector in the $x$ direction. Then, they considered the fluctuation from the ordered state, $\delta\rho = \rho - \rho_0$ and $\delta\mathbf{v} = (\sqrt{v_0^2-v_{\perp}^2}-v_0,v_{\perp})$, where $v_{\perp}$ is the velocity perpendicular to the direction of the steady-state velocity. The hydrodynamic equations can be modified into the equations of $\rho$ and the scalar version of $v_{\perp}/v_0$ (denoted as $h$) as follows:
\begin{eqnarray}
&{}& \partial_t \delta\rho + \rho_0\nabla_{\perp}^2 h + \lambda \nabla_{\perp} \cdot (\delta \rho \nabla_{\perp} h) = 0, \label{linear-toner-tu-analysis1}\\
&{}& \partial_t h + \frac{\lambda}{2} |\nabla_{\perp} h|^2 = - \nabla_{\perp} P + D_{\perp} \nabla_{\perp}^2 h + D_{\parallel} \partial_{\parallel}^2 h + \eta. \label{linear-toner-tu-analysis2}
\end{eqnarray}
Lastly, they conducted the dynamical renormalization analysis of this linearized model and concluded that the long-range order observed in active matter should survive in the infinite-size system. Unfortunately, the error was found in their earlier analysis by Toner themselves \cite{Toner2012}. The missing term in the Toner-Tu equations can be relevant, and thus the stability of the long-range order is marginal. However, this renormalization group analysis stimulates the subsequent theoretical studies of active hydrodynamics.

Numerical studies of the hydrodynamic equations have also reproduced the nonequilibrium features of active matter \cite{Bertin2009,Mishra2010}. Specifically, the active hydrodynamics exhibits a flocking-like behavior, i.e., ordered cluster formation. Furthermore, phase separations and band formulations observed in Vicsek-type models are also observed in hydrodynamics. Based on these theoretical and numerical results, the Toner-Tu equations are believed to capture some universal properties of active matter and widely used to study the phases and dynamics of active matter. In particular, most of the theoretical proposals of topological active matter are based on the Toner-Tu equations or their variants, as is discussed in Sec.~\ref{sec:5}.

Nevertheless, we mention that the hydrodynamic equations are not perfect; they fail to describe some features observed in experiments or even in numerical simulations of particle models. For example, the Vicsek model can selectively show micro-phase separation with band-shaped aggregations with a constant width \cite{Gregoire2004,Solon2015}. However, the hydrodynamics cannot explain the selectivity of the width. In more realistic situations, active fluids often show chaotic dynamics called active turbulence \cite{Wensink2012}. To analyze such turbulence-like phenomena, previous research has phenomenologically introduced a higher-order term of the Swift-Hohenberg-type \cite{Dunkel2013,Wensink2012,Cross1993} that is proportional to the fourth-order derivative of the velocity field, $(\nabla^2)^2 \mathbf{v}$. The effect of this correction term has not been investigated in topological active matter and thus should be an important issue to be solved for discussing topology of real biological active matter.

\subsubsection{From micro to hydrodynamic models}\label{sec:2c2}
While Toner and Tu derived the hydrodynamic equations phenomenologically, i.e., by only considering the symmetry of the particle model, one may also microscopically derive the hydrodynamics of active matter from the Boltzmann equations. In particular, by using the Boltzmann equations, one can reveal the relationships between the coefficients of the hydrodynamic equations and the parameters of particle models \cite{Chate2020,Bertin2006,Peshkov2014,Patelli2019}. In experimental setups, one can manipulate only the microscopic parameters of active particles, and the relationships between the hydrodynamic parameters and the microscopic parameters are important to bridge the analytical studies of active matter and the experimental results.

Another advantage of this derivation is that one can extend this method to any types of active-particle models and thus can systematically derive the corresponding hydrodynamics even if the model is too complicated to phenomenologically specify the leading-order terms. In fact, a previous study on topological active matter \cite{Sone2020} derived the hydrodynamics from a particle model and then discussed its connection to topological physics, which we introduce later in Sec.~\ref{sec:5c2}.

Since the Vicsek model turns out to be too complicated to directly derive the hydrodynamic equations, a simplified model has been used to derive the hydrodynamics via the Boltzmann-equation approach \cite{Bertin2006}. Such a simplified particle model includes the constant speed, alignment interactions, and fluctuations of the orientations that are seen in the Vicsek model, but ignores the three- or more-body interaction and considers a simplified alignment interaction. 

Starting from the simplified particle model, one can derive the dynamics of the distribution function of active particles. Generally, such differential equations of the distribution functions do not have a closed form. Thus, one should further assume the molecular chaos approximation, where the many-body distribution functions can be described by the products of the one-body distribution functions. Then, one can obtain the following Boltzmann equation \cite{Bertin2006,Peshkov2014},
\begin{eqnarray}
&{}&\partial_t f + v_0 \mathbf{n} \cdot \nabla f \nonumber \\
 &=& D_0 \Delta f + D_1 g^{\alpha\beta} \partial_{\alpha}\partial_{\beta} f + I_{dif} [f] + I_{col} [f], \label{Boltzmann_eq}
\end{eqnarray}
where $f$ is the one-body distribution function and 
\begin{eqnarray}
 &{}& I_{dif} [f] = \nonumber\\ &{}& 
 -\lambda f + \lambda \int d\theta' \int d\eta P_{\sigma}(\eta) \delta_{m\pi} (\theta'+\eta-\theta) f(\mathbf{r},\theta',t),\nonumber \\
\end{eqnarray}
\begin{eqnarray}
 I_{col} [f] &=& -f(\mathbf{r},\theta,t) \int d\theta' K(\theta'-\theta) f(\mathbf{r},\theta',t) \nonumber\\
 &{}& + \int d\theta_1 \int d\theta_2 \int d\eta P_{\sigma}(\eta) K(\theta_2-\theta_1) \nonumber\\
 &{}&\times f(\mathbf{r},\theta_1,t) f(\mathbf{r},\theta_2,t) \delta_{m\pi} (\Phi(\theta_1,\theta_2)+\eta-\theta)\nonumber\\ \label{collision_integral}
\end{eqnarray}
represent the dissipation due to the fluctuations of the orientation and the effect of two-body alignment interactions, respectively.  We consider the Fourier transform of $f$,
\begin{equation}
 f (\mathbf{r},\theta,t) = \frac{1}{2\pi} \sum_{k=-\infty}^{\infty} f_k (\mathbf{r},t) e^{-ik\theta},
\end{equation}
which leads to 
\begin{eqnarray}
 &{}& \frac{\partial f_k}{\partial t} + \frac{v_0}{2} (\nabla f_{k-1} + \nabla^{\ast} f_{k+1}) \nonumber\\
 &=& -(1-P_k)f_k + D_0 \Delta f_k + \frac{D_1}{4} (\nabla^2 f_{k-2} + (\nabla^{\ast})^2 f_{k+2}) \nonumber\\
 &{}& + \sum_{q=-\infty}^{\infty} (P_k I_{k,q} - I_{0,q} ) f_q f_{k-q}.
\end{eqnarray}
Here, we use the notation 
\begin{eqnarray}
I_{k,q} = (1/2\pi) \int d \Delta K(\Delta) e^{-iq\Delta +ikH(\Delta)},
\end{eqnarray} 
and $P_k = \int d\eta P_{\sigma}(\eta) e^{ik\eta}$. We also use $\nabla$ and $\nabla^{\ast}$ to represent the spatial derivative, $\nabla f = \partial_x f + i\partial_y f$ and $\nabla^{\ast} f = \partial_x f - i\partial_y f$. Finally, keeping the relevant terms, the equations corresponding to $k=0$ and $k=1$,
\begin{equation}
 \frac{\partial\rho^i}{\partial t} + v_0 {\rm Re} (\nabla^{\ast} f_1) = D_0 \Delta \rho + \frac{D_1}{2} {\rm Re} (\nabla^{\ast 2} f_2)
\end{equation}
\begin{eqnarray}
 &{}& \frac{\partial f_1}{\partial t}+ \frac{v_0}{2} (\nabla \rho + \nabla^{\ast} f_2) \nonumber\\
 &=& [P_1-1 + (P_1 I_{1,0} - I_{0,0} + P_1 I_{1,1} - I_{0,1} ) \rho ] f_1 \nonumber\\
 &{}& + D_0 \Delta f_1 + \frac{D_1}{4} \nabla^2 f_1^{\ast} \nonumber\\
 &{}& + (P_1 I_{1,2} - I_{0,2} + P_1 I_{1,-1} - I_{0,-1}) f_2 f_1^{\ast} + \cdots,
\end{eqnarray}
are equivalent to the hydrodynamic equations of dry aligning dilute active matter (see the next section for the definition of dry active matter).

While the Boltzmann approach can reproduce the Toner-Tu type hydrodynamics, it depends on the simplified model that ignores the three-body interactions and assumes molecular chaos. Several alternative approaches have been proposed to analytically derive the hydrodynamics of active particle models by using, e.g., the clustering kinetics \cite{Peruani2006,Harvey2013}. Another potentially promising direction is the computational approach using recent machine-learning techniques~\cite{Cichos2020}. Some studies \cite{Li2019,Colen2021,Supekar2023} have succeeded in identifying the hydrodynamic parameters from the results of the particle simulations or experiments. These approaches may overcome the imperfection of the conventional active hydrodynamics discussed in the previous section.

\begin{figure}[]
\centering
\includegraphics[width=8cm]{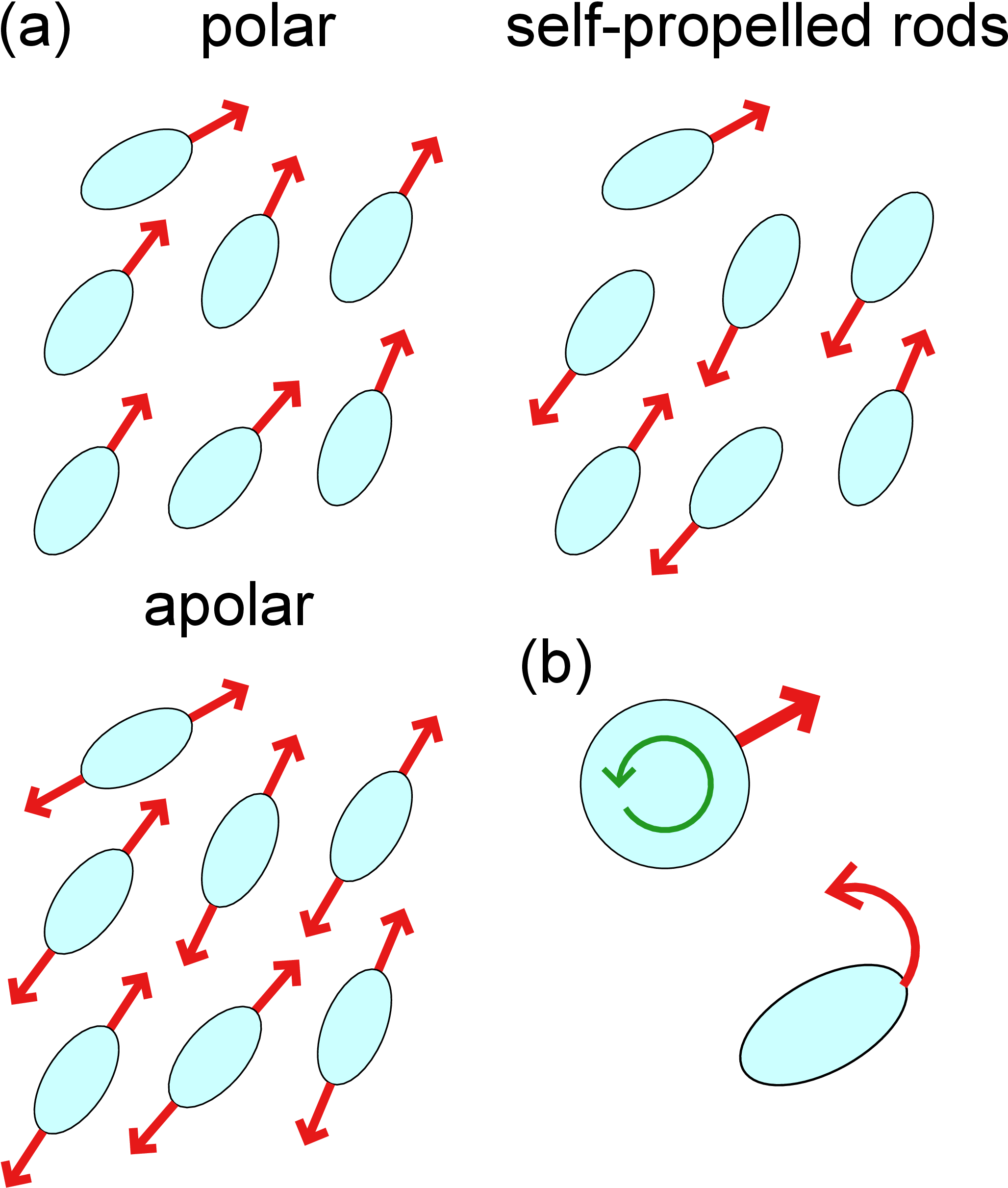}
\caption{\label{actmat_type_fig} (a) Classification of active matter by its symmetry and interaction. Polar active matter shows a directional motion and polar (i.e., ferromagnetic) interaction. Self-propelled rods also exhibit directional motions, while they only have aliging (nematic) interaction. There also exists apolar active matter that can move both forward and backward. (b) Chiral active particles. The chirality of active matter can be induced by self-rotation and/or curved motions.}
\end{figure}

\subsubsection{Hydrodynamics of various self-propelled particles}\label{sec:2c3}
While the pioneering work by Vicsek and coworkers studied self-propelled particles with polar (i.e., ferromagnetic) interactions, active matter can have various symmetries in its interaction and motion as shown in Fig.~\ref{actmat_type_fig}(a). A commonly appearing example is the self-propelled rods \cite{Bar2020}, which move unidirectionally but with nematic interactions, favoring both parallel and antiparallel states. These self-propelled rods can model the motion of, e.g., bacteria. Another type of active matter is active nematics \cite{Doostmohammadi2018}, which are typically seen in cultured cells that undergo flipping in their direction of motion. 

Since the modification of the interaction and motion of active matter constituents can alter the symmetry of the system, the modification also affects active hydrodynamics. In self-propelled rods and active nematics, there exist additional slow parameters, which are called the nematic tensor or the Q-tensor, described by a symmetric matrix as
\begin{eqnarray}
Q=\frac{S}{2}\left(
  \begin{array}{cc}
   \cos 2\theta & \sin 2\theta \\
   \sin 2\theta & -\cos 2\theta 
  \end{array}
  \right),
\end{eqnarray}
where $\theta$ is the mean orientation of the particles and $S$ represents the local nematic order. This tensor has two degrees of freedom, $S$ and $\theta$, and these two components and the density are the slow variables in the hydrodynamics of active nematics. We note that these nematic order terms correspond to the second Fourier components in the Boltzmann equation (\ref{Boltzmann_eq}). In active nematics, the symmetry prohibits the relevant velocity field and thus the hydrodynamics includes only the density and the nematic tensor. In self-propelled rods, both the velocity field and the nematic tensor are leading-order terms, and the hydrodynamic equation becomes
\begin{eqnarray}
&{}& \partial_t \rho + v_0 \nabla \cdot \mathbf{P} = 0, \label{self-propelled-eq1}\\
&{}& \partial_t \mathbf{P}  = (\alpha-\beta \tr(\tilde{Q}^2))\mathbf{P} - \frac{v_0}{2} (\nabla \rho + \nabla \cdot \tilde{Q})  \nonumber\\
&{}& + \gamma (\tilde{Q} \cdot \nabla) \cdot \tilde{Q} + \zeta \mathbf{P} \cdot \tilde{Q}, \label{self-propelled-eq2}\\
&{}& \partial_t \tilde{Q}  = (\mu-\xi \tr(\tilde{Q}^2)) \tilde{Q} - \frac{v_0}{2} \nabla \mathbf{P} + \nu \Delta \tilde{Q} + \omega \mathbf{PP}  \nonumber\\
&{}& + \tau ||\mathbf{P}||^2 Q - \chi \nabla \cdot (\mathbf{P}\tilde{Q}) - \kappa \mathbf{P} \cdot (\nabla \tilde{Q}), \label{self-propelled-eq3}
\end{eqnarray}
where $\mathbf{P}$ is the momentum field divided by the velocity of each active particle $v_0$, and $\tilde{Q}$ is the multiple of the density and the nematic tensor $\tilde{Q}=\rho Q$. As in polar active matter, the hydrodynamics includes activity terms $ (\alpha-\beta \tr(\tilde{Q}^2))\mathbf{P}$ and $(\mu-\xi \tr(\tilde{Q}^2)) \tilde{Q}$ \cite{Marchetti2013,Bar2020}.

Classification of active matter can be done from another point of view, namely, ``dry'' or ``wet'' active matter \cite{Marchetti2013,Bar2020}. A dry system refers to active particles that break momentum conservation by, e.g., ignoring hydrodynamic interactions via fluids surrounding active particles, which is mostly the case for cultured cells. A wet system, on the other hand, refers to a case where hydrodynamic interactions are significant and thus one needs to consider momentum-conserving hydrodynamics, such as in bacterial turbulence. So far, we have mainly discussed the hydrodynamics of dry active matter, while the hydrodynamics of wet active matter has been also derived in Ref.~\cite{Marchetti2013}. Such hydrodynamic equations include the active stress tensor that the self-propulsion of each active particle generates, which can be approximated by a force dipole (Stokeslet) far from each particle \cite{Drescher2011,Lauga2020}. Adding such hydrodynamic interaction makes the dynamics of wet active matter more difficult to analyze and calculate, but can reproduce more realistic behaviors in wet active matter \cite{Reinken2018}.

Active matter can also exhibit rotating motion, which further breaks the left-right symmetry of particles (cf.~Fig.~\ref{actmat_type_fig}(b)). Such chiral active matter can be found in various situations, such as bacteria \cite{DiLuzio2005} and molecular motors \cite{Sumino2012} (cf. Secs.~\ref{sec:2a2} and \ref{sec:2a3}). In hydrodynamics, the chirality of active particles adds two relevant terms, a Coriolis-like force and odd viscosity \cite{Banerjee2017,Furthauer2012}. The Coriolis-like force is described as $\omega_0 \nabla \times \mathbf{v}$, where $\omega_0$ is determined from the frequency of the self-rotation of the chiral active particles. This force acts on active matter perpendicularly to the velocity field in a similar manner to the conventional Coriolis force. The other chirality term, odd viscosity $\nu^o \nabla^2 \nabla \times \mathbf{v}$, is an unconventional viscosity that is antisymmetric under the time reversal. This term does not lead to energy dissipation but induces the Hall-like linear response perpendicular to the pressure. Both Coriolis-like force and odd viscosity play an essential role in topological active matter as discussed in Sec.~\ref{sec:5b}.

%
% -------------------------------------------------------------------------------------------------------------------------------------------------------------------------------------
%

\section{Hermitian band topology}\label{sec:3}
\subsection{Band theory and its classical counterpart}\label{sec:3a}
Historically, early studies of topological materials were motivated by electronic phenomena found in semiconductors \cite{Klitzing1980}, such as the quantum Hall effect. In general, the dynamics of an electron in a solid can be described by the Schr{\"o}dinger equation under a periodic potential \cite{Kittel2018},
\begin{eqnarray}
i\hbar\frac{\partial}{\partial t} \psi(\mathbf{r}) = H(\mathbf{r}) \psi(\mathbf{r}) := \left[ -\frac{\hbar^2}{2m} \nabla^2 + V(\mathbf{r}) \right] \psi(\mathbf{r}), \label{Schroedinger_eq}\\
V(\mathbf{r}+\mathbf{a}_i) = V(\mathbf{r}),\ \ (i\in\{1,2,3\}),
\end{eqnarray}
where $\mathbf{a}_i$ are lattice vectors that represent the translation symmetry of the crystalline structure of the solid. The overall Hamiltonian also satisfies the discrete translation symmetry 
\begin{equation}
H(\mathbf{r}+\mathbf{a}_i) = H(\mathbf{r}). \label{discrete-trans-sym}
\end{equation}
Assuming periodically oscillating state $\psi(\mathbf{r};t)=e^{-iEt}\psi(\mathbf{r})$, one also obtains the time-independent Schr{\"o}dinger equation $E\psi(\mathbf{r}) = H(\mathbf{r})\psi(\mathbf{r})$, which is an eigenvalue problem of the linear operator $H$. In the following, we focus on this time-independent Schr{\"o}dinger equation and discuss topological structures of eigenvalues and eigenvectors.

A Hamiltonian in solid-state physics is an operator that acts on a high-dimensional vector space, or equivalently a large-scale matrix, while the Bloch theorem helps us to reduce it to a small-scale operator by separating each wavenumber sector. We consider a translation-invariant Hamiltonian (\ref{discrete-trans-sym}) and introduce the translation operators $T_i$. Equation~(\ref{discrete-trans-sym}) then corresponds to the commutativity between the translation operators and the Hamiltonian. By utilizing this commutativity, the Bloch theorem shows that eigenfunctions of a Hamiltonian with discrete translation symmetries are also eigenfunctions of the translation operators and thus can be described as 
\begin{equation}
\psi(\mathbf{r}) = e^{i\mathbf{k}\cdot \mathbf{r}} u_{\mathbf{k}}(\mathbf{r}), \label{Bloch_wave}
\end{equation}
where $u_{\mathbf{k}}(\mathbf{r})$ is a translation-symmetric function $u_{\mathbf{k}}(\mathbf{r}+\mathbf{a}_i) = u_{\mathbf{k}}(\mathbf{r})$ and $\mathbf{k}$ is a corresponding wavenumber that is the inverse of the wavelength of the eigenfunction. By substituting the eigenvalue equation of the translation-symmetric Hamiltonian, we obtain the following reduced eigenvalue equation of each wavenumber sector,
\begin{eqnarray}
E(\mathbf{k}) u_{\mathbf{k}}(\mathbf{r}) &=& \tilde{H}_{\mathbf{k}}(\mathbf{r}) u_{\mathbf{k}}(\mathbf{r}) \nonumber\\
&{:=}& \left[ -\frac{\hbar^2}{2m} (\nabla + i\mathbf{k})^2 + V(\mathbf{r}) \right] u_{\mathbf{k}}(\mathbf{r}), \label{Bloch_Hamiltonian}
\end{eqnarray}
where $\tilde{H}_{\mathbf{k}}(\mathbf{r})$ is the Hamiltonian of each wavenumber sector that is called the Bloch Hamiltonian. We note that $u_{\mathbf{k}}(\mathbf{r})$ is defined on a unit cell of a periodic structure, and thus we only need to calculate the eigenvector in a smaller space than that used in the original eigenvalue equation.

By diagonalizing the Bloch Hamiltonian, one can obtain the dispersion relation $E_i(\mathbf{k})$, where $i\in\{1,2,\cdots, N\}$ represents the index of the dispersion. Such a dispersion relation, which forms a continuous curve as a function of the wavenumber $\mathbf{k}$, is called a band structure and plays crucial roles in condensed matter physics because it allows one to extract various information about materials such as the response coefficients.

Techniques of the Bloch Hamiltonian and the band structures are also useful in classical physics including active fluid. Basically, the Bloch Hamiltonian corresponds to the matrix obtained via the linear stability analysis of the Fourier modes in fluids \cite{Yang2015}. Then, its dispersion relation can be considered as a band structure of fluid. In particular, when all the Fourier modes are marginally stable, one obtains real dispersion relation and thus can assume the matrix obtained via the linear stability analysis as an effective (quasi-)Hermitian Hamiltonian. Recent studies \cite{Rechtsman2013,Khanikaev2013,Ozawa2019,Ma2016,Cummer2016} have utilized the notions in condensed matter physics to develop metamaterials that control the dynamics of photonic and phononic systems. Metamaterials typically possess a periodic structure of waveguides, which has a clear analogy to the periodic alignment of atoms in crystalline solids. Utilizing this method to connect the linear stability analysis of fluid and condensed matter physics, topological active matter is also proposed as we discuss later in Sec.~\ref{sec:5}.

In some classical systems including active fluids, one may consider a spatially uniform system, instead of one with discrete translation symmetry. In that case, we consider an effective Hamiltonian satisfying $H(\mathbf{r})=H(\mathbf{r}')$ for any pair of locations $(\mathbf{r},\mathbf{r}')$, instead of Eq.~(\ref{discrete-trans-sym}). Then, its eigenvector must be a plain wave $\psi(\mathbf{r}) = e^{i\mathbf{k}\cdot \mathbf{r}} u_{\mathbf{k}}$, which looks similar to Eq.~(\ref{Bloch_wave}) but has no location dependence because of the spatial uniformity. One can obtain the corresponding Bloch Hamiltonian just by replacing the derivative $\nabla$ in $H$ with the wavenumber $\mathbf{k}$. The difference between discrete and continuous periodicity is found in the range of the wavenumber; under discrete translation symmetry, there are an infinite number of equivalent wavenumbers because $e^{i\mathbf{k}\cdot \mathbf{a}_i}=e^{i\mathbf{k}'\cdot \mathbf{a}_i}$ in the case of $(\mathbf{k}-\mathbf{k}')\cdot \mathbf{a}_i = 0\  ({\rm mod}\,2\pi)$ and thus by taking $u(\mathbf{k}') = e^{i(\mathbf{k}-\mathbf{k}')\cdot \mathbf{r}}u(\mathbf{k})$ the wavefunctions $\psi(\mathbf{r})$ in Eq.~(\ref{Bloch_wave}) represent the same one. Therefore, one should only consider a restricted region of the wavenumber space that is called the Brillouin zone. In contrast, since we consider $u_{\mathbf{k}}$ without location dependence in a spatially uniform system, one should consider any real vectors $\mathbf{k}$. This difference induces a subtle difference in geometrical properties of the wavenumber space (cf.~Sec.~\ref{sec:5a1}). While in the remaining part of this section and the next section, we mainly focus on systems with the discrete translation symmetry, notions of band topology can also be applied to spatially uniform systems as we discuss in Sec.~\ref{sec:5}.

We also note that there is a subtle difference between quantum and classical systems when one considers the statistical properties of quantum many-body systems. In fermionic systems such as electrons in solids, each eigenstate is occupied by at most one electron because of Pauli's exclusion principle. Then, in a ground state of the many-body system, energy bands are occupied from the lowest-energy one, and the highest energy occupied is called the Fermi energy. In conventional topological insulators, we consider the case that the band structure opens a bulk band gap at the Fermi energy and discuss the existence of gapless edge modes at that gap as we discuss in the following. In contrast, classical systems cannot define the Fermi energy due to the absence of the fermionic statistics. Instead, one can discuss the existence of topological gapless modes in any bulk band gaps. To observe such topological gapless modes, one should excite them by applying oscillating external force with corresponding frequencies.

%
% -------------------------------------------------------------------------------------------------------------------------------------------------------------------------------------
%

\subsection{Hermitian topology}\label{sec:3b}
\subsubsection{Quantum Hall effect}\label{sec:3b1}
Topology is a property of matter that is invariant under a continuous deformation. For example, considering a genus, the number of holes, we cannot either increase or decrease the genus without cutting the geometry or sticking different surfaces together. Thus, the genus is a topological invariant, and we cannot continuously transform objects with different genera into each other, as exemplified by a ball and a donut. Condensed matter physics employs the notion of topology into a wavenumber space by considering the topology of band structures \cite{Hasan2010,Qi2011}. To discuss the topology of band structures, one must define the prohibited deformation of the Hamiltonian. In topological insulators, we first consider an energy gap at the Fermi energy, where no bands exist. Then, we exclude the deformation under which band structures cross the Fermi energy. Under such continuous deformation (i.e., adiabatic deformation), the bulk is always insulating. Then, one can find defective structures in a wavenumber space, and the number of such defects must be preserved under the continuous deformation in a similar manner as in the preservation of the genus. Thus, insulating phases of matter can be classified by such a topological number, which leads to the notion of topological insulators. Below we shall delineate the definition and physical meanings of the defective structures in band structures.

\begin{figure}[]
\centering
\includegraphics[width=8cm]{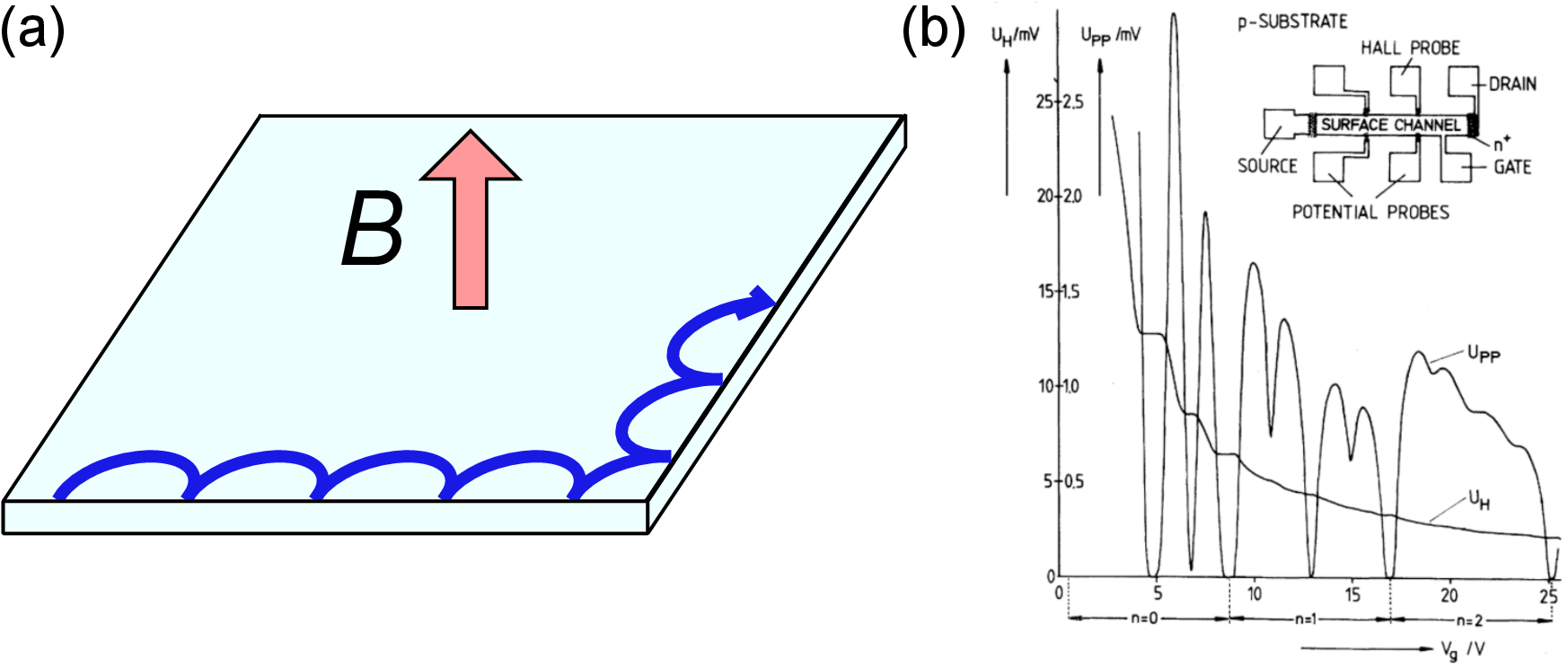}
\caption{\label{QHE_fig} Quantum Hall effect. (a) Schematic of the quantum Hall effect. Cycrotolon motions of electrons are prevented at the edge of the sample, and instead, the skipping orbit can emerge, which induces the chiral edge current. (b) Experimental results of the quantized Hall voltage. The inset shows the top view of the device where the quantized Hall conductance was first observed. Panel (b) is adapted from K. v. Klitzing, G. Dorda, and M. Pepper, ``New Method for High-Accuracy Determination of the Fine-Structure Constant Based on Quantized Hall Resistance'' Phys. Rev. Lett. 45, 494 (1980) \cite{Klitzing1980}, Copyright \copyright\  1980 by the American Physical Society.}
\end{figure}

A prototypical example of topological materials is a quantum Hall system, which was experimentally discovered by von Klitzing et al.~in 1980 \cite{Klitzing1980}. They measured the Hall conductivity in semiconductors under the existence of the external magnetic field, and found its quantization. We first introduce some basic notions to understand this experiment. Under a uniform magnetic field perpendicular to a (effectively) two-dimensional semiconductor $\mathbf{B} = (0,0,B)$, the Hamiltonian is described as
\begin{eqnarray}
i\hbar\frac{\partial}{\partial t} \psi(\mathbf{r}) &=& H(\mathbf{r}) \psi(\mathbf{r}) \nonumber\\
&{:=}& \left[ -\frac{\hbar^2}{2m} \left(\nabla-\frac{ie}{\hbar c}\mathbf{A}(\mathbf{r}) \right)^2 + V(\mathbf{r}) \right] \psi(\mathbf{r}). \label{Schroedinger_eq2}
\end{eqnarray}
by using the vector potential $\mathbf{A}(\mathbf{r}) = (0,Bx,0)$ (we here adopt the Landau gauge). Solving the eigenequation of this Hamiltonian, one obtains discrete eigenenergies like the Hamiltonian under a harmonic potential, which is called the Landau level. Meanwhile, a uniform magnetic field induces the Lorentz force perpendicular to the velocity of an electron. As one can infer from classical electromagnetics, electronic and Lorentz forces are balanced when the electron moves perpendicular to both of them, and thus the corresponding current, i.e., the Hall current occurs. The Hall conductivity is defined as the Hall current divided by the electric field $\sigma_{xy} = j_y / E_x$. As we will discuss below, since each Landau level contributes to the Hall conductivity in a quantized way, the quantum Hall effect appears.

Von Klitzing et al. tuned the Fermi energy by changing the gate voltage (voltage between the gate and source in the inset of Fig.~\ref{QHE_fig}(b)) and obtained the contributions of different Landau levels. If the Fermi energy exists between Landau levels, the longitudinal conductivity becomes zero, and the Hall conductivity takes a constant value until the Fermi energy is increased to the higher Landau level. Thus, the Hall conductivity exhibits plateaus as shown in Fig.~\ref{QHE_fig}. Furthermore, the Hall conductivity is proportional to the number of occupied Landau levels and thus is quantized. 

Thouless et al.~\cite{Thouless1982} revealed that the quantization of the Hall conductance originates from the nontrivial topology of the wave functions. Their theoretical analysis starts from the Hamiltonian of an electron under the external magnetic field in Eq.~(\ref{Schroedinger_eq2}). 
%\begin{eqnarray}
%i\hbar\frac{\partial}{\partial t} \psi(\mathbf{r}) &=& H(\mathbf{r}) \psi(\mathbf{r}) \nonumber\\
%&{:=}& \left[ -\frac{\hbar^2}{2m} \left(\nabla-i\frac{e}{\hbar c}\mathbf{A} \right)^2 + V(\mathbf{r}) \right] \psi(\mathbf{r}). \label{Schroedinger_eq2}
%\end{eqnarray}
Here, we assume that the Fermi energy exists in the band gap. By considering the uniform electronic field $\vec{E}=(E,0)$ as a perturbation and using the Kubo formula, one can calculate the Hall current $\langle j_y \rangle = \langle e[\hat{y},H] \rangle$ at zero temperature \cite{Ando2013}. This leads to the following expression of the Hall conductance
\begin{eqnarray}
\sigma_{xy} = \frac{\langle j_y \rangle}{E} = \frac{e^2}{h}C, \label{quantized_hall_conducatance} \\
C = \sum_{n:{\rm filled}}C_n, \label{filled_Chern}
\end{eqnarray}
where $e$ and $h$ represent the elementary charge and the Planck constant. $C_n$ is the so-called Chern number (or TKNN number) defined as
\begin{eqnarray}
C_n &=& \frac{1}{2\pi} \int _{\rm BZ} \mathbf{B}_n (\mathbf{k}) \cdot d\mathbf{S},\label{chern} \\
\mathbf{B}_n (\mathbf{k}) &=& \nabla_{\mathbf{k}} \times \mathbf{A}_n (\mathbf{k}), \\
\mathbf{A}_n (\mathbf{k}) &=& i\mathbf{u}_n(\mathbf{k}) \cdot (\nabla_{\mathbf{k}} \mathbf{u}_n(\mathbf{k})).
\end{eqnarray}
with $\mathbf{u}_n(\mathbf{k})$ being the eigenvector of the $n$th band of the Bloch Hamiltonian $H(\mathbf{k})$. The Chern number $C_n$ can be defined for any gapped (i.e., nondegenerate) band and must be an integer. Since the Chern number is also a continuum functional of the eigenvectors $\mathbf{u}_n(\mathbf{k})$, the Chern number is unchanged under the continuous deformation of the Hamiltonian until the band structure becomes gapless.

While the Chern number and the quantized Hall conductance are the bulk properties of topological materials, nontrivial band topology also accompanies unconventional localized modes at the edge of the sample. The appearance of such edge modes can be intuitively understood as follows; since electrons under a uniform magnetic field exhibit cyclotron motions, each electron circularly moves in the bulk of the quantum Hall system. Meanwhile, near the boundary of the sample, the circular motions are interfered by the potential barrier, and instead, electrons bounce at the boundary. The iteration of the curved motions and the bounds leads to a skipping orbit of an electron (cf.~the blue curved allows in Fig.~\ref{QHE_fig}(a)), where electrons unidirectionally flow along the edge of the sample. Therefore, the quantum Hall system exhibits edge-localized modes that exhibit unidirectional edge flow. We note that a similar boundary flow can occur in rotating active particles, which is an origin of nontrivial topology in chiral active matter in Sec.~\ref{sec:5b1}.

This intuitive picture of the emergence of the edge modes can be justified by various theoretical arguments. One of the explanations is the argument proposed by Laughlin \cite{Laughlin1981}. He considered electrons in a cylindrical system where a magnetic flux is inserted inside of the cylinder. During the insertion of the magnetic field, an induced electronic field emerges in the circumferential direction. Such an induced electronic field leads to a Hall current from one edge to the other. The insertion of a flux quantum $\Delta \phi = h/e$ maps the system back into that without a magnetic flux, and thus the number of transferred electrons must be an integer $n$. Then, the current is assessed by $I=neE/(\Delta\phi)=ne^2E/h$, where $neE$ is the energy increase by the magnetic flux. One can further show that the integer $n$ must be equal to the Chern number of the system \cite{Hatsugai1993b,Hatsugai1993}. Since the insertion of the magnetic field increases and decreases the number of electrons at the edge of the sample, there must be gapless modes that exist around the Fermi energy. Since the bulk bands remain gapped, eigenfunctions of such gapless modes are localized at the edge of the sample.

In general, the correspondence between the bulk topological invariant and the gapless boundary modes is believed as the bulk-boundary correspondence. To mathematically formulate this in two-dimensional systems, one considers the left semi-infinite boundary condition, under which the open boundary exists at $x=0$, and the system extends to the $x>0$ region and is periodic in the $y$ direction. Then, the bulk-boundary correspondence indicates
\begin{equation}
 C = n_L - n_R,
\end{equation}
where $C$ is the sum of the Chern numbers of occupied bands in Eq.~(\ref{filled_Chern}) and $n_L$ and $n_R$ are the numbers of gapless edge modes with left and right chirality (i.e., propagating in the upper and lower direction), respectively. 
One can also extend this bulk-boundary correspondence to general dimensions and various boundary conditions, and in many cases, the bulk-boundary correspondence is mathematically proved.

Since topology is unchanged under continuous deformations of Hamiltonian, the associated phenomena are also robust against disorders. In fact, if one calculates the eigenstates under the existence of perturbative random on-site potentials $H_{\rm topo}+\Delta(x)$, one still obtains localized zero modes. We note that the Chern number in Eq.~(\ref{chern}) is defined in a system with discrete translation symmetries, while a disordered system has no translation symmetries. Nevertheless, one can still define the Chern number in disordered systems by utilizing twisted boundary conditions \cite{Niu1985}. The noncommutative geometry \cite{Bellissard1994} has also revealed the real-space representation of the Chern number and shown its correspondence to the quantized Hall conductance in disordered two-dimensional systems. In this way, the bulk-boundary correspondence has been mathematically proved even under the existence of disorders, which guarantees the robustness of the topological edge modes.

As we see in Eq.~(\ref{quantized_hall_conducatance}), the quantized Hall conductance and associated gapless edge modes are induced by the nontrivial topology of band structures. As such, topologically nontrivial insulators can exhibit the quantum Hall effect without external magnetic fields provided that the bands possess nontrivial topology, which is known as the quantum anomalous Hall effect \cite{Haldane1988}. Since the chiral edge current in edge modes of a quantum Hall system breaks the time-reversal symmetry, the quantum anomalous Hall effect requires broken time-reversal symmetry by utilizing {\it internal} magnetic fields. Such internal magnetic fields have been experimentally realized by using ferromagnetic insulators, such as Cr-doped Bi2Se3 and MnBi2Te4 thin flake \cite{Chang2023}.

\begin{figure}[]
\centering
\includegraphics[width=8cm]{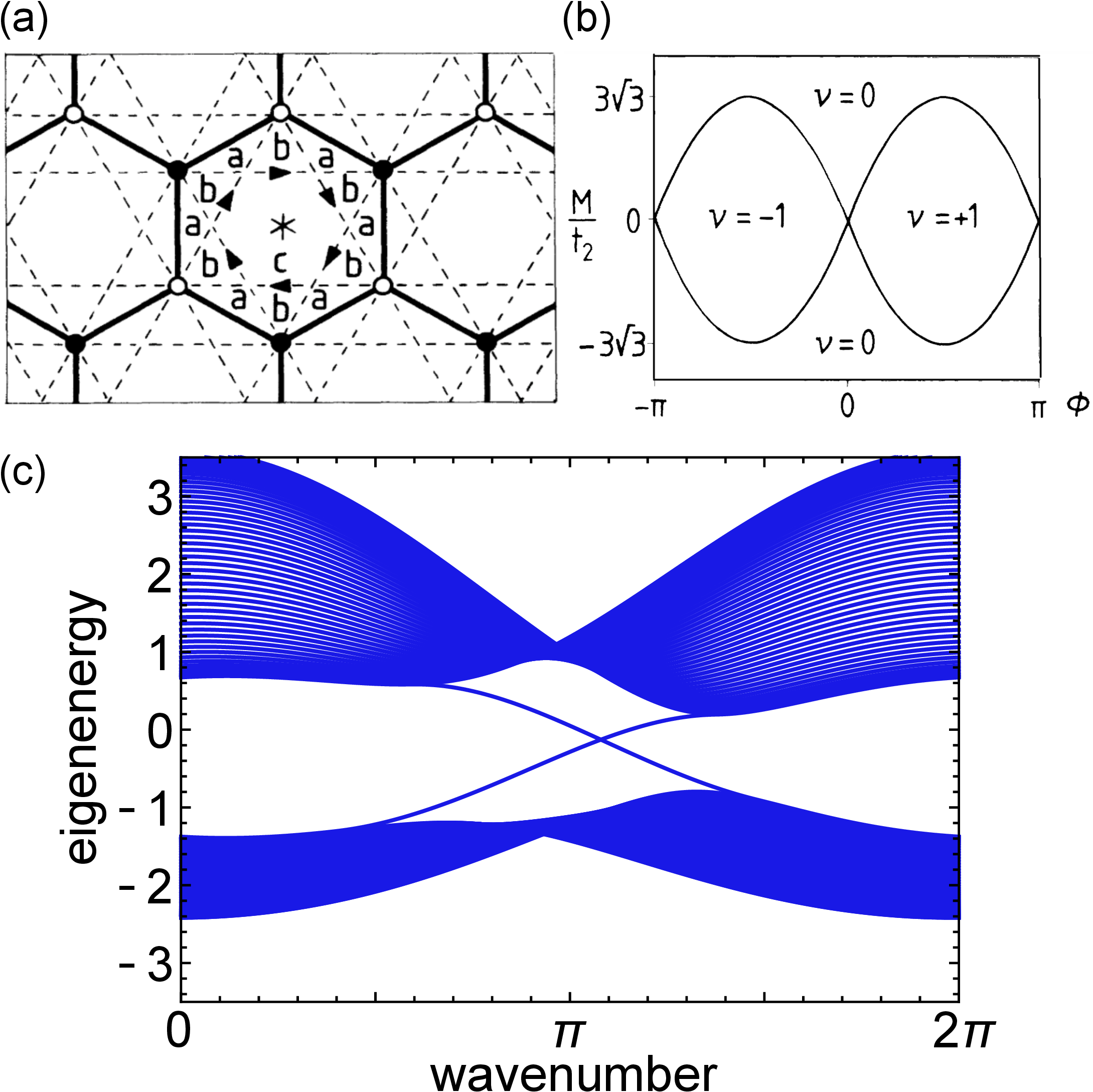}
\caption{\label{Haldane_fig} Haldane's honeycomb model. (a) Schematic of the Haldane model is shown. The arrow represents the direction of the gauge field, i.e., the vector potential. (b) The phase diagram of the Haldane model. $\nu$ corresponds to the Chern number in each phase. The panels (a) and (b) are adapted from F. D. M. Haldane, ``Model for a Quantum Hall Effect without Landau Levels; Condensed-Matter Realization of the ``Parity Anomaly''' Phys. Rev. Lett. 61, 2015 (1988) \cite{Haldane1988}, Copyright \copyright\  1988 by the American Physical Society. (c) We can obtain gapless edge modes at the zigzag boundary of the Haldane model. }
\end{figure}

To demonstrate the quantum anomalous Hall effect in a concrete setting, Haldane \cite{Haldane1988} introduced a honeycomb-lattice model. Before moving to the details of the model, we shall mention the role of lattice models in condensed matter physics. As described in the previous section, the Hamiltonians of electrons in condensed matter are continuous operators, whose eigenequations are in general hard to solve. Meanwhile, lattice Hamiltonians are often solvable by analytical or numerical calculations because one can assume it to be a matrix. Condensed matter physics often uses such lattice Hamiltonians as simple models. If one assumes that electrons are bound to atomic orbitals in most of the time and sometimes jump to neighbor ones, the lattice Hamiltonian can be a good approximation of the continuous Hamiltonian with periodic potentials generated by the atoms in a solid. This correspondence between continuous and discrete models is analogous to the stochastic dynamics of passive and active colloids \cite{VanKampen1992,Klumpp2016,dAlessandro2021}, which can be described by both Langevin equations and Markov jump processes.

The honeycomb-lattice Haldane model basically imitates the electrons in graphene, while it also introduces an internal magnetic field represented as next-nearest complex-valued hoppings with phase factors. Specifically, the lattice Hamiltonian is described as 
\begin{eqnarray}
H_{\rm Haldane} &=& \sum_{\mathbf{r}} (-1)^{\sigma(\mathbf{r})} M | \mathbf{r} \rangle \langle \mathbf{r}| + \sum_{\langle \mathbf{r},\mathbf{r}' \rangle} (t_1 | \mathbf{r} \rangle \langle \mathbf{r}' | + {\rm H.c.}) \nonumber\\
&{}& + \sum_{\langle\langle \mathbf{r},\mathbf{r}'' \rangle\rangle} (t_2 e^{i\phi} | \mathbf{r} \rangle \langle \mathbf{r}'' | + {\rm H.c.}), \label{Haldane_model}
\end{eqnarray}
where $\langle \mathbf{r},\mathbf{r}' \rangle$ and $\langle\langle \mathbf{r},\mathbf{r}'' \rangle\rangle$ represent the pairs of nearest and next-nearest neighbor sites, respectively, and ${\rm H.c.}$ represents the Hermitian conjugate of the previous term. $t_1$, $t_2$ are real-valued parameters determining the hopping amplitudes, and $\phi$ is the phase gained via the hopping to the next-nearest site. $\sigma(\mathbf{r})$ determines the sign of the on-site potential and depends on the sublattice. Such phase factor plays the role of a vector potential (in other words, a gauge field) in a lattice system, and the direction of the effective vector potential is shown in Fig.~\ref{Haldane_fig}(a). Since the vorticity of the vector potential $\nabla\times \mathbf{A}$ corresponds to the magnetic field, one can calculate the local flux in the Haldane model by counting the number of winding of the arrows in the figure. Then, the flux with a strength $-B$ appears around each lattice point, while each hexagonal plaquette has $2B$ flux inside. Therefore, the sum of the magnetic fluxes in each unit cell is zero, which means the absence of the external magnetic field. The internal magnetic field, however, breaks the time-reversal symmetry as one can see by confirming that the reversal of the arrows does not reproduce the same figure as in Fig.~\ref{Haldane_fig}(a), which leads to the chiral edge current of the quantum anomalous Hall effect.

One can numerically confirm the nontrivial Chern number and the existence of the associated edge modes in the quantum anomalous Hall system. On the one hand, to calculate the Chern number, one should obtain the Bloch Hamiltonian of the model (\ref{Haldane_model}) and numerically diagonalize it. Substituting the obtained eigenvectors into Eq.~(\ref{chern}), one can calculate the Chern number of the Haldane model as is shown in Fig.~\ref{Haldane_fig}(b). On the other hand, to confirm the existence of the gapless edge modes, one should consider a supercell system that aligns $L$ unit cells of the Haldane model in the $x$ direction and imposes the periodic boundary condition in the $y$ direction. Then, assuming it as an effectively one-dimensional system and deriving the Bloch Hamiltonian, one can calculate the band structure that has gapless edge modes whose energies lie in the bulk band gap as in Fig.~\ref{Haldane_fig}(c).

In this section, we have introduced key concepts in band topology, i.e., topological invariants such as the Chern number and their bulk-boundary correspondence. We note that both the Chern number and corresponding edge modes are properties of eigenvectors and thus are not related to the quantumness of condensed matter. Thus, one can extend these notions to classical systems including active matter by considering effective Hamiltonians via the linear stability analysis. We will discuss this point later in Sec.~\ref{sec:5}.

%
% -------------------------------------------------------------------------------------------------------------------------------------------------------------------------------------
%

\subsubsection{Band topology protected by symmetries}\label{sec:3b2}
\begin{figure}[]
\centering
\includegraphics[width=8cm]{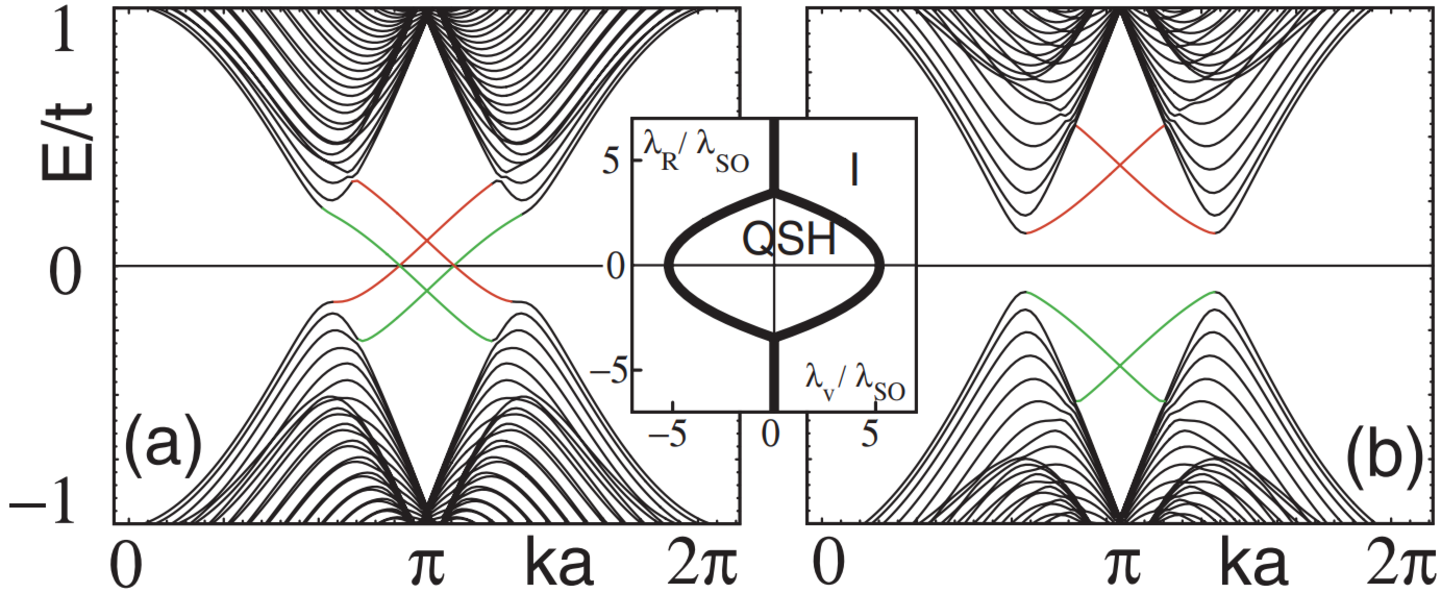}
\caption{\label{Kane_Mele_fig} The band structure and phase diagram of the Kane-Mele model. The inset shows the phase diagram of the model, where the vertical axis represents the nearest neighbor Rashba term that introduces spin couplings and breaks mirror symmetry, and the horizontal axis represents a staggered potential. The topological phase (a) exhibits gapless dispersions. The colors of gapless bands show at which side the gapless modes are localized. In contrast, the nontrivial phase (b) exhibits a gapped band structure, and thus topological modes are absent. These figures are adapted from C. L. Kane and E. J. Mele, ``$Z_2$ Topological Order and the Quantum Spin Hall Effect'' Phys. Rev. Lett. 95, 146802 (2005) \cite{Kane2005b}, Copyright \copyright\  2005 by the American Physical Society.}
\end{figure}

While the quantum Hall effect requires the breakdown of the time-reversal symmetry by magnetic flux because of the chiral edge current, the presence of symmetries can enrich the topological phases of matter. One typical example of such symmetry-protected topological phenomena is the quantum spin Hall effect \cite{Kane2005a,Kane2005b}. In the quantum spin Hall systems, one assumes the time-reversal symmetry, which implies that the exchange between spin-up and spin-down electrons only reverses the direction of motion (this situation is natural in materials without internal and external magnetic fields). Then, both spin-up and spin-down electrons can be localized at the edge of the sample and exhibit helical edge currents that flow in the opposite direction. Without symmetry protections, perturbative interactions between different spins can open the energy gap and break the gapless modes, while the time-reversal symmetry protects edge modes because it guarantees that edge current and its time-reversal counterpart must appear in a pair.

To theoretically show the possibility of the quantum spin Hall effect, Kane and Mele proposed a honeycomb-lattice system that imitates electrons in graphene \cite{Kane2005a,Kane2005b}. The lattice model includes nearest-neighbor hoppings and next-nearest-neighbor hoppings as in the Haldane model, while the phase obtained via the next-nearest-neighbor hoppings have opposite signs depending on the direction of spin of an electron. The Hamiltonian of this model takes the form of the combination of the Haldane model and its time-reversal counterpart,
\begin{equation} 
H = \left(
  \begin{array}{cc}
   H_{\rm Haldane} & C \\
   C^{\dagger} & H^{\ast}_{\rm Haldane}
  \end{array}
  \right),
\end{equation} 
where $C$ introduces interactions between spin-up and spin-down electrons that preserve the time-reversal symmetry. Calculating the dispersion relation of this Hamiltonian under the open boundary condition in the $x$ or $y$ direction, one can confirm the existence of a pair of gapless edge modes localized at the same side of the open system (cf.~Fig.~\ref{Kane_Mele_fig}). Since each gapless band supports a unidirectional current along the edge of the sample, one can observe a helical current under the open boundary conditions, which is a characteristic behavior of the quantum spin Hall effect. We note that the spins of electrons may not be fully polarized at each gapless band (i.e., the eigenstate at each gapless band is not an eigenstate of the spin operator $S_z$) because of the spin coupling term $C$, and thus the spin Hall conductivity is not quantized. However, both the quantum Hall effect and quantum spin Hall effect still share similar topological origins and topological edge modes.

\begin{figure}[]
\centering
\includegraphics[width=8cm]{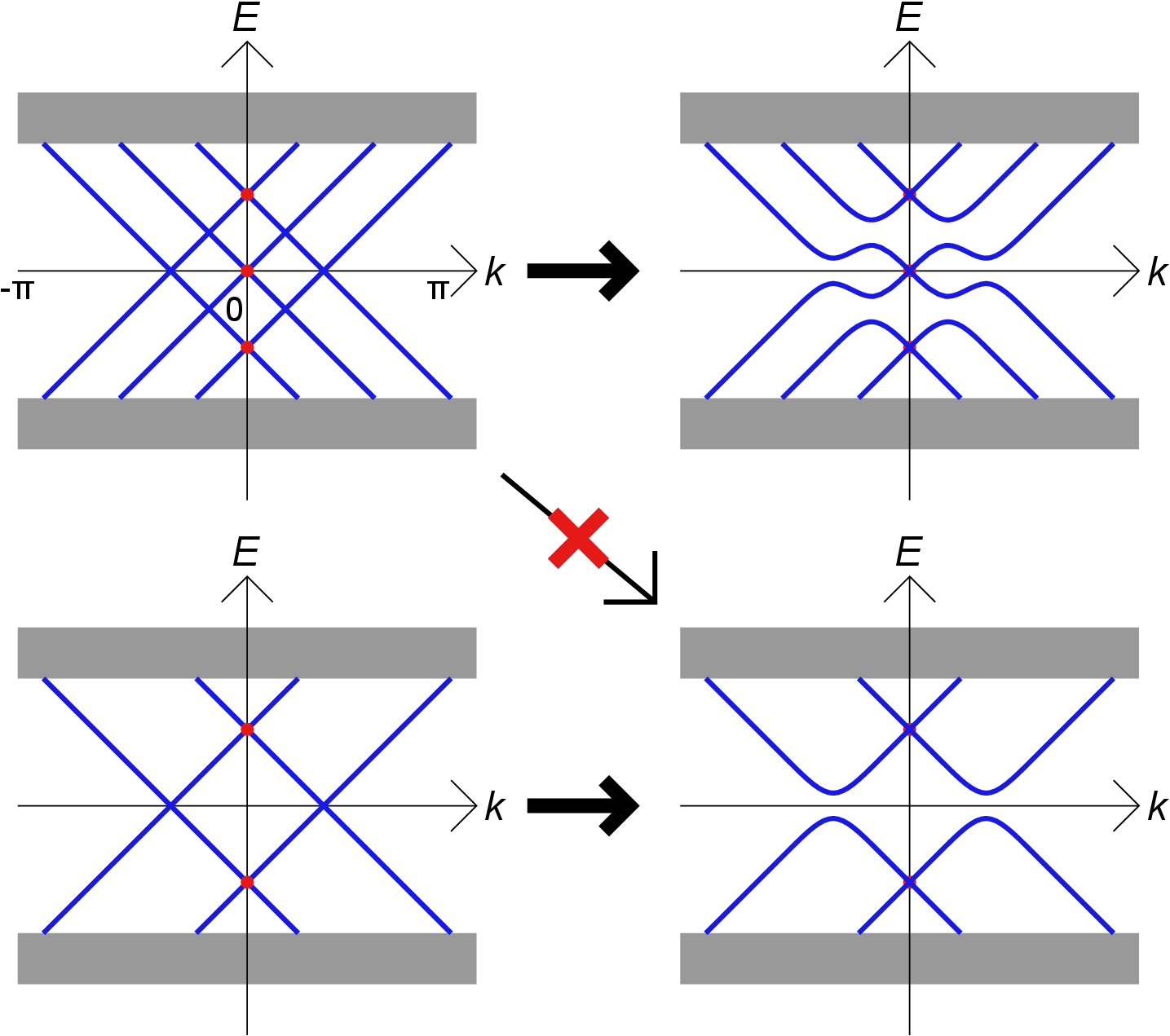}
\caption{\label{Z2protection_fig} Topological protection in the quantum spin Hall system. The gray squares represent the bulk bands, and the blue curves are dispersions of edge modes. The degeneracies at the time-reversal symmetric moment, $k=0,\pm \pi$ (the red points) are protected by the time-reversal symmetry (i.e., the Kramers degeneracy). Then, the parity of the number of gapless edge modes cannot be changed by the continuous deformation, which is the origin of $Z_2$ protection of the quantum spin Hall system.}
\end{figure}

We shall explain the mechanism of the symmetry protection in the quantum spin Hall effect in more detail. Using a unitary matrix $T$, the time-reversal symmetry is described as $TH(\mathbf{k})T^{-1}=H^{\ast}(-\mathbf{k})$ with $H(\mathbf{k})$ being the Bloch Hamiltonian. If we focus on time-reversal invariant momenta where $\mathbf{k}_{\rm TRIM}$ and $-\mathbf{k}_{\rm TRIM}$ are equivalent in the Brillouin zone, the eigenvalues of $H(\mathbf{k}_{\rm TRIM})$ must be degenerate, which is known as the Kramers degeneracy. Thus, dispersions of a pair of gapless edge modes must cross at a time-reversal invariant momentum as shown in Fig.~\ref{Z2protection_fig}, which protects the edge modes from gap opening under continuous deformations of band structures. Meanwhile, when the band structures have two pairs of gapless modes, they can open a bandgap at wavenumbers which are variant under the time reversal. Therefore, topology of the quantum spin Hall system depends on whether the number of gapless modes is odd or even \cite{Fu2007}. This $Z_2$-type classification is different from the topological classification of quantum Hall systems that are simply characterized by the number of gapless modes.

\begin{figure}[]
\centering
\includegraphics[width=8cm]{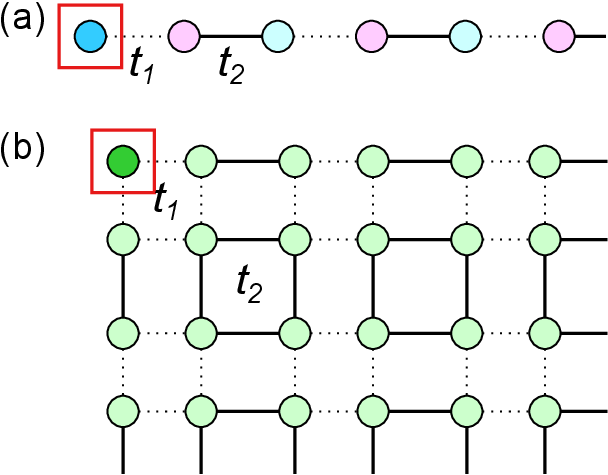}
\caption{\label{SSH_BBH_fig} Edge modes and corner modes in the SSH model and the BBH model. The black lines and dotted lines represent the nearest neighbor hoppings with strengths $t_2$ and $t_1$, respectively. (a) When we consider the fully dimerized case ($t_1=0$) in the SSH model, we can analytically obtain a zero mode localized at the left edge site surrounded by the red square. (b) In the BBH model, a corner mode appears at the site surrounded by the red square.}
\end{figure}

Topological insulators can be found in one-dimensional systems under another type of symmetry called chiral symmetry (sometimes called sublattice symmetry). The Su-Schriefer-Heeger (SSH) model is a prototypical model of such one-dimensional topological insulators, which was originally introduced to explain the electronic properties of polyacetylene \cite{Su1980} and later used as a platform to study fundamental theories of topological insulators (see Sec.~\ref{sec:4c3} examples of studies in the non-Hermitian topology). The model is a lattice model whose Hamiltonian is described as
\begin{eqnarray} 
H_{\rm SSH} = \sum_x \left( t_1 |x,B \rangle \langle x,A | + t_2|x,B \rangle \langle x+1,A | \right) + {\rm H.c.}, \nonumber \\ 
\label{SSHmodel}
\end{eqnarray} 
where $\rm H.c.$ represents the Hermitian conjugate. In general, the chiral symmetry is defined as $\Gamma H({k})\Gamma^{-1} = -H^{\dagger}({k})$ with $H({k})$ being the Bloch Hamiltonian, and $H(k)$ and $\Gamma$ are described under a proper unitary transformation as
\begin{eqnarray}
&{}& H\left( k \right) = \left( \begin{array}{cc}
O                     & h^{\dagger}(k) \\
h(k) & O
\end{array}\right),
\label{chiral_sym_BlochHamiltonian}\\
&{}& \Gamma = \left( \begin{array}{cc}
I                     & O \\
O & -I
\end{array}\right),
\label{chiral_sym_operator}
\end{eqnarray}
where $I$ represents an identity matrix, and $O$ represents a zero matrix. In the SSH model, $h(k)$ becomes a scalar function $h(k) = t_1 + t_2 e^{ik}$. Then, the topological invariant is the winding number of $h(k)$,
\begin{equation}
 w = \frac{1}{2\pi i} \int_0^{2\pi} dk \frac{d}{dk} \log (\det h(k)). \label{winding_Hermitian}
\end{equation}
In the SSH model, the winding number becomes $w=0$ for $|t_1|>|t_2|$ and $w=1$ for $|t_1|<|t_2|$. Corresponding to the nonzero winding number, localized zero modes appear at the edge of the open chain of the SSH model. The appearance of gapless edge modes is obvious in the fully dimerized case, $t_1=0$ (cf.~Fig.~\ref{SSH_BBH_fig}(a)). In this case, there exist isolated sites at the open boundaries, which are absent in the case of $t_2=0$. The fully localized modes at such isolated sites have zero eigenvalues and thus can be considered as gapless edge modes. If $t_1$ becomes nonzero but small compared to $t_2$, the interaction between dimerized pairs can still be considered as a perturbation, and gapless edge modes remain. In general chiral-symmetric systems, the winding number corresponds to the difference between the numbers of left-localized edge modes $|\psi_L \rangle$ with different chirality (i.e., $\Gamma|\psi_L \rangle = |\psi_L \rangle$ or $\Gamma|\psi_L \rangle = -|\psi_L \rangle$), which is the bulk-boundary correspondence of one-dimensional systems. We note that the winding number is equivalent to the Zak phase \cite{Zak1989,Xiao2010} of the occupied band divided by $2\pi$, which is defined as
\begin{equation}
 \frac{\phi}{2\pi} = \frac{1}{2\pi i} \int_0^{2\pi} dk\, \vec{u}(k) \cdot \left[ \partial_k \vec{u}(k)\right]  
\end{equation}
with $\vec{u}(k)$ being the eigenvector of the occupied band.

In the last part of this section, we shall briefly discuss the classification of the symmetry-protected topological phases. To discuss such topological classification, we need to introduce the particle-hole symmetry, which is defined as $CH(\mathbf{k})C^{-1} = -H^T(-\mathbf{k})$ where $H(\mathbf{k})$ is a Bloch Hamiltonian and $C$ is a unitary charge-conjugate operator. Based on the particle-hole, time-reversal, and chiral symmetry, Altland-Zirnbauer (AZ) symmetry classes \cite{Altland1997} classify the Hermitian system into 10 classes. Then, the classification table of topological phases \cite{Kitaev2009,Ryu2010} indicates the existence or absence of topological phases in each symmetry class and dimension of the system. The classification table also tells us which types of topological invariants classify the topological phases; examples include the integer type such as the Chern number and the $Z_2$ type found in the quantum spin Hall systems. The winding number discussed in the previous paragraph is also included in the classification table as the $Z$ classification of the AI\hspace{-.1em}I\hspace{-.1em}I class in one dimension. If we further consider spatial symmetries such as rotational symmetries, topological phases can be enriched \cite{Song2017,Kruthoff2017}. Another advantage of introducing spatial symmetries is that one can calculate symmetry indicators \cite{Fu2007,Po2017}, whose calculation is simpler than conventional topological invariants. 

We note that most of classical linear dynamics are described as $\partial_t \psi = A\psi$ by using a real matrix $A$ and real state vectors $\psi$. By assuming $iA$ as the effective Hamiltonian $H$, such a linear dynamics must have the particle-hole symmetry $H=-H^{\ast}$. The time-reversal and chiral symmetries also widely exist by using only reciprocal elements or preparing time-reversal counterparts (effective spins). Therefore, symmetry-protected topological phases are ubiquitous in classical systems such as active matter. In fact, we will discuss the non-Hermitian topology of active matter protected by the chiral symmetry in Sec.~\ref{sec:5c}.
%
% -------------------------------------------------------------------------------------------------------------------------------------------------------------------------------------
%

\subsubsection{Higher-order topological insulator}\label{sec:3b3}
Extending the SSH model to a two-dimensional system, one can find more exotic phases of matter where the gapless modes are localized at the corner \cite{Benalcazar2017}. Such topological insulators with corner-localized modes are called second-order topological insulators because the gapless modes appear in two lower dimensions than that of the bulk system. To construct a second-order topological insulator, here we consider $2L_y$ chains of the SSH model. The overall system has $2L_y$ edge modes localized at the left and right edge for each. Then, introducing SSH-like staggered interaction into these chains of the SSH model, the $2L_y$ edge modes are coupled in the same manner as in the SSH model. Therefore, we obtain ``edge modes'' of the edge modes at the upper and lower sides of the system, which are localized at the corners of the stacked chains. The stacked chains of SSH models are a typical model of a second-order topological insulator protected by the chiral symmetry and called the Benalcazar-Bernevig-Hughes (BBH) model \cite{Benalcazar2017}. One can also intuitively understand the emergence of gapless corner modes by considering the disconnected case as in the SSH model (cf.~Fig.~\ref{SSH_BBH_fig}(b)). In general, if a Bloch Hamiltonian is described as $H(\mathbf{k}) = H_{\rm topo}(k_x) \otimes \Gamma_y + I \otimes H'_{\rm topo}(k_x)$ with $H_{\rm topo}$, $H'_{\rm topo}$ being Hamiltonians of chiral symmetric topological insulators and $\Gamma_y$ being the chiral operator satisfying $\Gamma_y H'_{\rm topo} \Gamma_y^{-1}=-H'_{\rm topo}$, it is a second-order topological insulator whose localized corner modes are tensor products of the edge modes of $H_{\rm topo}$ and $H'_{\rm topo}$, denoted by $\psi_x \otimes \psi_y$ \cite{Okugawa2019b}.

The notion of the second-order topological insulator can be further extended to higher dimensions and orders. In three dimensions, gapless hinge modes (second order) and corner modes (third order) can appear. In particular, topological hinge states have been observed in condensed matter such as bismuth \cite{Schindler2018}. Higher-order topology is also feasible in various classical systems as we will discuss later (cf. Sec.~\ref{sec:5c1}).

%
% -------------------------------------------------------------------------------------------------------------------------------------------------------------------------------------
%

\section{Non-Hermitian band topology}\label{sec:4}
\subsection{Non-Hermitian system}\label{sec:4a}
Non-Hermitian systems, namely, nonequilibrium open systems effectively described by non-Hermitian operators \cite{Ashida2020}, can exhibit a variety of phenomena without Hermitian counterparts. One of the origins of non-Hermiticity is gain and/or loss through exchanging energies with the external environment. Non-Hermitian operators can in general be nondiagonalizable, and the corresponding eigenvectors coalesce at a degenerate point of the energy eigenvalues, which is known as an exceptional point \cite{Kato1966,Bergholtz2021,Ding2022}. An exceptional point can be interpreted as a topological object in gapless non-Hermitian systems, and its topological nature can lead to anomalous bulk-edge correspondence as explained below. Another possible origin of non-Hermiticity is nonreciprocity caused by, e.g., the unidirectional flow of energies and particles, which can lead to numerous localized modes. This localization phenomenon, also known as a non-Hermitian skin effect \cite{Yao2018,Zhang2022,Okuma2022,RLin2023}, can be attributed to nontrivial non-Hermitian topology originating from a unique complex-valued energy gap. In the following sections, we shall explain the unconventional bulk-edge correspondence between a topological invariant defined by complex-valued energy bands and the aggregation of the bulk modes at the boundary. As such, studies of non-Hermitian topology might deepen our understanding of the robust behavior in nonequilibrium open systems including active matter.

Non-Hermitian physics has been experimentally studied in optics and photonics in controllable settings \cite{Feng2017,El-Ganainy2018,Miri2019,Ozdemir2019,Wang2021}. More recently, non-Hermitian systems have been also explored in the fields of atomic physics  \cite{Zhang2018} and magnonics \cite{Hurst2022,Tao2023}. Meanwhile, various classical systems, such as acoustics  \cite{Ma2016,Cummer2016,Zangeneh2019,Gu2021}, can also emulate non-Hermitian models to realize and observe various types of non-Hermitian phenomena. In recent years, the concept of non-Hermitian topology has also found applications to active matter; owing to its inherently nonequilibrium nature, active matter can naturally be described as non-Hermitian systems as detailed below (cf.~Sec.~\ref{sec:5c}). For instance, unidirectional flow induced by self-propelled force can lead to nonreciprocity, which gives rise to non-Hermitian dynamics.

%
% -------------------------------------------------------------------------------------------------------------------------------------------------------------------------------------
%

\subsection{Point gap and line gap}\label{sec:4b}
\begin{figure}[]
\centering
\includegraphics[width=8.5cm]{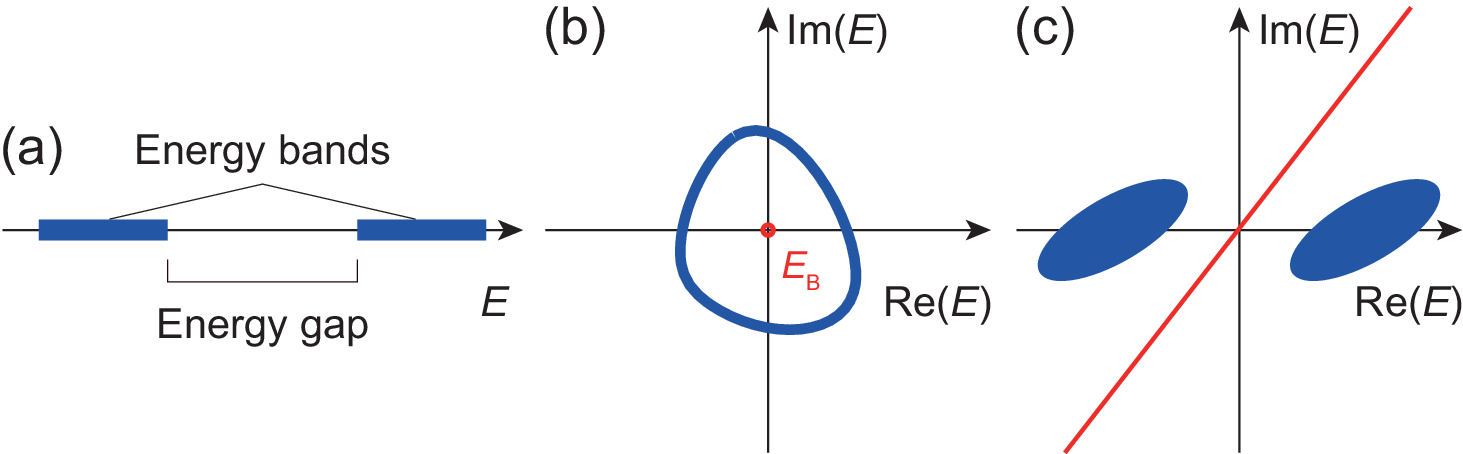}
\caption{\label{bandgapfig}Schematic figures of energy gaps. (a) Energy gap in Hermitian systems. (b) Point gap in non-Hermitian systems. The base energy $E_{\rm B}$ is located inside the energy band on the complex-valued energy plane. (c) Line gap in non-Hermitian systems. The red line divides the two sets of the energy bands.}
\end{figure}
The key ingredient of defining topological invariants is an energy gap, the range in which no energy eigenvalues exist. There is no ambiguity in the definition of the energy gap for Hermitian systems [Fig.~\ref{bandgapfig}(a)] because the energy eigenvalue is always real-valued, for which the corresponding topological invariant is well-defined as explained in Sec.~\ref{sec:3b}. Meanwhile, in non-Hermitian systems, the fact that the energy eigenvalue can take complex values leads to several possible definitions of energy gaps \cite{Ashida2020}. There are two common ways to do so; first, non-Hermitian systems are said to have a {\it point gap} when the energy eigenvalue opens a gap at the base energy on the complex-valued energy plane. As shown in Fig.~\ref{bandgapfig}(b), when the energy band forms a closed curve, the point gap opens inside and outside the curve, closing on the curve \cite{Gong2018}. Accordingly, the topological invariant related to the point gap changes its values depending on the base energy, as discussed below. The other possible definition is a {\it line gap} where two sets of energy eigenvalues are separated by a line determining the gap on the complex-valued energy plane [Fig.~\ref{bandgapfig}(c)] \cite{Kawabata2019}. The topology associated with the line gap is essentially equivalent to the Hermitian topology because any non-Hermitian Hamiltonian can be continuously deformed to a Hermitian Hamiltonian while keeping the line gap if any \cite{Ashida2020,Kawabata2019}. The line-gap topology can give rise to localization modes with discrete energy eigenvalues similar to Hermitian cases.

%
% -------------------------------------------------------------------------------------------------------------------------------------------------------------------------------------
%

\subsection{Non-Hermitian skin effect}\label{sec:4c}
\subsubsection{Non-Hermitian skin effect in one-dimensional systems}\label{sec:4c1}
\begin{figure}[]
\centering
\includegraphics[width=6.5cm]{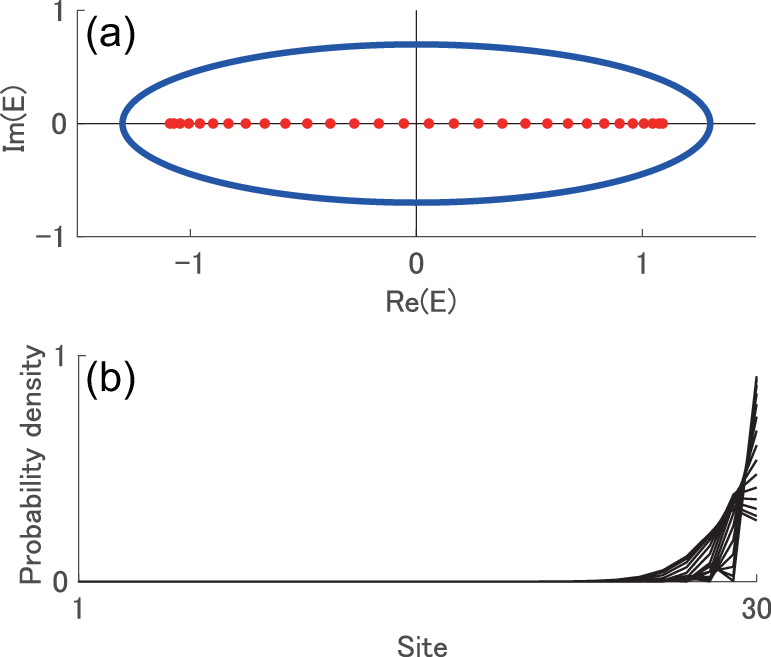}
\caption{\label{HNmodelfig}Energy eigenvalues and spatial profile of the bulk eigenstates of the Hatano-Nelson model. (a) Energy band obtained from Eq.~(\ref{HNBlochHamiltonian}) (blue) and energy eigenvalues of the finite-size system with an open boundary condition (red). (b) Localization of the bulk eigenstates. The system parameters are set to be $t_L=0.3$ and $t_R=1$.}
\end{figure}

We first illustrate the notion of non-Hermitian topology on the basis of the bulk-edge correspondence in one-dimensional (1D) non-Hermitian tight-binding systems. The key idea is that the winding of the complex-valued energy band can characterize the point-gap topology \cite{Borgnia2020,Okuma2020,Zhang2020}. Specifically, for a non-Hermitian Bloch Hamiltonian denoted by $H\left(k\right)$, the topological invariant detecting the point-gap topology is defined as
\begin{equation}
W_{\rm P}=\frac{1}{2\pi i}\int_0^{2\pi}dk\frac{d}{dk}\log\det\left[H\left(k\right)-E_B\right],
\label{energywinding}
\end{equation}
which is called an energy winding number. The base energy $E_{\rm B}$ can be any one point on the complex-valued energy plane. To see a nonzero energy winding number, we consider the Hatano-Nelson model~\cite{Hatano1996}, a prototypical tight-binding model with a nontrivial non-Hermitian topology. Its real-space Hamiltonian reads
\begin{equation}
{\cal H}_{\rm HN}=\sum_x \left(t_R\left|x+1\right\rangle\left\langle x\right|+t_L\left|x\right\rangle\left\langle x+1\right|\right),
\label{HNHamiltonian}
\end{equation}
and the Bloch Hamiltonian is given by
\begin{equation}
H_{\rm HN}\left(k\right)=t_Le^{ik}+t_Re^{-ik},
\label{HNBlochHamiltonian}
\end{equation}
where $t_L$ and $t_R$ are real-valued parameters, which express asymmetric hopping amplitudes. We show the energy band of the Hatano-Nelson model in the case of $t_L<t_R$ in Fig.~\ref{HNmodelfig}(a). From Eq.~(\ref{energywinding}), one can immediately check that the energy winding number takes the value of either $0$ or $-1$ depending on whether the base energy is located outside or inside the energy band. Namely, the energy winding number changes when the point gap closes.

We next explain the physical meaning of the energy winding number. For a non-Hermitian Hamiltonian ${\cal H}$, we can define the extended Hermitian Hamiltonian as
\begin{eqnarray}
\tilde{\cal H}=\left( \begin{array}{cc}
O                    & {\cal H}-E \\
{\cal H}^\dag-E^\ast & O
\end{array}\right),
\label{extendedHH}
\end{eqnarray}
where $E$ is an arbitrary complex number. We note that $\tilde{\cal H}$ preserves the chiral symmetry expressed by
\begin{equation}
\Gamma\tilde{\cal H}\Gamma^{-1}=-\tilde{\cal H},
\label{chiralsym}
\end{equation}
where $\Gamma$ is defined in the form of Eq.~(\ref{chiral_sym_operator}). Remarkably, the classification of the point-gap topology of the non-Hermitian Hamiltonian is equivalent to the topological classification of the extended Hermitian Hamiltonian with the chiral symmetry \cite{Gong2018}. For example, since the Hatano-Nelson model has no symmetries, the extended Hermitian Hamiltonian obtained from Eq.~(\ref{HNHamiltonian}) belongs to Class AI\hspace{-.1em}I\hspace{-.1em}I in the conventional AZ symmetry class. The Hermitian topology in Class AI\hspace{-.1em}I\hspace{-.1em}I can be characterized by a $Z$ topological invariant, as discussed in the SSH model (cf. Sec.~\ref{sec:3b2}). Accordingly, the point-gap topology of the Hatano-Nelson model can also be characterized by a $Z$ topological invariant, which is nothing but the energy winding number.

Section~\ref{sec:3b} shows that the Hermitian topology allows us to establish the bulk-edge correspondence between a topological invariant defined in bulk and the existence of spatially localized in-gap states. The topological equivalence between non-Hermitian systems and extended Hermitian systems above indicates that a similar bulk-edge correspondence might be also established in non-Hermitian systems. Indeed, one can show that the nonzero energy winding number corresponds to the localization of the {\it bulk} eigenstate to the edge \cite{Okuma2020,Zhang2020}. Such an unconventional localization phenomenon is called the non-Hermitian skin effect \cite{Yao2018}. For example, in the Hatano-Nelson model, all the eigenstates are localized at one edge [Fig.~\ref{HNmodelfig}(b)] due to the nonzero winding number. The localization can be understood as the accumulation of the eigenstates under unidirectional flow owing to the asymmetric hopping amplitudes. We note that the corresponding energy eigenvalues under an open boundary condition take real values, which are distinct from the energy band obtained from the Bloch Hamiltonian~(\ref{HNBlochHamiltonian}). So far, many physical implementations of the non-Hermitian skin effect have been reported in various areas of research, such as photonics \cite{Weidemann2020,Xiao2020,Xiao2021,Gao2022,Liu2022}, ultracold atoms \cite{Liang2022}, acoustics \cite{Gu2022}, mechanics \cite{Chen2021,Brandenbourger2019,Ghatak2020}, and electrical circuits \cite{Hofmann2020,Helbig2020}. On another front, a recent previous work has proposed that the boundary sensitivity of the energy eigenvalues is applicable for highly sensitive sensors~\cite{Budich2020}.

Symmetries can enrich a classification of the point-gap topology because the topological classification of an extended Hermitian Hamiltonian depends on symmetries that the corresponding non-Hermitian Hamiltonian preserves \cite{Gong2018,Kawabata2019,Zhou2019}. As an example, we here focus on 1D non-Hermitian systems with the pseudo-time-reversal symmetry \cite{Okuma2020} defined as
\begin{equation}
{\cal T}H^{\rm T}{\cal T}^{-1}=H,~{\cal T}{\cal T}^\ast=-1.
\label{pTRS}
\end{equation}
We note that this operation is not equivalent to the conventional time-reversal symmetry because of $H^{\rm T}\neq H^\ast$ in non-Hermitian systems. This symmetry acts as an inverter of the spin direction corresponding to the self-rotation of particles, leading to the Kramers degeneracy. The corresponding extended Hermitian Hamiltonian then belongs to Class D, and it is classified by a $Z_2$ topological invariant. The physical meaning of this classification is that odd numbers of Kramers pairs appear as topological edge states. Accordingly, the $Z_2$ point-gap topology corresponds to the unconventional non-Hermitian skin effect, where the bulk eigenstates with spin-up are localized at one end, and those with spin-down are localized at the other, in contrast to the localization nature induced by the $Z$ point-gap topology. Thus, the symmetry-protected non-Hermitian skin effect can be understood as an analog of a quantum spin Hall insulator (cf. Sec.~\ref{sec:3b2}).

The point-gap topology plays a crucial role also in stochastic processes, which can be described by a master equation given as follows:
\begin{equation}
\frac{\partial{\bm P}}{\partial t}=W{\bm P},
\label{mastereq}
\end{equation}
where ${\bm P}$ and $W$ represent a vector of occupation probabilities of states and a transition matrix, respectively. We consider a 1D Markov system with a spatial translation symmetry, in which the transition matrix can be diagonalized by employing the representation $W\left(k\right)$ in the Fourier basis. One can then extract the point-gap topology of the transition matrix. For example, the transition matrix akin to the Hatano-Nelson model has a nonzero winding number. This simply means that a random walker described by the transition matrix flows in one direction due to an asymmetric hopping rate and eventually reaches a steady state which is localized at the edge. Indeed, Ref.~\cite{Dasbiswas2018} has established the correspondence of the nonzero topological invariant and the existence of the localized steady state. We will give a specific definition of the winding number and a qualitative description of the localized steady state in Sec.~\ref{sec:4c4}. Thus, the point-gap topology allows us to predict the nature of localization in stochastic processes.

%
% -------------------------------------------------------------------------------------------------------------------------------------------------------------------------------------
%

\subsubsection{Non-Hermitian skin effect in two-dimensional systems}\label{sec:4c2}
\begin{figure}[]
\centering
\includegraphics[width=8cm]{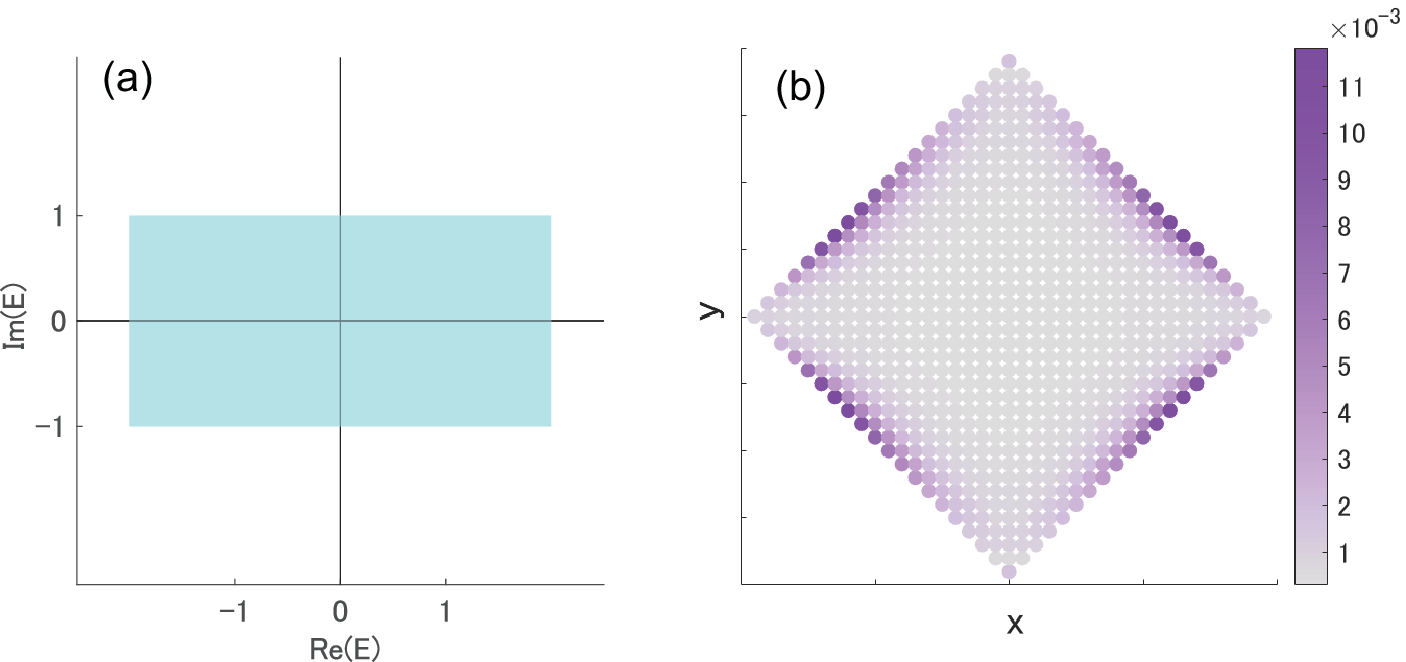}
\caption{\label{GDSEfig}Geometry-dependent skin effect of the model (\ref{GDSEBlochHamiltonian}). (a) Energy band. (b) Density of state on the diamond geometry. The bulk eigenstates are localized at the edges.}
\end{figure}
We next review the non-Hermitian topology and the associated non-Hermitian skin effect in two-dimensional (2D) non-Hermitian tight-binding systems. We first consider a 2D square-lattice system composed of horizontal and vertical Hatano-Nelson chains. One can immediately infer that this system can be characterized by two energy winding numbers defined in Eq.~(\ref{energywinding}). Indeed, the corresponding non-Hermitian skin effect localizes all the bulk eigenstates at one corner \cite{Lee2018}. The localization can be readily understood as the flow of the eigenstates toward the corner due to the asymmetric hopping amplitudes in all the directions. Such localization of the bulk eigenstate and its manipulation have been experimentally realized in acoustics \cite{Zhang2021} and photonics \cite{Lin2023}. However, the system does not exhibit phenomena unique to 2D systems in the sense that the observed non-Hermitian skin effect merely inherits 1D nature.

Recent works have pointed out that the point-gap topology corresponds not only to the Hermitian first-order topology but also to the second-order topology \cite{Kawabata2020b,Okugawa2020,Fu2021}, which we have discussed in Sec.~\ref{sec:3b3}. Specifically, one can define a topological invariant of a non-Hermitian system via an extended Hermitian Hamiltonian which exhibits a second-order topological phase. When the topological invariant takes a nonzero value, several eigenstates are then localized at the corners. This phenomenon is called a second-order topological non-Hermitian skin effect, and it is unique to 2D systems since the localization originates from second-order topology which can appear in more than 2D systems. Remarkably, it is possible to realize the second-order non-Hermitian skin effect without nonreciprocal interaction or unidirectional flow in contrast to 1D non-Hermitian systems discussed in Sec.~\ref{sec:4c1}.

The additional mechanism to localize eigenstates has been proposed by combining the point-gap topology and the first-order Hermitian topology. This type of localization is called a hybrid skin-topological effect. The key idea is to manipulate the topological edge state so that the winding number (\ref{energywinding}) takes a nonzero value. For example, in the Chern insulator (cf. Sec.~\ref{sec:3b1}), adding gain and loss to the bulk \cite{Li2022,Zhu2022,Schindler2023} or the edges \cite{Nakamura2023} induces the topological edge state localized at the corners. Additionally, Ref.~\cite{Okugawa2020} has proposed a model constructed by stacking the Hatano-Nelson model so that the net flow becomes zero and by introducing the SSH-model-like hopping amplitude along the stacking direction. As a result, the non-Hermitian skin effect localizes the topological edge states induced by the SSH-model-like hopping amplitude at the corners. Said differently, the model construction gives rise to hybridizing the topological edge states from the SSH model and the skin states from the Hatano-Nelson model. The hybrid skin-topological effect has been experimentally observed in electrical circuits \cite{Zou2021} and active matter \cite{Palacios2021} as we will review in Sec.~\ref{sec:5c1}.

Finally, we briefly review a novel non-Hermitian skin effect found in recent years. We consider a one-band model described by a Bloch Hamiltonian $H\left({\bm k}\right)$ and assume that all hopping amplitudes of the model exhibit reciprocity. Reference \cite{Zhang2022} has proposed that, for any ${\bm k}_{\rm B}$ in the Brillouin zone, one can define the winding number as follows:
\begin{equation}
W_{\rm 2D}=\frac{1}{2\pi i}\oint_{\Gamma_{{\bm k}_{\rm B}}}d{\bm k}\cdot{\bm\nabla}_{\bm k}\log\det\left[H\left({\bm k}\right)-E_{\rm B}\right],
\label{2Dwinding}
\end{equation}
where $\Gamma_{{\bm k}_{\rm B}}$ represents an infinitesimal counterclockwise loop enclosing ${\bm k}_r$, and $E_{\rm B}=H\left({\bm k}_{\rm B}\right)$. The notable point is that the energy band has a finite area on the complex-valued energy plane when Eq.~(\ref{2Dwinding}) equals nonzero. The existence of the spectral area indicates that one can take an integral path inside the area so that the contour integral defined in Eq.~(\ref{energywinding}) can take nonzero values, which leads to the non-Hermitian skin effect. Said differently, the finite spectral area gives rise to the non-Hermitian skin effect even though the system does not have nonreciprocal hopping amplitudes. The key point is that the occurrence of the non-Hermitian skin effect depends on the shape of the system. 

To be concrete, we consider the model with the Bloch Hamiltonian
\begin{equation}
H\left({\bm k}\right)=2\cos k_x+i\cos k_y.
\label{GDSEBlochHamiltonian}
\end{equation}
One can immediately confirm that the winding number (\ref{2Dwinding}) is $1$ in the finite area occupied by the energy band [Fig.~\ref{GDSEfig}(a)]. The bulk eigenstates are then localized at the edges of the diamond geometry [Fig.~\ref{GDSEfig}(b)], and the localization states are qualitatively distinct from the corner-localized states introduced above. We remark that the non-Hermitian skin effect does not occur on the square geometry. This phenomenon is called a geometry-dependent skin effect, which has been experimentally observed in phononics \cite{Zhou2023}.

%
% -------------------------------------------------------------------------------------------------------------------------------------------------------------------------------------
%

\subsubsection{Non-Bloch band theory}\label{sec:4c3}
\begin{figure}[]
\centering
\includegraphics[width=6.5cm]{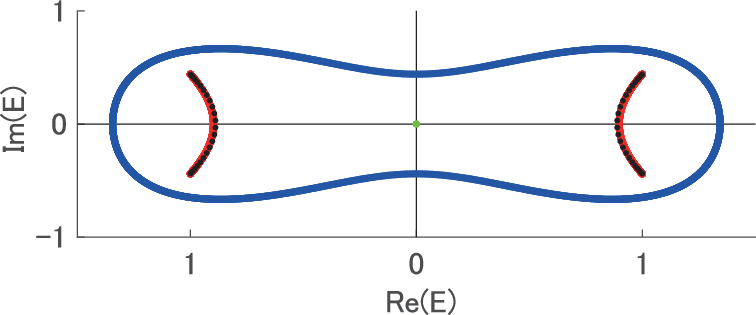}
\caption{\label{nHSSHmodelfig}Energy eigenvalues of the non-Hermitian Su-Schrieffer-Heeger model. We show the energy bands obtained from the conventional Bloch band theory (blue curve), the continuum energy band calculated from Eq.~(\ref{conditionGBZ}) (red curve), and the discrete energy eigenvalue of the finite-size system with an open boundary condition (black dots). The green dots around $E=0$ represent the energies of the topological edge states. The system parameters are set to be $t_1=0.5,t_2=1$, and $\gamma=2/3$.}
\end{figure}
As shown in Fig.~\ref{HNmodelfig}(a), the energy eigenvalues of the open-boundary system can be considerably different from the energy band obtained by the conventional Bloch Hamiltonian~(\ref{HNBlochHamiltonian}). Indeed, Eq.~(\ref{HNBlochHamiltonian}) reproduces the energy eigenvalues under periodic boundary conditions. Importantly, if one invokes the usual Bloch band theory, the extreme boundary sensitivity can induce the breakdown of the bulk-edge correspondence associated with the line-gap topology \cite{Lee2016,Leykam2017,Xiong2018,Kunst2018,Lee2019a}. To investigate the topological nature of open-boundary non-Hermitian systems, it is necessary to accurately calculate the energy band that reproduces the energy eigenvalue under open boundary conditions.

Motivated by such observation, the novel type of the Bloch band theory called a non-Bloch band theory has been established as a powerful tool to study non-Hermitian lattice systems \cite{Yao2018,Yokomizo2019,Yokomizo2020a,Yokomizo2021}. The non-Bloch band theory allows us to calculate the generalized Brillouin zone $\beta=e^{ik}$ for a complex-valued Bloch wavenumber $k$, which gives the continuum sets of the energy eigenvalues under open boundary conditions. The complex-valued Bloch wavenumber indicates that all the eigenstates are localized at the edge due to the non-Hermitian skin effect. Meanwhile, in the conventional Bloch band theory, the Bloch wavenumber takes real values, which means that all the bulk eigenstates extend over a whole system. To be concrete, we illustrate the main idea of the non-Bloch band theory by considering a non-Hermitian SSH model. This model is a non-Hermitian extension of the SSH model (cf. Fig.~\ref{SSH_BBH_fig}) by introducing intracell asymmetric hopping amplitudes, and its real-space Hamiltonian reads
\begin{eqnarray}
{\cal H}_{\rm nHSSH}=\sum_x\left[\left(t_1-\gamma\right)\left|x,{\rm B}\right\rangle\left\langle x,{\rm A}\right|+t_2\left|x,{\rm B}\right\rangle\left\langle x+1,{\rm A}\right|\right. \nonumber\\
\left.+\left(t_1+\gamma\right)\left|x,{\rm A}\right\rangle\left\langle x,{\rm B}\right|+t_2\left|x+1,{\rm A}\right\rangle\left\langle x,{\rm B}\right|\right],
\label{nHSSHHamiltonian}
\end{eqnarray}
where $t_{1,2}$ and $\gamma$ take real values. From a general theory of difference equations, one can take an ansatz of the real-space Schr\"{o}dinger equation, ${\cal H}_{\rm nHSSH}\left|\psi\right\rangle=E\left|\psi\right\rangle$. Specifically, the $x$-site component of the wavefunction can be written as a linear combination of the solutions of the eigenvalue equation, $\det\left[H_{\rm nHSSH}\left(\beta\right)-E\right]=0$, where $H_{\rm nHSSH}\left(\beta\right)$ is given by
\begin{eqnarray}
H_{\rm nHSSH}\left(\beta\right)=\left( \begin{array}{cc}
0                     & R_+\left(\beta\right) \\
R_-\left(\beta\right) & 0
\end{array}\right), \\
R_\pm\left(\beta\right)=\left(t_1\pm\gamma\right)+t_2\beta^{\mp1}.
\label{nHSSHnonBlochHamiltonian}
\end{eqnarray}
Conventionally, the energy eigenvalue of the finite-size system is calculated by combining the eigenvalue equation and open boundary conditions that the wavefunction satisfies. Meanwhile, the non-Bloch band theory allows us to readily calculate the asymptotic continuum sets of the energy eigenvalues in the thermodynamic limit, avoiding the cumbersome calculation process. To explain the method, let $\beta_1$ and $\beta_2$ denote the solutions of the eigenvalue equation, a quadratic equation of $\beta$. Notably, one can calculate the continuum energy band by combining the eigenvalue equation and the continuum set of $\beta_1$ and $\beta_2$, satisfying
\begin{equation}
\left|\beta_1\right|=\left|\beta_2\right|.
\label{conditionGBZ}
\end{equation}
The closed curve determined by Eq.~(\ref{conditionGBZ}) is called the generalized Brillouin zone, and it forms the circle with the radius $\sqrt{\left|\left(t_1-\gamma\right)/\left(t_1+\gamma\right)\right|}$ enclosing the origin of the $\beta(=e^{ik})$ plane. We note that the generalized Brillouin zone becomes a unit circle in Hermitian cases because the Bloch wavenumber takes real values. We show the continuum energy bands calculated from Eq.~(\ref{conditionGBZ}), the energy bands obtained from the conventional Bloch band theory, and the energy eigenvalues of the finite-size system with an open boundary condition in Fig.~\ref{nHSSHmodelfig}. One can confirm that the continuum energy bands (red curve) truly reproduce the energy eigenvalues (black dots) except for the energies of the topological edge states around $E=0$ (green dots). We remark that the continuum energy bands are distributed inside the energy band calculated from the real-valued Bloch wavenumber (blue curve).

\begin{figure}[]
\centering
\includegraphics[width=6.5cm]{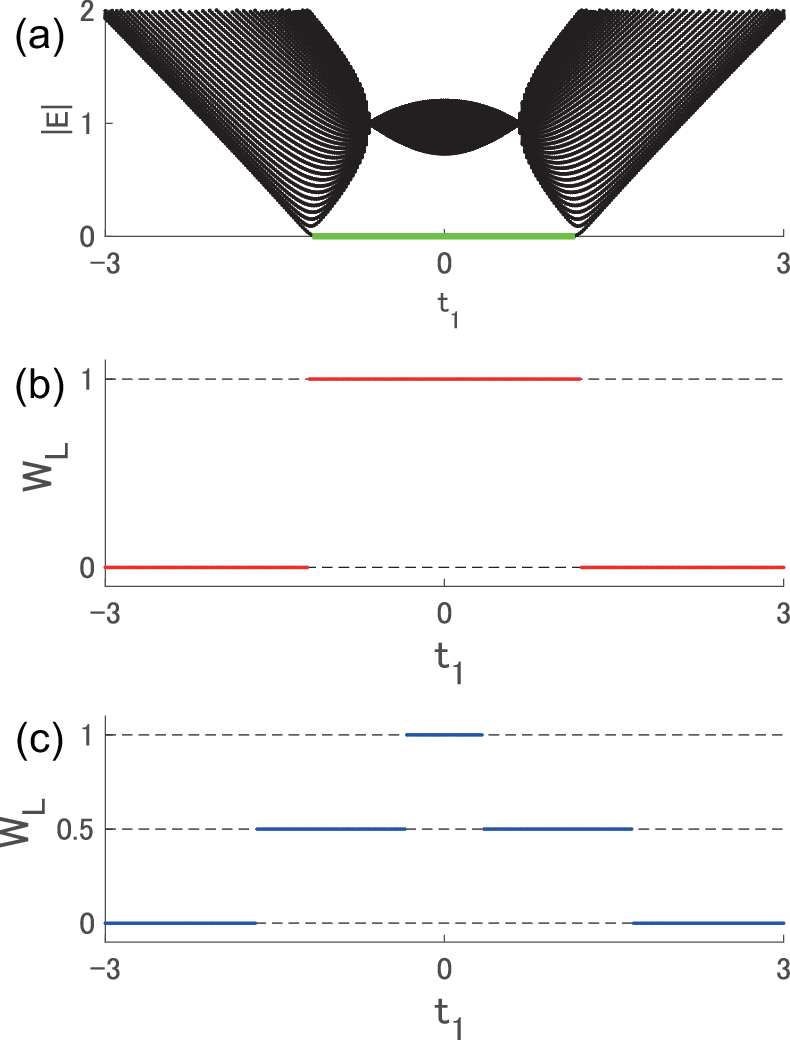}
\caption{\label{WNfig}Energy eigenvalue and winding numbers of the non-Hermitian Su-Schrieffer-Heeger model. (a) Energies of the bulk modes (black) and energies of the topological edge states (green) in the finite-size system with an open boundary condition. (b) Winding number defined in Eq.~(\ref{windingnumber2}). (c) Winding number calculated from the conventional Brillouin zone $e^{ik}~\left(k\in R\right)$. The system parameters are set to be $t_2=1$ and $\gamma=2/3$.}
\end{figure}

The continuum energy bands of the non-Hermitian SSH model open the line gap, whose topology can be characterized by the $Z$ topological invariant, similar to the conventional SSH model. Meanwhile, the point-gap topology for the continuum energy band becomes trivial because the energy winding number always takes zero. We next review how to investigate the line-gap topology from the viewpoint of the non-Bloch band theory. According to Refs.~\cite{Yao2018,Yokomizo2019}, the winding number can be defined from the generalized Brillouin zone as follows:
\begin{equation}
W_{\rm L}=-\frac{1}{2\pi}\frac{\left[{\rm arg}\;R_+\left(\beta\right)\right]_{C_\beta}-\left[{\rm arg}\;R_-\left(\beta\right)\right]_{C_\beta}}{2},
\label{windingnumber2}
\end{equation}
where $\left[{\rm arg}R_\pm\left(\beta\right)\right]_{C_\beta}$ means the change of ${\rm arg}R_\pm\left(\beta\right)$ as $\beta$ goes along the closed curve formed by $\beta=e^{ik}$ in a counterclockwise way. As shown in Figs.~\ref{WNfig}(a) and (b), the nonzero winding number fully corresponds to the existence of the topological edge states around $E=0$. Meanwhile, one can also define another winding number from the conventional Brillouin zone $e^{ik}~\left(k\in R\right)$, similar to Eq.~(\ref{winding_Hermitian}). However, as can be seen in Figs.~\ref{WNfig}(a) and (c), the bulk-edge correspondence appears to be broken in this case, and the winding number calculated from the conventional Brillouin zone cannot be used to predict the appearance of the topological edge states. Thus, the non-Bloch band theory is essential to investigate the line-gap topology of non-Hermitian systems.

We comment on the generalization of the main result of the non-Bloch band theory. The eigenvalue equation of a 1D non-Hermitian tight-binding system can be written as an algebraic equation for $\beta$ of $2M$ degrees. We denote the solutions of the eigenvalue equation by $\beta_1,\dots,\beta_{2M}$, which satisfy $\left|\beta_1\right|\leq\cdots\leq\left|\beta_{2M}\right|$. Reference \cite{Yokomizo2019} has shown that the generalized Brillouin zone is formed by the trajectories of $\beta_M$ and $\beta_{M+1}$, satisfying
\begin{equation}
\left|\beta_M\right|=\left|\beta_{M+1}\right|.
\label{generalcase}
\end{equation}
Additionally, Ref.~\cite{Yang2020} has shown that the generalized Brillouin zones are composed of the multiple curves formed by Eq.~(\ref{generalcase}), which become closed curves encircling the origin of the $\beta(=e^{ik})$ plane and can possess some cusps where the curves are indifferentiable. We remark that the existence of the cusps indicates a novel type of topological phase transition. Indeed, Ref.~\cite{Yokomizo2020b} has found that a topological semimetal phase appears as an intermediate phase between topologically trivial and nontrivial phases, and gap-closing points (exceptional points) are stable against perturbations in the topological semimetal phase due to the cusps. Equation~(\ref{generalcase}) can be understood as the condition that the localization lengths of ``plane waves'' expressed by $\beta_M$ and $\beta_{M+1}$ are equivalent. Said differently, the condition for the generalized Brillouin zone means the formation condition of the standing wave by the plane waves. It is worth noting that the generalized Brillouin zone and the corresponding continuum energy band are independent of boundary conditions of open-boundary systems despite the spatial localization of the bulk eigenstates at the edge.

%
% -------------------------------------------------------------------------------------------------------------------------------------------------------------------------------------
%

\subsubsection{Application of the non-Bloch band theory}\label{sec:4c4}
The non-Bloch band theory allows us to investigate fundamental non-Hermitian physics and topological properties of various non-Hermitian systems. We review several application examples of the non-Bloch band theory in the following. First, Ref.~\cite{Kawabata2020a} has proposed constructing the non-Bloch band theory under the symmetry-protected non-Hermitian skin effect, where the system preserves the pseudo-time-reversal symmetry defined in Eq.~(\ref{pTRS}). The eigenvalue equation of the system becomes an algebraic equation for $\beta$ of $4M$ degrees. The solutions of the characteristic equation are denoted by $\beta_1,\dots,\beta_{2M},\beta_{2M}^{-1},\dots,\beta_1^{-1}$ with $\left|\beta_1\right|\leq\dots\leq\left|\beta_{2M}\right|\leq1\leq\left|\beta_{2M}^{-1}\right|\leq\dots\leq\left|\beta_1^{-1}\right|$ due to the symmetry. As a result, the previous work has shown the conditions for the generalized Brillouin zone written as
\begin{equation}
\left|\beta_{2M-1}\right|=\left|\beta_{2M}\right|,~\left|\beta_{2M}^{-1}\right|=\left|\beta_{2M-1}^{-1}\right|,
\label{pTRScase}
\end{equation}
which are distinct from the condition (\ref{generalcase}). The physical meaning of Eq.~(\ref{pTRScase}) is that the standing wave formed by $\left|\beta_{2M-1}\right|=\left|\beta_{2M}\right|$ is localized at the edge opposite to one formed by $\left|\beta_{2M}^{-1}\right|=\left|\beta_{2M-1}^{-1}\right|$. Thus, this result is consistent with the localization nature of the symmetry-protected non-Hermitian skin effect.

We next explain the construction of the non-Bloch band theory of spatially periodic continuous systems. Reference \cite{Yokomizo2022a} has considered 1D systems described by the Strum-Liouville-type wave equation as follows:
\begin{eqnarray}
\left[-\frac{d}{dx}p\left(x\right)\!\frac{d}{dx}\!-\frac{i}{2}\!\left(\lambda_1\!\left(x\right)\!\frac{d}{dx}\!+\!\frac{d}{dx}\lambda_2\!\left(x\right)\!\right)\!+v\left(x\right)\right]\!\psi\left(x\right) \nonumber\\
=\omega^2\psi\left(x\right),
\label{SLeq}
\end{eqnarray}
where $\psi\left(x\right)$ is a wave function, and $p\left(x\right)$, $\lambda_{1,2}\left(x\right)$, and $v\left(x\right)$ are complex-valued periodic functions with a period $a$. The first-order derivative terms in Eq.~(\ref{SLeq}) essentially contribute to realizing the non-Hermitian skin effect since the terms can be regarded as an imaginary gauge potential~\cite{Hatano1996}. The main result of the previous work is that the generalized Brillouin zone becomes a circle with a radius
\begin{equation}
r=\exp\left(\frac{1}{2}\int_0^adx\;{\rm Im}\left[ \frac{\lambda_1\left(x\right)+\lambda_2\left(x\right)}{p\left(x\right)}\right]\right),
\label{continuousGBZ}
\end{equation}
which means that the imaginary part of the complex-valued Bloch wavenumber is obtained by $-\frac{1}{a}\log r$. The generalized Brillouin zone allows us to investigate the bulk-edge correspondence, for example, in photonic crystals with anisotropy and optical loss. The electromagnetic wave in the photonic crystal obeys the wave equation written in the form of Eq.~(\ref{SLeq}) and exhibits the non-Hermitian skin effect. The topological edge state appears when the Zak phase defined by the generalized Brillouin zone takes a nonzero value. Recently, a general extension of the framework has been proposed in Ref.~\cite{Hu2024v2}. Additionally, Ref.~\cite{Yokomizo2024} has investigated the construction of the non-Bloch band theory in spatially periodic continuous systems described by generalized eigenvalue equations.

\begin{figure}[]
\centering
\includegraphics[width=8cm]{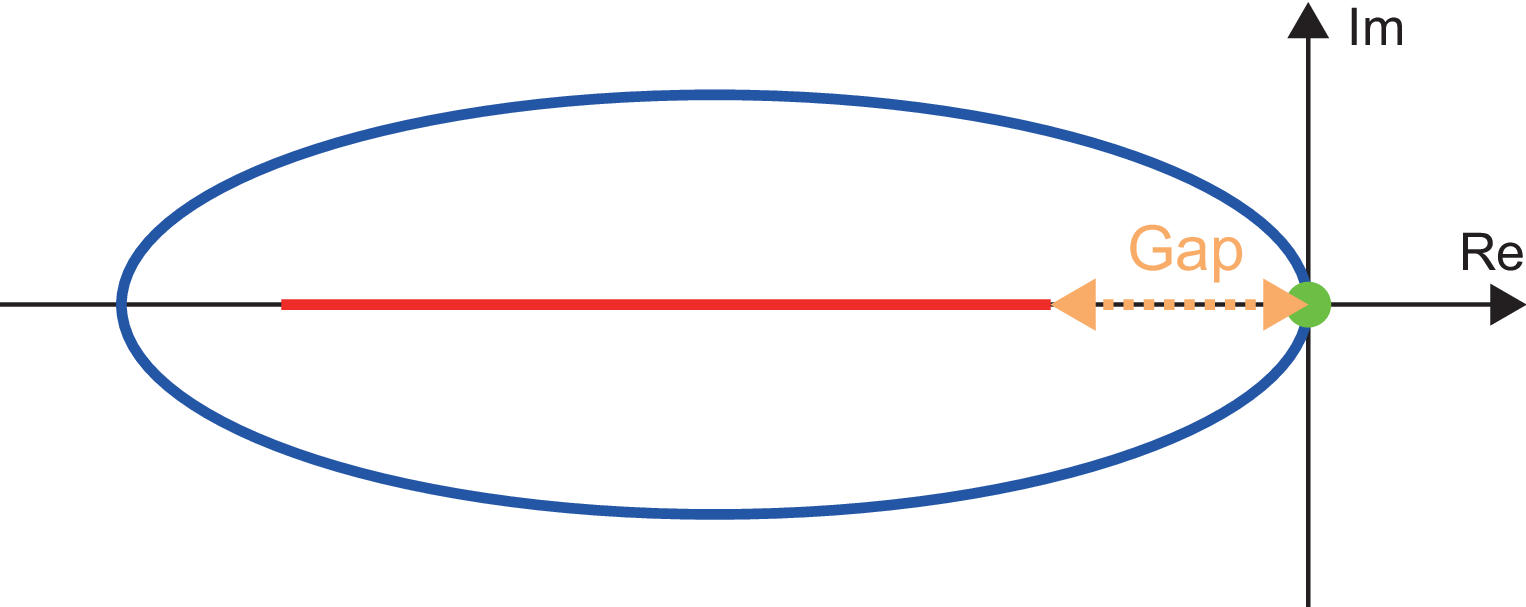}
\caption{\label{sawadafig}Schematic figure of the spectrum of a transition matrix with a nonzero winding number. We show the spectrum obtained from the real-valued Bloch wavenumber (the non-Bloch band theory) in the blue curve (the red line). The green dot represents the zero eigenvalue of the steady state.}
\end{figure}

The relaxation dynamics from initial states to a steady state have been studied in stochastic processes described by Eq.~(\ref{mastereq}). The steady state uniquely corresponds to the eigenvector of the transition matrix whose eigenvalue is equal to $0$. We remark that the non-Bloch band theory is useful for investigating the role of topology in the relaxation dynamics. Let us suppose that the system has the spatial translation symmetry, and the transition matrix is denoted by $W_{nm}$, where $n$ and $m$ are the site indices. Furthermore, let $W\left(k\right)$ denote the diagonalized form of the transition matrix in the Fourier basis. For the sake of convenience, we introduce the scale-transformation form of the transition matrix as follows: $\left(W^\lambda\right)_{nm}\equiv W_{nm}e^{\lambda\left(n-m\right)}$. Reference~\cite{Sawada2024} has proposed that the point-gap topology can be characterized by the winding number defined as
\begin{eqnarray}
w=w_+ +w_-, \\
w_\pm=\lim_{\lambda\rightarrow\pm0}\frac{1}{2\pi i}\int_0^{2\pi} dk \frac{d}{dk}\log\left(\det W^\lambda\left(k\right)\right).
\label{eqsawada}
\end{eqnarray}
The main claim of the previous work is that, when the winding number takes a nonzero value, the non-Hermitian skin effect occurs, and the spectrum obtained by the non-Bloch band theory has a finite gap below the zero eigenvalue [Fig.~\ref{sawadafig}]. Therefore, this claim indicates that an eigenvalue of the transition matrix is either within the spectrum or equal to zero in a finite-size system with arbitrary open boundary conditions. Importantly, the existence of the gap leads to a finite relaxation time in the stochastic process in an intuitive sense. However, the relaxation time also depends on the overlap between initial states and the eigenvectors of the transition matrix \cite{Mori2020,Haga2021}. One can immediately infer that the overlap decreases because the eigenvectors are localized due to the non-Hermitian skin effect. As a result, both the inverse of the gap and of the overlap contribute to the relaxation time, and the relaxation time becomes proportional to the system size. This system-size dependence of the relaxation time is still different from that when the winding number vanishes, because it is proportional to the square of the system size in a topologically trivial system.

Finally, we comment on recent efforts in extending the non-Bloch band theory to 2D non-Hermitian systems. Reference~\cite{Yokomizo2023} has shown that, in 2D tight-binding systems with specific symmetries, one can determine the generalized Brillouin zone and calculate the continuum energy band as in the case of 1D tight-binding systems. However, the  previous work has pointed out that the effect of systems' shapes on the formation of the standing wave leads to the difficulty of calculating the continuum energy band. Beyond the previous work, Ref.~\cite{Wang2024} has proposed an amoeba formulation to calculate the continuum energy band in higher-dimensional tight-binding systems on arbitrary geometry. Importantly, the amoeba formulation allows us to successfully establish the bulk-edge correspondence in the Chern insulator with asymmetric hopping amplitudes. While the previous work has clarified the calculation method for the continuum energy band of the bulk skin mode, the construction of the non-Bloch band theory for higher-order skin mode is still an open problem.

%
% -------------------------------------------------------------------------------------------------------------------------------------------------------------------------------------
%

\begin{figure}[]
\centering
\includegraphics[width=8cm]{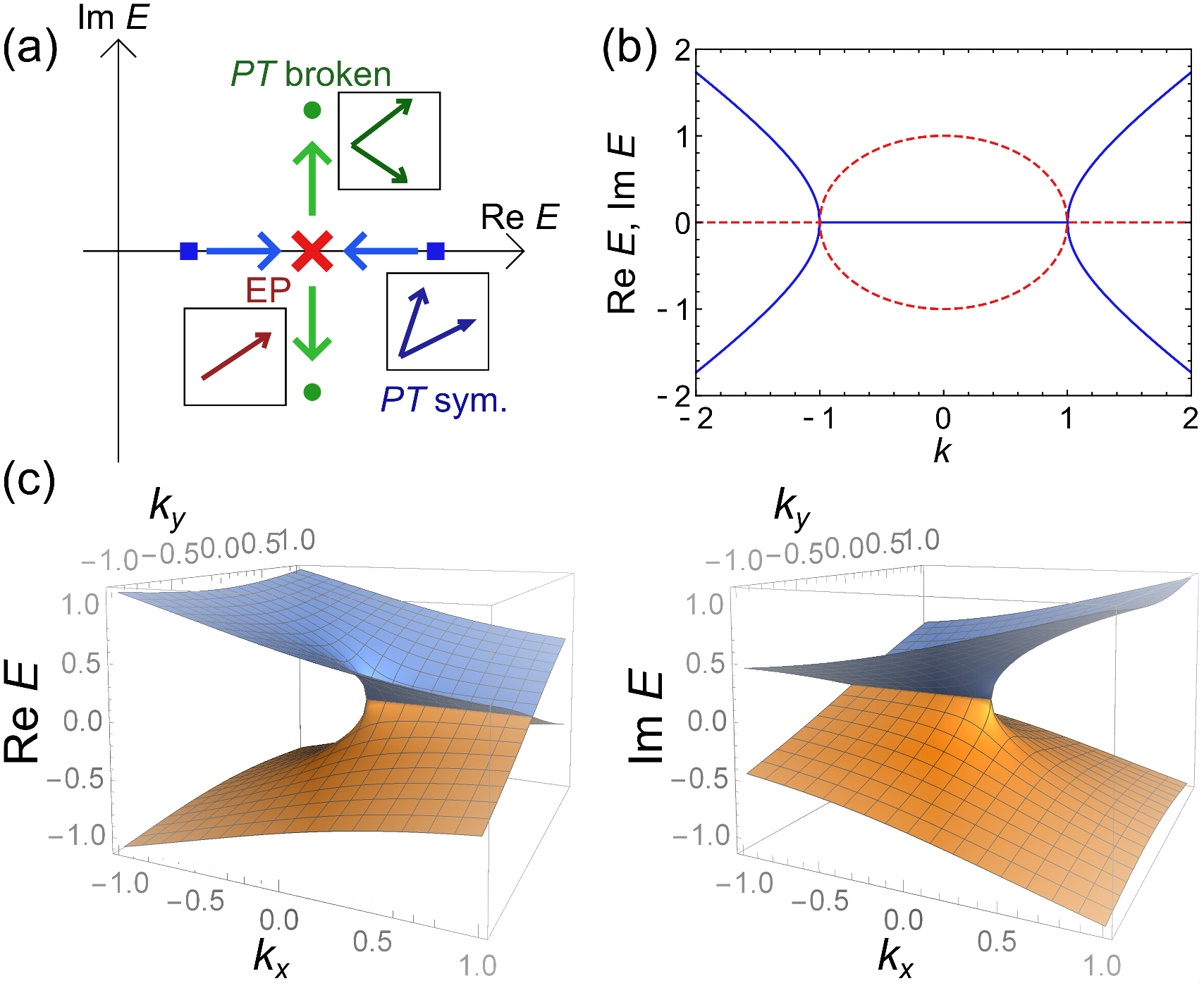}
\caption{\label{exceptional_point_fig} Typical band structures accompanying exceptional points. (a) Changes in the eigenvalues and eigenvectors around an exceptional point in zero dimension. In the $PT$ symmetric phase, eigenvalues are real. As a parameter is changed, these two eigenvalues approach and they coalesce at the exceptional point. After the coalescence, the eigenvalues are split again, while they become a pair of complex conjugates. The arrows show the schematics of the eigenvectors at each phase. Two eigenvectors become parallel at the exceptional point, and thus the Hamiltonian becomes nondiagonalizable. (b) Exceptional points in one dimension. The blue (red dashed) curve represents the real (imaginary) part of the dispersion. Exceptional points exist at $k=\pm1$. (c) Exceptional point in two dimensions. A Riemann-surface structure appears around the exceptional point, which is the origin of its topological protection.}
\end{figure}

\subsection{Exceptional point}\label{sec:4d}
Non-Hermitian point-gap topology also plays an important role in gapless phases, which is relevant to the stability of exceptional points. An exceptional point is a gapless point in a parameter or wavenumber space where two or more eigenvectors coalesce as well as eigenvalues \cite{Kato1966}. The coalescence of the eigenvectors indicates that the Hamiltonian is nondiagonalizable at the exceptional point. Nondiagonalizable Hamiltonian is associated with a Jordan form such as
\begin{equation}
H_{\rm Jordan} = \left(
  \begin{array}{cc}
   E & 1 \\
   0 & E
  \end{array}
  \right).
\end{equation}
While a unidirectional hopping term in the Jordan form may seem to be hardly realizable in experimental setups, exceptional points are ubiquitous in nature because they can appear by using simple gain and loss or dissipative couplings. A minimal Hamiltonian of the exceptional point is
\begin{equation}
H = \left(
  \begin{array}{cc}
   i\gamma & a \\
   a & -i\gamma
  \end{array}
  \right), \label{minimal_EP}
\end{equation}
which depends on two real parameters $\gamma$ and $a$. $i\gamma$ represents the on-site gain and loss, which leads to the non-Hermiticity of the Hamiltonian. If one multiplies the Hamiltonian by the imaginary unit $i$, $\gamma$ corresponds to the natural frequency and $a$ represents the dissipation between two components, the latter of which can be introduced by, e.g., aligning interaction between active particles as we will discuss in Sec.~\ref{sec:5c2}. This Hamiltonian has exceptional points at $\gamma=\pm a$, where one obtains only one eigenvector $\psi\propto(i,\pm 1)^T$. Figure \ref{exceptional_point_fig}(b) shows the eigenvalues of the Hamiltonian at different $\gamma$. We can see the singular behavior around the exceptional points, which plays important roles both in the topological protection of the exceptional point and its application to enhanced sensitivity as we will discuss below.

Exceptional points are topologically protected in a similar sense to Weyl points \cite{Armitage2018,Kawabata2019b}. To understand the topological protection of exceptional points, we consider the following Hamiltonian,
\begin{equation}
H = \left(
  \begin{array}{cc}
   0 & k_x + ik_y \\
   a & 0
  \end{array}
  \right).
\end{equation}
This Hamiltonian has exceptional points at $k_x=k_y=0$. Figure \ref{exceptional_point_fig}(c) shows the complex spectra around the exceptional point, which exhibit a characteristic branchpoint structure. The exceptional point is protected by the nontrivial topology of this branchpoint structure around it, and thus cannot be removed until two or more exceptional points coalesce and pair-annihilate. To clearly show its topological feature, one should consider the eigenvalues on a circular path $C$ around the exceptional point. Moving along the path $C$, a pair of eigenvalues are swapped to each other. This braiding cannot be disentangled until other gapless points cross the path $C$, which is similar to the fact that a M\"{o}bius band cannot be deformed to a normal band as long as one does not cut the band. Meanwhile, the swapping of eigenvalues is unique to gapless systems because if spectra have a gap, one can distinguish and separate each spectrum. Therefore, exceptional points are protected by the topology of the spectra around them including the braiding of eigenvalues. One can also define the topological invariant of exceptional points as 
\begin{equation}
w = \oint_C d\mathbf{k} \cdot \nabla_{\bf k} \log \det [H(\mathbf{k}) - E_0], \label{EP_winding}
\end{equation}
where $E_0$ is the eigenvalue at the exceptional point. We note that this definition of the topological invariant is analogous to that of a one-dimensional non-Hermitian Hamiltonian~(\ref{energywinding}) and indeed it has the same origin.

Due to the topological property of exceptional points characterized by the winding number, exceptional points have robustness against perturbations to the Hamiltonian. On the other hand, when two or more exceptional points coalesce, the gapless path $C$ in Eq.~(\ref{EP_winding}) cannot be defined around each exceptional point, and thus the band topology can be altered. In fact, if the sum of the winding numbers is zero, exceptional points can be removed via the coalescence, which resembles the annihilation of topological defects in the real space \cite{Muhlbauer2009,Yu2010}.

As in Hermitian topological insulators, exceptional points can also be protected by symmetries. To see this, we again consider the Hamiltonian (\ref{minimal_EP}). When we fix $a$ and assume $\gamma$ as a wavenumber $\gamma=k$, the Hamiltonian is considered as a non-Hermitian Bloch Hamiltonian. This one-dimensional Hamiltonian is symmetric under the combination of the parity inversion (i.e., space inversion) and the time reversal, which is known as the parity-time ($PT$) symmetry \cite{Bender1999,Mostafazadeh2002}. By using a unitary matrix $PT$, one can define the $PT$ symmetry of the Hamiltonian as $PT H(k) (PT)^{-1} = H^{\ast}(k)$. The $PT$ symmetry guarantees that eigenvalues are real or have complex-conjugate counterparts. To continuously modify the complex-conjugate pair of eigenvalues into real ones, the Hamiltonian must be degenerate at the transition point as demonstrated in Fig.~\ref{exceptional_point_fig}(a). Such a degenerate point corresponds to an exceptional point and cannot be removed until two exceptional points coalesce, which indicates the topological protection of exceptional points under the $PT$ symmetry. In fact, one can define the topological invariant as
\begin{equation}
\nu = {\rm sgn} (\det H(k_0-\delta) H(k_0 + \delta)), \label{1d-EP-topo-inv}
\end{equation}
where $k=k_0$ is the exceptional point, and $\delta$ is a sufficiently small real number \cite{Kawabata2019b}.

In the above argument, we have focused on isolated exceptional points in two band models, while exceptional points can exhibit a variety of structures, including node structures and highly degenerate ones. An exceptional ring \cite{Xu2017,Cerjan2019} is a typical example of exceptional node structures, where exceptional points are aligned on a ring-shaped path in a wavenumber space. In addition, at an exceptional point, three or more bands can degenerate, which is known as a higher-order exceptional point \cite{Graefe2008,Demange2011}. Such higher-order exceptional points can stably exist under the proper symmetry \cite{Delplace2021,Mandal2021}. In addition, the parameter dependence of the eigenvalues around a higher-order exceptional point is different from that of an ordinary exceptional point, which might be of practical advantage \cite{Hodaei2017,Wang2019} in, e.g., enhanced sensitivity discussed in the following paragraph.

As in the gapped topological phases, a topological classification of exceptional points has also been proposed in Ref.~\cite{Kawabata2019b}. Topological classification of $d$-dimensional exceptional structures (cf. $d=0$ for the exceptional point and $d=1$ for the exceptional ring) in $D$-dimensional systems corresponds to the topological classification of gapped phases in $D-d-1$ dimension. However, we need to consider $PT$ and $CP$ symmetries instead of the conventional time-reversal and charge-conjugate symmetries, as we have already seen in the one-dimensional case.

Nevertheless, the presence of the non-Hermitian skin effect can alter the stability of the exceptional points in a nontrivial manner. In fact, exceptional points in non-Bloch bands can be stable even if the classification table predicts the absence of topologically protected exceptional points \cite{Yokomizo2020b}. The emergence of stable exceptional points in non-Bloch bands can be related to so-called non-Bloch $PT$ symmetry breaking \cite{Xiao2021,Longhi2019,Hu2024}, which is the non-Bloch-band version of the $PT$-symmetry breaking. One can define the non-Bloch $PT$ symmetry as $PT H(\beta) (PT)^{-1} = H^{\ast}(\beta)$ with $\beta$ satisfying the condition in Eq.~(\ref{conditionGBZ}). Then, the spontaneous symmetry breaking of the non-Bloch Hamiltonian occurs at the exceptional point, and one can show the stability of the exceptional point as in conventional Bloch bands. However, the comprehensive understanding of the stability of the exceptional points in non-Bloch bands still remains an open problem.

The branchpoint structures around exceptional points can induce unconventional non-Hermitian phenomena in various physical systems, including optics \cite{Ruschhaupt2005,El-Ganainy2007,Ruter2010}. In early experimental works \cite{Dembowski2001}, the topological structure of exceptional points has been observed as a dynamical interchanging of the eigenmodes in optical systems when one changes the parameters around an exceptional point. On another front, singular structures of the spectrum around exceptional points can be utilized in sensors \cite{Hodaei2017,Liu2016,Wiersig2020}. Typically, the spectrum around an exceptional point behaves as  $E=E_0+|\mathbf{k}-\mathbf{k}_0|^{1/\nu}$, where $E_0$ and $\mathbf{k}_0$ are the eigenvalue and the parameters at the exceptional point, respectively, and $\nu$ is determined by the order of the exceptional point (often being $\nu=2$). The slope of the eigenvalue is proportional to $|\mathbf{k}-\mathbf{k}_0|^{1/\nu-1}$ and diverges at $\mathbf{k}=\mathbf{k}_0$. Therefore, the eigenvalue drastically changes around the exceptional point and thus enhances the signal, which is known as enhanced sensitivity. Exceptional points can also be found in the $S$ matrix in photonic lattices. In particular, by utilizing the Jordan form structure, one can realize unidirectional invisibility \cite{Lin2011} where light comes from only one side can be reflected, and coherent perfect absorption that entirely absorbs light at discrete frequencies. Exceptional points in band structures, known as spectral singularities, have also been found in various setups including photonics \cite{Zhou2018,Zhen2015} and acoustics \cite{Zhou2023}. The spectral singularities often accompany the so-called bulk Fermi arc \cite{Zhou2018}, where the real parts of two eigenvalues coalesce.

\begin{figure}[]
\centering
\includegraphics[width=8cm]{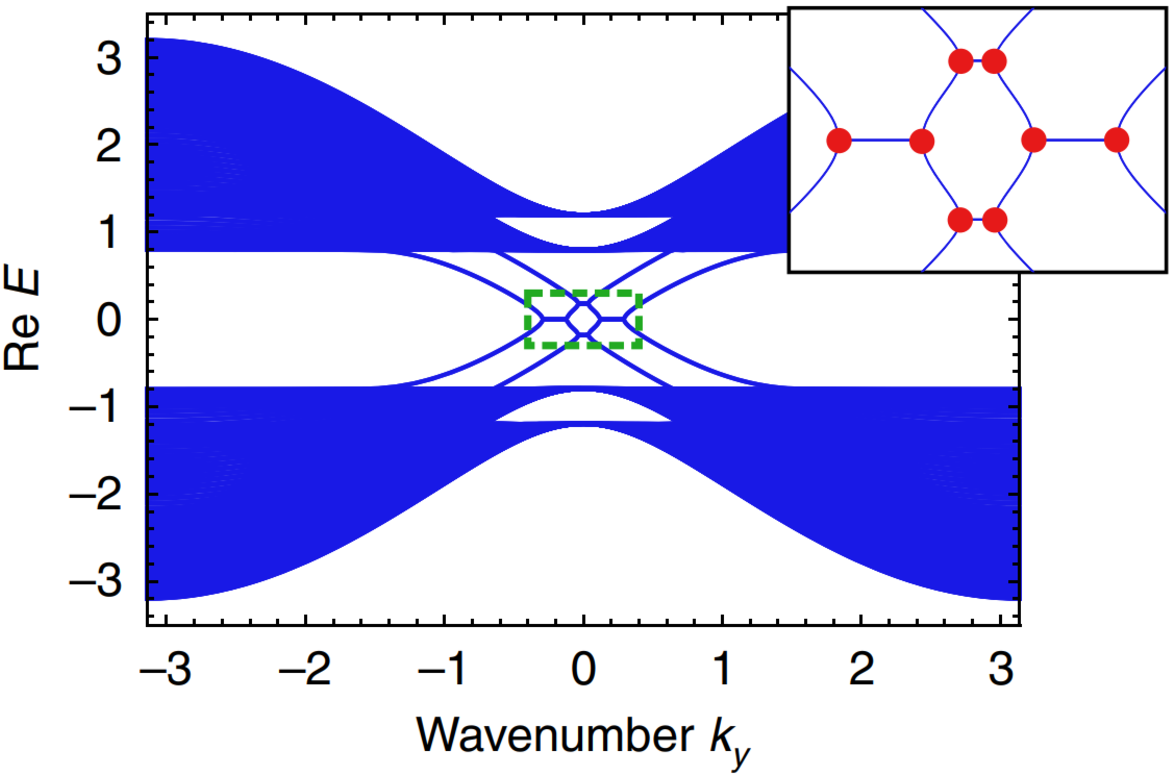}
\caption{\label{exceptional_edgemode_fig} Band structure of exceptional edge modes. The inset shows the enlarged view of the band structure corresponding to that in the green dashed box. The red points are the exceptional points that protect the gapless modes. This figure is adapted from K. Sone, Y. Ashida, and T. Sagawa, ``Exceptional non-Hermitian topological edge mode and its application to active matter'' Nat. Commun. 11, 5745 (2020) \cite{Sone2020}, licensed under a Creative Commons Attribution 4.0 International License (http://creativecommons.org/licenses/by/4.0/).}
\end{figure}

While exceptional points are associated with the topology of the band structure around them, they can also interact with gapless boundary modes seen in topological insulators. In particular, exceptional points can protect gapless edge modes even with a topologically trivial bulk by using the nontrivial topology of the edge dispersions. Thus, such gapless modes imply the breakdown of the bulk-edge correspondence in a manner different from the one caused by the non-Hermitian skin effect, and those edge modes are termed exceptional edge modes \cite{Sone2020}. Figure \ref{exceptional_edgemode_fig} shows the typical band structure of the exceptional edge modes, where one can find a pair of exceptional points in gapless bands. Since these exceptional points cannot be removed until they coalesce, one cannot take apart upper and lower bands as if these edge bands are stuck by glue. The exceptional edge modes are typically realized by introducing non-Hermitian couplings into a Chern insulator and its time-reversal counterpart. One can understand the emergence of the exceptional edge modes in such coupled Chern insulators from the effective Hamiltonian of low-energy dispersion,
\begin{equation}
H = \left(
  \begin{array}{cc}
   k & -ia \\
   -ia & -k
  \end{array}
  \right),
\end{equation}
where the diagonal terms $\pm k$ represent the linear dispersion of edge modes in each layer of a Chern insulator and the off-diagonal terms $-ia$ correspond to the non-Hermitian coupling. Since this effective Hamiltonian is equal to the Hamiltonian (\ref{minimal_EP}) multiplied by $-i$, the Hamiltonian exhibits gapless dispersions with exceptional points. One can also confirm the topological protection of gapless bands in the same way as exceptional points in one-dimensional bulk bands. The exceptional boundary modes can be realized in photonic systems \cite{Sone2022}, and one can also find them in active matter \cite{Sone2020} as discussed later in Sec.~\ref{sec:5c2}. 

%
% -------------------------------------------------------------------------------------------------------------------------------------------------------------------------------------
%

\section{Topological active matter}\label{sec:5}
\subsection{Topological sounds in classical systems}\label{sec:5a}
\subsubsection{Topology in passive fluid}\label{sec:5a1}

While early studies of topological physics were motivated by electronic band structures, recent studies have also revealed the existence of topological edge modes in classical systems. One of the platforms for exploring such classical topological waves is fluid, such as air and water in engineered structures \cite{Yang2015,Ma2016,Cummer2016,He2016}. In topological fluids, one can observe the robust propagation of density waves, i.e., sound, along the boundary of the system, which is analogous to the edge current in topological insulators (cf.~Fig.~\ref{QHE_fig}(a)). 
Below we review the theoretical basis of topological fluids, whose extension is also utilized in theoretical proposals of topological active matter.

\begin{figure}[]
\centering
\includegraphics[width=8cm]{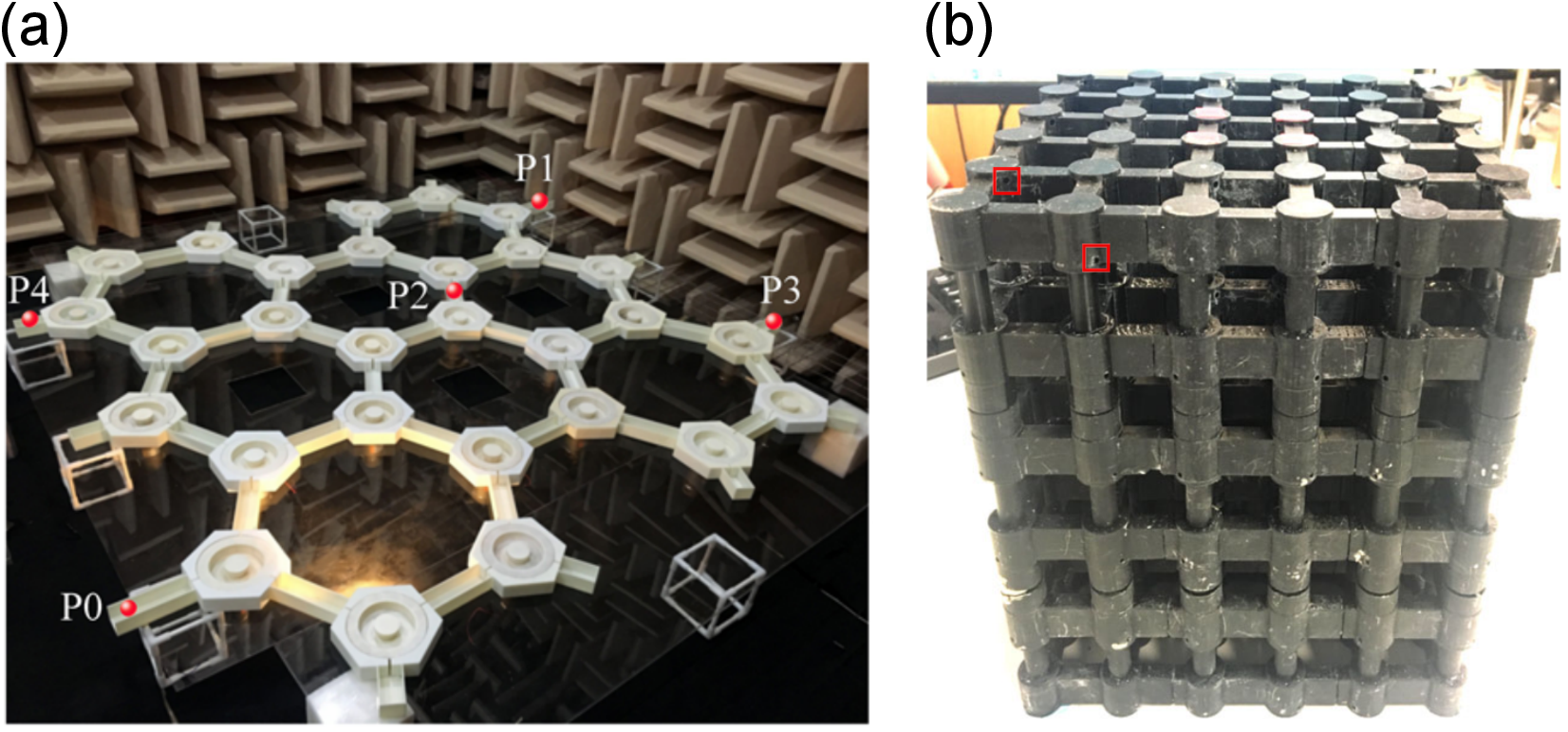}
\caption{\label{topological_acoustic_fig} Acoustic topological insulators. (a) Fluid realization of the quantum Hall effect. At each lattice point of the honeycomb lattice, rotors are arranged, which generates the effective magnetic field. This figure is adapted from Y. Ding et al., ``Experimental Demonstration of Acoustic Chern Insulators'' Phys. Rev. Lett. 122, 014302 (2019) \cite{Ding2019}, Copyright \copyright\  2019 by the American Physical Society. (b) Acoustic third-order topological insulator. The periodic structure imitates the three-dimensional extension of the BBH model. This figure is adapted from X. Ni et al., ``Demonstration of a quantized acoustic octupole topological insulator'' Nat. Commun. 11, 2108 (2020) \cite{Ni2020}, licensed under a Creative Commons Attribution 4.0 International License (http://creativecommons.org/licenses/by/4.0/).}
\end{figure}

To make an analogy between topological insulators and hydrodynamics, the crucial point is that there naturally exists nonlinearity in hydrodynamics equations, while the band topology discussed in condensed matter physics is based on linear Schr\"{o}dinger equations. One way to resolve this issue is to consider linearization of the hydrodynamic equations around a steady-state solution. Such linearized equations can describe the response to the perturbation and the fluctuation dynamics, which are also experimentally observable. Specifically, one can start from the conventional hydrodynamics described by the Navier-Stokes equations in (effectively) two-dimensional systems,
\begin{eqnarray}
&{}& \partial_t \rho + \nabla \cdot (\rho \mathbf{v}) = 0, \label{Navier-Stokes1}\\
&{}& \partial_t \mathbf{v} + (\mathbf{v} \cdot \nabla) \mathbf{v} =
-\nabla P + D_T \nabla^2 \mathbf{v} + \mathbf{f}.
\label{Navier-Stokes2}
\end{eqnarray}
To obtain the linearized equation, the steady-state density $\rho_0$ and velocity $\mathbf{v}_0$ and their deviation from the steady-state solution, $\delta \rho(t) = \rho(t) - \rho_0$ and $\delta \mathbf{v}(t) = \mathbf{v}(t) - \mathbf{v}_0$, are introduced. The leading-order terms of $\delta \rho(t)$ and $\delta \mathbf{v}(t)$ then give the linearized equation,
\begin{eqnarray}
&{}& \partial_t \delta \rho + \rho_0 \nabla \cdot \delta \mathbf{v} + \mathbf{v}_0 \cdot \nabla \delta\rho = 0, \label{linearized-Navier-Stokes1}\\
&{}& \partial_t \delta\mathbf{v} + (\mathbf{v}_0 \cdot \nabla) \delta \mathbf{v}  + (\delta\mathbf{v} \cdot \nabla) \mathbf{v}_0  \nonumber\\
&=& -p_0 \nabla \delta\rho + D_T \nabla^2 \delta\mathbf{v}.
\label{linearized-Navier-Stokes2}
\end{eqnarray}
To further simplify the equation and make its connection to the Schr\"{o}dinger equation clear, it is useful to assume that the viscosity and heat flow are negligible. Applying $(\partial_t+\mathbf{v}_0\cdot\nabla)$ to the left-hand side of Eq.~(\ref{linearized-Navier-Stokes1}) and rewriting $(\partial_t+\mathbf{v}_0\cdot\nabla) \delta\mathbf{v}$ by using Eq.~(\ref{linearized-Navier-Stokes2}) without the viscosity term $D_T \nabla^2 \delta\mathbf{v}$ and the term related to heat flow $(\delta\mathbf{v} \cdot \nabla) \mathbf{v}_0$, one obtains
\begin{equation}
(\partial_t+\mathbf{v}_0\cdot\nabla)^2 \delta \rho =\rho_0 p_0 \Delta \delta \rho. \label{Schroedinger_like_eq1}
\end{equation}
Then, one rewrites $p_0$ in terms of the sound velocity $c$ as $p_0 = c^2 / \rho_0$ and assumes $|\mathbf{v}/c| \ll 1$. By considering the Fourier component $\delta \rho(t) = e^{-i\omega t}\delta \rho'$, one can obtain the following Schr\"{o}dinger-like equation:
\begin{equation}
\frac{\omega^2}{c^2} \delta \rho = (-i\nabla+\mathbf{A}_{\rm eff})^2 \delta \rho. \label{Schroedinger_like_eq2}
\end{equation}
This equation is reminiscent of the conventional Schr\"{o}dinger equation (cf.~Eqs.~(\ref{Schroedinger_eq}) and (\ref{Schroedinger_eq2})) without a scaler potential. Comparing this equation with the Schr\"{o}dinger equation, one can assume $\mathbf{A}_{\rm eff}=\omega \mathbf{v}_0 / c^2$ as an effective vector potential.

One can also derive another Schr\"{o}dinger-like equation via the following sound master equation \cite{Yang2015}
\begin{equation}
\frac{1}{\rho}\nabla \cdot \rho\nabla\phi-\frac{1}{c^2}(\partial_t+\mathbf{v}_0\cdot\nabla)^2 \phi =0, \label{sound_master_eq}
\end{equation}
where $\phi$ is the velocity potential that satisfies $\mathbf{v} = -\nabla \phi$ and $c$ corresponds to the sound velocity. This sound master equation is derived from the Navier-Stokes equation under the same assumption (the viscosity and heat flow are negligible) as in Eq.~(\ref{Schroedinger_like_eq1}). After the Fourier transformation, the time derivative $\partial_t$ is replaced by $i\omega$, where $\omega$ represents the frequency of a standing wave. By assuming $|\mathbf{v}/c| \ll 1$ and thus ignoring the higher-order terms as in Eq.~(\ref{Schroedinger_like_eq1}), one finally obtains the following Schr\"{o}dinger-like equation \cite{Yang2015}:
\begin{equation}
\frac{\omega^2}{c^2} \Psi = [(-i\nabla+\mathbf{A}'_{\rm eff})^2 + V] \Psi
\end{equation}
where $\Psi$ is the modified velocity potential $\Psi=\sqrt{\rho} \phi$, and the effective vector and scalar potentials are described as
\begin{eqnarray}
\mathbf{A}'_{\rm eff} &=& -\omega\mathbf{v}_0/c^2, \\ 
V &=& -|\nabla \ln \rho|^2/4-\Delta \ln \rho/2.
\end{eqnarray}
By assuming $\Psi$ as an effective wave function in fluid, one can construct an analogy to the quantum (anomalous) Hall effect.

The derived Schr\"{o}dinger-like equations tell us that the steady-state flow plays a role of the effective vector potential, whose vorticity corresponds to the effective (local) magnetic flux that is necessary to realize a counterpart of the quantum (anomalous) Hall effect. Therefore, if the steady-state flow in a phononic system imitates the effective vector potential in quantum (anomalous) Hall systems, such a system should exhibit topological boundary waves. One straightforward way of constructing a topological fluid is globally rotating fluid by using circulators or external forces to a container of fluid \cite{Delplace2017,Souslov2019}. The global circular motion of fluid induces a net effective magnetic field. Since the nonzero net magnetic field is a key ingredient of the quantum Hall effect (cf.~Sec.~\ref{sec:3b1}), the rotating system constructs a fluidic counterpart of the quantum Hall effect. This correspondence can also be understood from the rotating coordinate system; if we consider the rotating coordinate that cancels the rotational motion of fluid, the fluid feels a Coriolis force. The Coriolis force acts perpendicularly to the direction of flow, which is the same as a (homogeneous) magnetic force.  We can thus again assume that the fluid feels the effective magnetic field, and such effective magnetic force leads to the phononic counterpart of the quantum Hall effect.

Topological fluid has been also realized without a net effective magnetic field by using an analogy to the quantum anomalous Hall effect. We note that the quantum anomalous Hall effect also requires the breakdown of the time-reversal symmetry. Therefore, purely passive fluids cannot construct a fluid counterpart of the quantum anomalous Hall effect, and one must utilize external driving devices such as circulators \cite{Ding2019}. One of the early studies \cite{Yang2015} utilized periodically aligned circulators, where circulators are placed on each lattice point of a triangular lattice. Then, the engineered vector potential imitates that of Haldane's honeycomb model \cite{Haldane1988} in Fig.~\ref{Haldane_fig}(a) and thus constructs the counterpart of the quantum anomalous Hall effect.

Topological fluids are not restricted to counterparts of the quantum (anomalous) Hall effect. Without breaking the time-reversal symmetry (i.e., no external drivings), one can realize acoustic quantum spin Hall systems. Reference~\cite{He2016} has realized such a quantum spin Hall system, by using a honeycomb-lattice structure and assuming left and right-rotational modes of sound waves as effective spins, which looks like the Kane-Mele model discussed in Fig.~\ref{Kane_Mele_fig}. One can also realize higher-order topological insulators of fluids. Since the width of a channel corresponds to the strength of a hopping, one can realize staggered hoppings that are akin to the SSH and BBH models (cf.~Fig.~\ref{SSH_BBH_fig}). By using such a technique, topological corner modes have been proposed and observed in acoustic metamaterials \cite{Ni2020,Serra2018} (cf.~Fig.~\ref{topological_acoustic_fig}(b)).

To derive the sound master equation (\ref{sound_master_eq}), we have ignored dissipative terms originating from the viscosity; however, such dissipative terms can in general make the effective Hamiltonian non-Hermitian. Some papers also explore the combination of such non-Hermiticity and topology of fluids \cite{Zhu2018,Zhang2019}. In particular, the non-Hermitian skin effect (cf.~Sec.~\ref{sec:4c}) has been experimentally realized \cite{Zhang2021} by using feedback controls.

Some of the theoretical proposals of topological phononics are realized in experiments of acoustic metamaterials. In particular, chiral edge modes using circulation devices are observed in a metamaterial with a honeycomb-lattice structure \cite{Ding2019} in Fig.~\ref{topological_acoustic_fig}(a). Another intriguing example of applications of topological phononics can be seen in geophysics \cite{Delplace2017}. Around the equator, unidirectional waves called the Yanai wave and Kelvin wave have been observed. A recent study has revealed that those equatorial waves can be a consequence of the nontrivial topology of the fluid (water or atmosphere) on a sphere. As discussed above, the Coriolis force can be regarded as an effective magnetic force, and its direction changes when fluid crosses the equator. Therefore, the seawater on the north and the south hemispheres exhibits opposite Chern numbers, and the topological modes appear at the boundary. i.e., the equator.

In the last part of this section, we mention the possible breakdown of the bulk-boundary correspondence in continuum systems including fluids. Unlike condensed matter systems with discrete translation symmetries (defined in Eq.~(\ref{discrete-trans-sym})), fluids often exhibit continuum translation symmetries if there are no periodic structures that confine the fluids and thus the system is spatially homogeneous. In such a case, one cannot consider a finite-size wavenumber space (i.e. a Brillouin zone) but should assume an infinite wavenumber space, where the wavenumber can go to infinity. Mathematically, the finite and infinite wavenumber spaces can be distinguished by their compactness. Thus, the conventional topological band theory can be broken in continuum systems, as theoretically predicted in certain phononic systems by numerically confirming the inconsistency between the topological invariant and the number of gapless modes \cite{Tauber2020}. The key to recover the bulk-boundary correspondence is the compactification of the wavenumber space; one should construct an effective Hamiltonian $H(\mathbf{k})$ that exhibits the asymptotically same behavior in the limit of $|\mathbf{k}|\rightarrow \infty$ independently of the direction of $\mathbf{k}$. This compactification can be done by utilizing, e.g., odd-viscosity \cite{Souslov2019}.

\subsubsection{Other topological phononics}\label{sec:5a2}
Topological waves can be found in various classical systems including optics \cite{Rechtsman2013,Khanikaev2013,Ozawa2019}, electrical circuits \cite{Ningyuan2015,Albert2015}, and mechanical lattices \cite{Huber2016}. In particular, mechanical lattices (cf. mass-spring systems) are other setups than fluids to realize localized phononic excitations with a topological origin. For example, if one considers a one-dimensional chain of masses and springs with a staggered structure, one can realize a mechanical counterpart of the SSH model \cite{Kane2014}. Chiral and helical edge modes have also been experimentally realized by using gyroscopic elements \cite{Nash2015,Susstrunk2015}. The rotation of a gyroscopic element induces a Coriolis force, which imitates the Lorentz force and thus leads to an effective magnetic field. 

Non-Hermiticity is also ubiquitous in classical systems because they are typically open systems exchanging energies with environments, and thus their fundamental equations include dissipative terms. In mechanical lattices, friction is a major origin of the non-Hermiticity, while its interplay with topological edge modes is largely unexplored. Another possibility to introduce the non-Hermitian interaction is a nonreciprocal force generated by elaborate devices such as fans attached to a mass point whose direction depends on the distance between two mass points \cite{Scheibner2020b}. By introducing nonreciprocal forces into mechanical lattices, some previous studies \cite{Chen2021,Brandenbourger2019,Ghatak2020} have investigated the non-Hermitian skin effect in classical systems.

\subsection{Hermitian topology in active systems}\label{sec:5b}
\subsubsection{Classical analogs of quantum Hall effect}\label{sec:5b1}

As in passive fluid, one can construct an analogy between the linearized hydrodynamics of self-propelled particles and the Schr{\"o}dinger equation. Recent studies \cite{Souslov2017,Dasbiswas2018,Souslov2019,Shankar2017} have proposed active-matter counterparts of topological insulators based on such linearized hydrodynamics. Starting from the Toner-Tu equations (Eqs.~(\ref{toner-tu-eq1}) and (\ref{toner-tu-eq2})), one can obtain the following linearized equation,
\begin{eqnarray}
\partial_t \delta \rho + (\mathbf{v}_{0} \cdot \nabla)\delta \rho &=& -\rho_{0} \nabla \cdot \delta \mathbf{v}, \label{linear-toner-tu-eq1}
\\
\partial_t \delta \mathbf{v} + \lambda (\mathbf{v}_{0} \cdot \nabla)\delta \mathbf{v} \! &=& \! -2\beta(\mathbf{v}_{0}\cdot \delta\mathbf{v})\mathbf{v}_{0} -c^2 \frac{\nabla \delta \rho}{\rho_{0}}.
\label{linear-toner-tu-eq2}
\end{eqnarray}
where $\delta \rho = \rho - \rho_0$ and $\delta \mathbf{v} = \mathbf{v} - \mathbf{v}_0$ are the fluctuations from the steady state density $\rho_0$ and velocity field $\mathbf{v}_0$, respectively. Here, we ignore the advection terms including $\lambda_2,\lambda_3$ since their effect can be included by renormalization of $\lambda$ \cite{Toner1995}. We also ignore the diffusion terms to focus on the Hermitian topology. If we consider the short-term dynamics of fluctuations compared to the inverse of the decaying rate, we can assume that such dynamics can be captured by a Hermitian matrix without dissipative terms, though nonreciprocity or exceptional points can break such an assumption as we will see later. Furthermore, by assuming that the speed of active particles is $v_0 = \sqrt{\alpha/\beta}$, the first term in the right-hand side of the second equation exhibits no $\alpha$ dependence. In this linearized hydrodynamics, the effect of the activity appears in the additional parameter $\lambda$ and the higher-order term $-2\beta(\mathbf{v}_{\rm ss}\cdot \delta\mathbf{v})\mathbf{v}_{\rm ss}$ of the steady flow. However, if we further assume that the steady-state vector field $\mathbf{v}_{\rm ss}$ is small enough compared to the effective sound speed $c$ (cf.~Eq.~(\ref{toner-tu-eq2})), one can derive the Schr{\"o}dinger-like equation
\begin{equation}
(-i\nabla-\mathbf{V}_{0})^2 \delta \tilde{\rho}=\omega^2 \delta \tilde{\rho},
\label{eq:sch}
\end{equation}
as in passive fluids (cf.~Eq.~(\ref{Schroedinger_like_eq2})). In this equation, $\mathbf{V}_{0} = \omega (\lambda+1)\mathbf{v}_{0}/2$ is the effective vector potential that is proportional to the steady-state flow $\mathbf{v}_{0}$. The advantage of the active system is that the steady-state flow appears spontaneously, i.e., without using external drivings. 

\begin{figure}[]
\centering
\includegraphics[width=8cm]{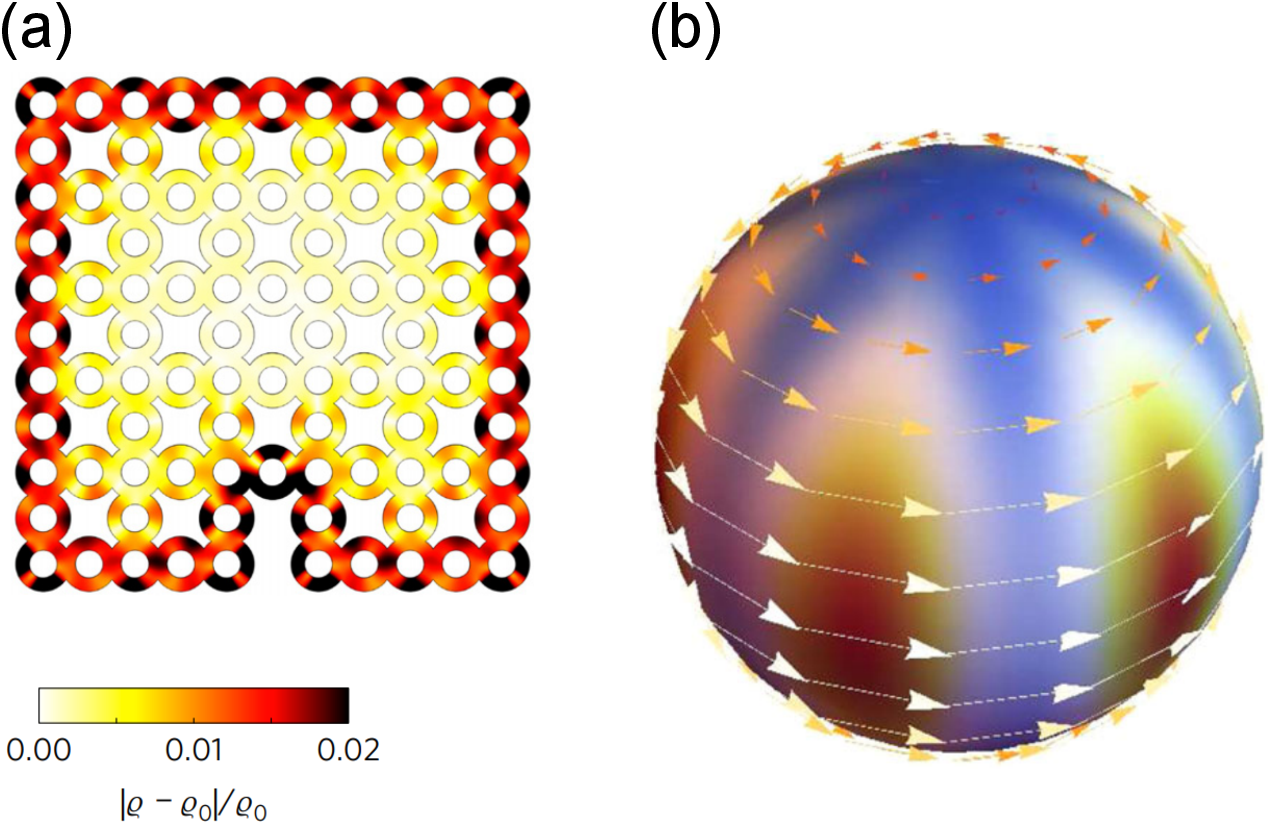}
\caption{\label{active_QHE_fig} Analogs of the quantum Hall effect in active matter. (a) Lieb-lattice structure realizes a nonzero net effective magnetic field and exhibits the topological edge modes. This figure is adapted from A. Souslov, B. C. van Zuiden, D. Bartolo, and V. Vitelli, ``Topological sound in active-liquid metamaterials'' Nat. Phys. 13, 1091--1094 (2017) \cite{Souslov2017}, Springer Nature. Copyright \copyright\  2017. (b) Active matter on a sphere can exhibit chiral flow along the equator. This is because the steady flow induces effective magnetic fields whose signs are opposite between the north and south hemispheres. This figure is adapted from S. Shankar, M. J. Bowick, and M. C. Marchetti, ``Topological Sound and Flocking on Curved Surfaces'' Phys. Rev. X 7, 031039 (2017) \cite{Shankar2017}, licensed under a Creative Commons Attribution 4.0 International License (http://creativecommons.org/licenses/by/4.0/).}
\end{figure}

To construct active-matter counterparts of the quantum Hall effect, one needs to realize the effective magnetic field as we have discussed in Sec.~\ref{sec:3b1} for hard condensed matter and in Sec.~\ref{sec:5a1} for passive fluids. Unlike passive fluids, effective vector potentials can be created by the spontaneous flow of active particles. However, if only a simple structure is available, the effective magnetic fields must be globally canceled, and analogs of the quantum Hall effect are prohibited \cite{Sone2019}. To avoid the cancellation of local effective magnetic fields, a pioneering work of topological active matter \cite{Souslov2017} has proposed to utilize channels aligned on a Lieb-lattice structure that has defective sites [Fig.~\ref{active_QHE_fig}(a)]. In the Lieb-lattice channels, clockwise and anticlockwise flows are alternately generated by the spontaneous motion of active matter, and they create local magnetic fields with opposite signs. If we consider a simple square lattice, the numbers of clockwise and anticlockwise flows are the same, which implies the cancellation of the effective magnetic field. Meanwhile, the defects in the Lieb lattice lead to the imbalance of the number of the clockwise and anticlockwise vortices, and thus the net effective magnetic field remains nonzero. As a consequence of the nontrivial topology due to the nonzero effective magnetic field, localized boundary modes are observed in the numerical calculation of the eigenvectors of the effective Hamiltonian of the active metamaterial. To experimentally observe such localized boundary modes, one should oscillate the active fluid at the source point, and confirm the existence of the chiral edge wave. Such a chiral edge mode has been also confirmed in the numerical simulation of the Lieb-lattice model.

\begin{figure}[]
\centering
\includegraphics[width=8cm]{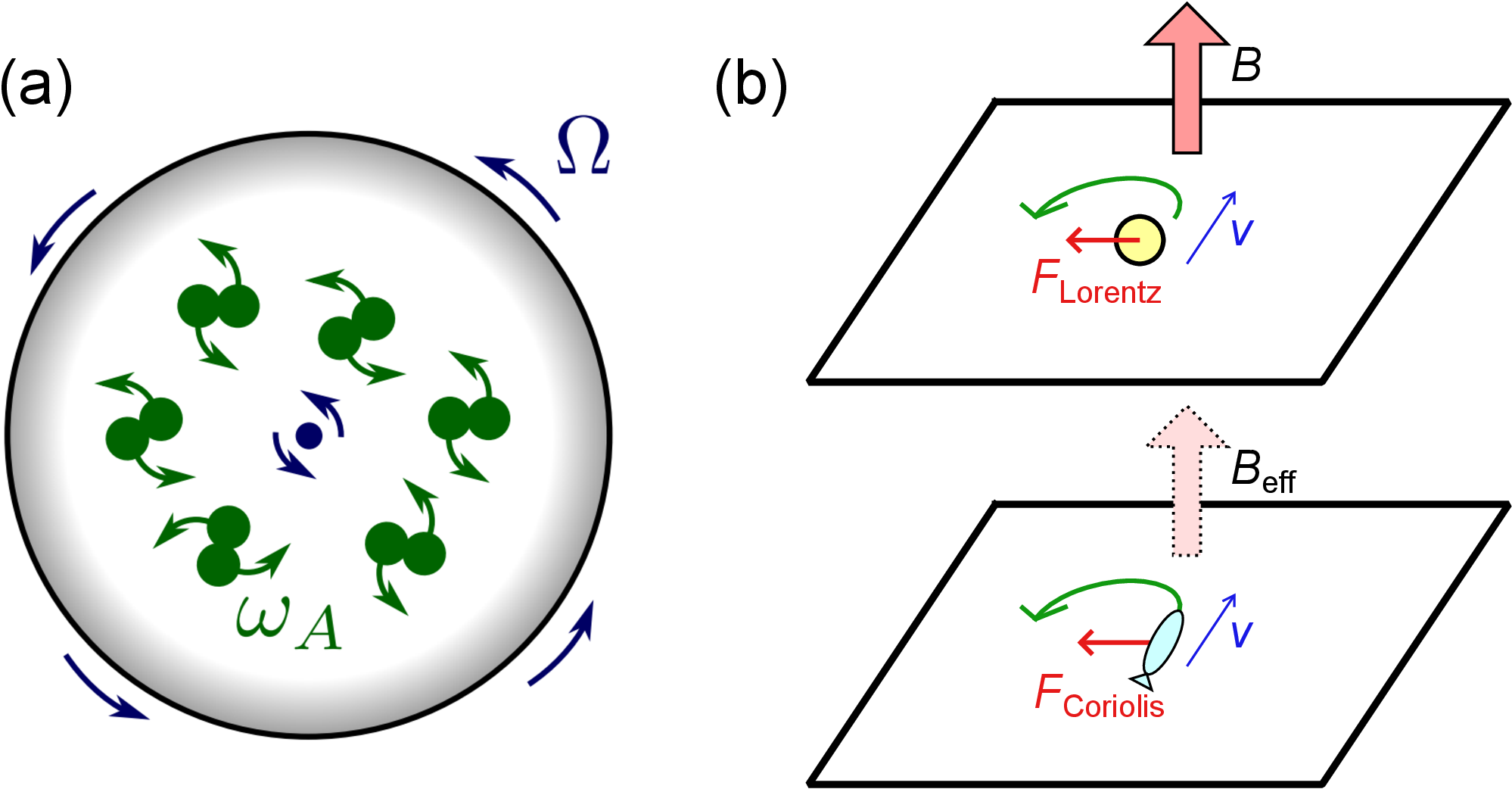}
\caption{\label{chiral_actmat_QHE_fig} Chiral active matter realizing analogs of the quantum Hall effect. (a) Collections of self-rotating particles exhibit topological boundary modes. This figure is adapted from A. Souslov et al., ``Topological Waves in Fluids with Odd Viscosity'' Phys. Rev. Lett. 122, 128001 (2019) \cite{Souslov2019}, Copyright \copyright\  2019 by the American Physical Society. (b) Chiral motion of active matter induces a Coriolis-like force. Since both the Coriolis-like force and the Lorentz force act on particles perpendicularly to the velocity, one can make an analogy between the self-rotation of active matter and cyclotron motion under a magnetic field.}
\end{figure}

The counterpart of the quantum Hall effect can also be constructed from active matter on curved surfaces \cite{Shankar2017,Green2020}. As in a Lieb-lattice model, spontaneous flows of active matter induce effective magnetic fields without the help of, e.g. the external rotation (cf.~the self-rotation of a sphere). For example, active matter on a sphere goes around and creates north and south poles and hemispheres reflecting on the direction of rotation. Then, the localized modes can be observed at the boundary of hemispheres [Fig.~\ref{active_QHE_fig}(b)], which is analogous to the Yanai and Kelvin waves on the Earth discussed in Sec.~\ref{sec:5a1}. One surprising point is that the spontaneous flow is stabilized by the activity and interactions of active particles, and thus the topological boundary modes can emerge independently of an initial condition.
Topological boundary modes can also emerge on more complex surfaces, such as a torus and a helicoidal sphere \cite{Green2020}. When such a complex surface is described as $(x,y,z) = (r(\theta)\cos \phi, r(\theta)\sin \phi, \zeta(\theta))$, the direction of the effective magnetic force depends on the sign of $dr/d\theta$ as the direction of the Coriolis force differs in the north and south hemisphere. Therefore, the Chern number is also changed by $dr/d\theta$, and topological modes appear at the boundary.

\begin{figure}[]
\centering
\includegraphics[width=8cm]{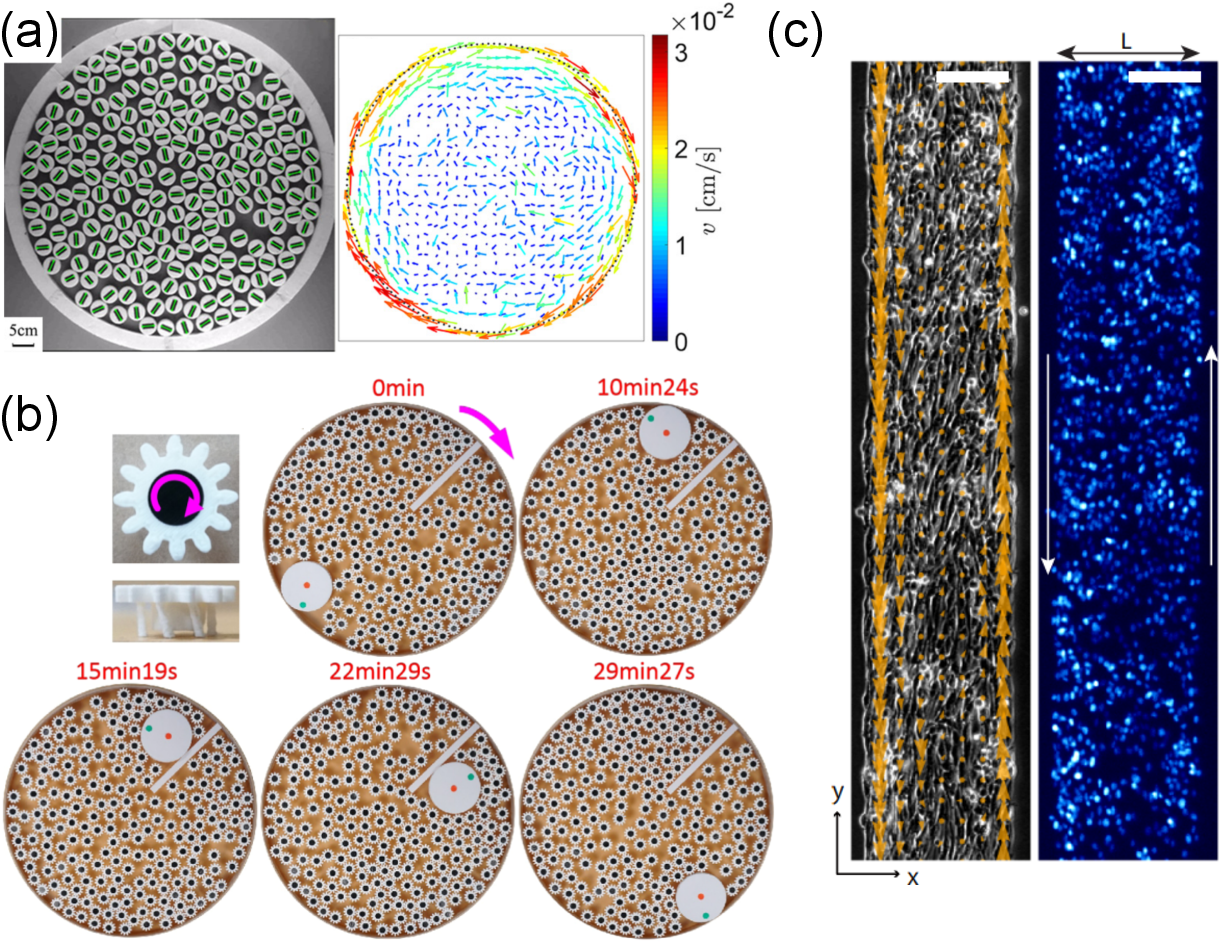}
\caption{\label{experiment_topoactmat_fig} Experimental realizations of topological chiral active matter. (a) Collections of self-rotating robots exhibit unidirectional flows along the circular boundary. This figure is adapted from X. Yang, C. Ren, K. Cheng, and H. P. Zhang, ``Robust boundary flow in chiral active fluid'' Phys. Rev. E 101, 022603 (2020) \cite{Yang2020b}, Copyright \copyright\  2020 by the American Physical Society. (b) Active spinners can transport cargo along the boundary. Even if a wall exists at the boundary, the cargo avoids it, which indicates the robustness of the topological boundary flow. This figure is adapted from Q. Yang et al., ``Topologically Protected Transport of Cargo in a Chiral Active Fluid Aided by Odd-Viscosity-Enhanced Depletion Interactions'' Phys. Rev. Lett. 126, 205502 (2021) \cite{Yang2021}, Copyright \copyright\  2021 by the American Physical Society. (c) Neural progenitor cells exhibit chiral motions and induce topological edge flows. This figure is adapted from L. Yamauchi et al., ``Chirality-driven edge flow and non-Hermitian topology in active nematic cells'' arXiv:2008.10852 (2020) \cite{Yamauchi2020}.}
\end{figure}

Chiral active matter, i.e., active matter exhibiting rotating motions (cf.~Sec.~\ref{sec:2c3}) is also useful to realize topological active matter \cite{Dasbiswas2018,Souslov2019} because its chirality naturally leads to the effective Lorentz force [Fig.~\ref{chiral_actmat_QHE_fig}]. Since chiral active matter often shows self-rotation, when it goes to, e.g., the left direction at a certain moment, it tends to move in the upper or lower direction at the next moment. This implies that chiral active matter feels a rotational force that is perpendicular to the velocity of particles, which is analogous to the Lorentz force. The effective Lorentz force can appear as the term $\omega_0\nabla \times \mathbf{v}$ in the hydrodynamic equations. By linearizing the hydrodynamics of chiral active matter, we obtain the following equation,
\begin{eqnarray}
&{}& \partial_t \delta\rho + \rho_0 \nabla \cdot \delta \mathbf{v} = 0, \label{linearized-chiral-active-hydrodynamics1}\\
&{}& \partial_t \delta\mathbf{v} = -\alpha \delta\mathbf{v}-\nabla \delta\rho + \omega_0\nabla \times \delta\mathbf{v}
\label{linearized-chiral-active-hydrodynamics2}
\end{eqnarray}
where we assume the homogeneous steady state without aligning orders ($\mathbf{v}_0=0$) and what we have imposed to derive the effective Hamiltonian of active matter in the first part of this section. The term $\omega_0\nabla \times \mathbf{v}$ is unique to chiral active matter with $\omega_0$ determining the speed of self-rotation or the radius of the circular orbit. This chirality term breaks the time-reversal symmetry and opens a band gap around $\omega=0$. The Chern numbers of the upper and lower dispersions become $C = \pm 2$, which leads to the corresponding boundary modes.

The subtle point in the bulk-boundary correspondence in fluids mentioned above (Sec.~\ref{sec:5a1}) can be also seen in active hydrodynamics. Specifically, without the odd-viscosity term, the gapless boundary modes disappear \cite{Green2020} even if the bulk bands have nontrivial topology. Fortunately, the odd viscosity can ubiquitously emerge in chiral active matter as discussed in Sec.~\ref{sec:2}, and thus we can expect the conventional bulk-boundary correspondence between the Chern number and edge modes in active fluid.

Topological active matter utilizing chirality seems to be more amenable to experimental realizations since there is no need to prepare engineered structures. Indeed, some studies have utilized artificial chiral active matter, such as robotic rotators \cite{Yang2020b} and vibrated granular gears \cite{Yang2021}, as well as active spinners \cite{Van2016,Soni2019}, which are introduced in Sec.~\ref{sec:2a4}. The robotic rotators have an asymmetric structure and self-vibration using motors, leading to their rotational movements. After preparing the collection of such robotic rotators, one can observe the localization of the distribution of the rotators as discussed in Ref.~\cite{Dasbiswas2018}. Vibrated granular gears have also asymmetric legs, and external vibration leads to their active motion. If one puts a passive cargo in the collection of the granular gears, the robust transport of the cargo can be observed along the edge of the system, which is analogous to the topological boundary modes discussed in Ref.~\cite{Souslov2019}.

Since chirality is abundant in biological active matter as is discussed in Secs.~\ref{sec:2a2} and \ref{sec:2a3}, topological chiral active matter is also realizable in biological systems. An experimental study \cite{Yamauchi2020} shows the chiral collective motion of neural progenitor cells. Specifically, when confining such cells in a wide channel, a chiral boundary flow of cells can be observed. Boundary modes are also found in power spectra of fluctuations of the flow field, which directly confirms the existence of topological edge modes in the linearized hydrodynamics. The neural progenitor cells are self-propelled rods that involve nematic tensors when described in the hydrodynamic equations; their hydrodynamic equations look like those in Eqs.~(\ref{self-propelled-eq1}--\ref{self-propelled-eq3}) with the Coliolis-like force and the odd-viscosity term. One can derive linearized equations similar to those obtained from active hydrodynamics of polar active matter and numerically confirm its nonzero Chern number. The importance of odd-viscosity terms in the bulk-boundary correspondence in neural progenitor cells has also been discussed~\cite{Yamauchi2020}. More recently, another group has found chiral flows at the boundary of bacterial colonies \cite{Li2024}, which also seems to have a topological origin.

In addition, the classical analog of the quantum Hall effect can also be implemented in a model of human group dance. Reference \cite{Du2023} proposed an algorithm to simulate a chiral edge current in the Harper-Hofstadter model \cite{Harper1955,Hofstadter1976}, a lattice Hamiltonian under the existence of an external magnetic field, by using discrete dynamics, corresponding to dance moves. The emergence of a chiral edge current has been demonstrated in the dance performed by tens of people aligned in a square lattice.

\begin{figure}[]
\centering
\includegraphics[width=8cm]{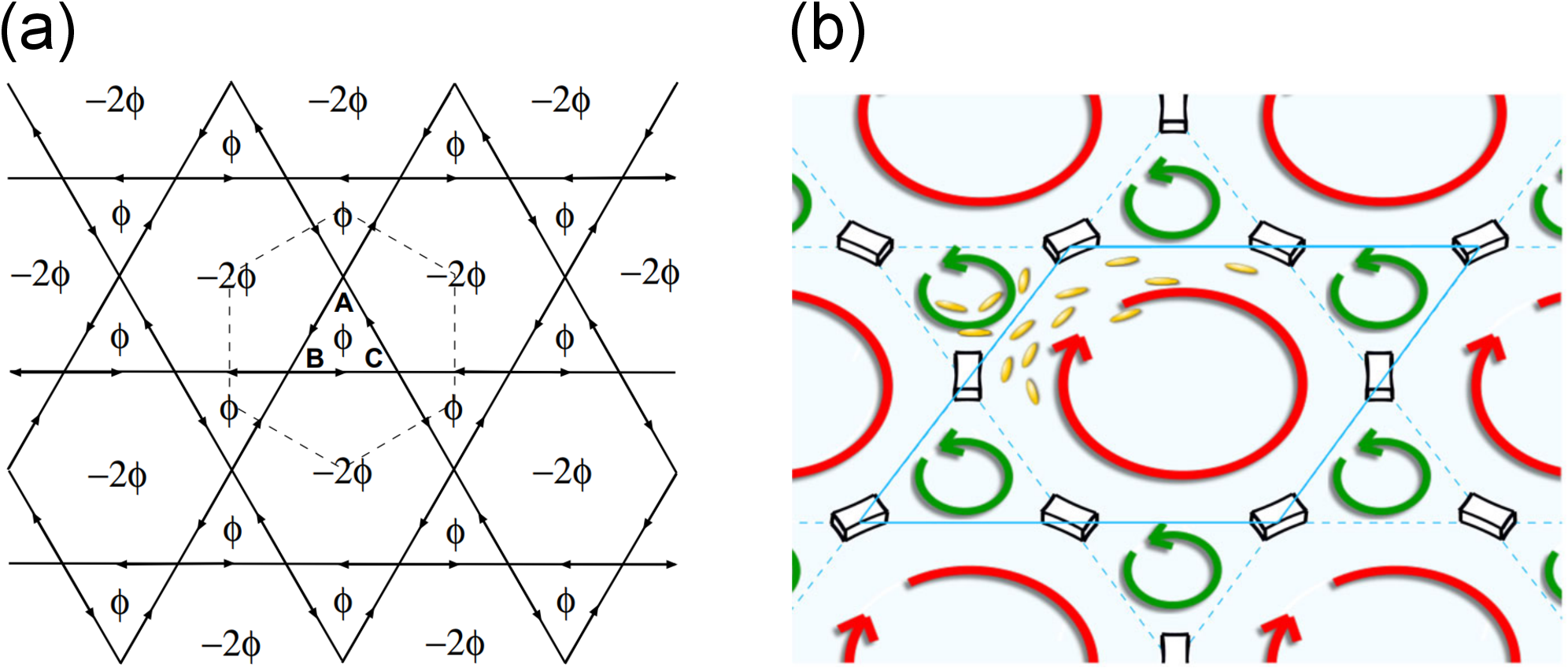}
\caption{\label{active_QAHE_fig} (a) Kagome-lattice model of the quantum anomalous Hall effect in condensed matter physics. The arrows show the direction of the vector potential. This figure is adapted from K. Ohgushi, S. Murakami, and N. Nagaosa, ``Spin anisotropy and quantum Hall effect in the kagomé lattice: Chiral spin state based on a ferromagnet'' Phys. Rev. B 62, R6065(R) (2000) \cite{Ohgushi2000}, Copyright \copyright\  2000 by the American Physical Society. (b) Active-matter counterpart of the quantum anomalous Hall effect. Steady flows denoted by the red and green curved arrows generate the effective vector potential similar to that in panel (a). This figure is adapted from K. Sone and Y. Ashida, ``Anomalous Topological Active Matter'' Phys. Rev. Lett. 123, 205502 (2019) \cite{Sone2019}, Copyright \copyright\  2019 by the American Physical Society.}
\end{figure}

\subsubsection{Classical analogs of quantum anomalous Hall effect}\label{sec:5b2}
As is discussed in Sec.~\ref{sec:3}, a net (effective) magnetic field is not a requirement to realize a topological band structure. In particular, counterparts of the quantum anomalous Hall effect can also be realized in active matter. To construct such a topological active-matter system, one can utilize a kagome-lattice structure in Fig.~\ref{active_QAHE_fig} \cite{Sone2019}, which has been also considered in a model of the quantum Hall effect [Fig.~\ref{active_QAHE_fig}(b)] in a ferromagnetic material \cite{Ohgushi2000}. We note that the kagome lattice is a dual lattice of the honeycomb lattice, and thus the same physics explains the nontrivial topology of the kagome-lattice model and the Haldane model in Fig.~\ref{Haldane_fig}. The theoretical proposal of an active-matter counterpart of the quantum anomalous Hall effect utilized the experimental technique to control bacterial turbulence by poles (discussed in Sec.~\ref{sec:2a2}), and rectify the steady-state flow so that it can imitate the vector potential in the kagome-lattice model of the quantum anomalous Hall effect. Then, it was numerically shown that the linearized dynamics (Eqs.~(\ref{linear-toner-tu-eq1}) and (\ref{linear-toner-tu-eq2})) of the proposed system exhibit gapless edge modes.

Analogs of the quantum anomalous Hall effect may be advantageous in that they remove some intricate structures utilized in active-matter counterparts of the quantum Hall effect. Specifically, the quantum-Hall-like system of active matter contains defects, curved surfaces, or extra degrees of freedom corresponding to the rotational force, while the topological active matter analyzed in Ref.~\cite{Sone2019} requires no such structures. One can prove that the net effective magnetic field must disappear without these intricate structures as follows; one can write the net vorticity as an integral of the rotation of the effective vector potential $\mathbf{V}_{\rm ss}$ over the unit cell. Using the Stokes' theorem, one can change the integral into a line integral on the boundary of the unit cell,
\begin{equation}
\int_{\Omega} (\nabla \times \mathbf{V}_{\rm ss}) \cdot d \mathbf{S} = \oint_{\partial \Omega} \mathbf{V}_{\rm ss} \cdot d\mathbf{r},
\label{eq:stokes}
\end{equation}
where we assume the absence of defects. However, the line integral above must be canceled in a periodic system. Since the vorticity of $\mathbf{V}_{\rm ss}$ is a unique resource of the effective magnetic field in nonchiral active matter, the net effective magnetic field must be zero. This conclusion implies the importance of utilizing analogs of the quantum anomalous Hall effect if we want to realize topological active matter without the intricate structures.

Analogs of the quantum anomalous Hall effect can also be utilized in active biological networks. For example, the game-theoretic model of competing agents on a kagome-lattice network can exhibit topological edge flows \cite{Knebel2020,Yoshida2021,Yoshida2022}. In more detail, one considers agents that play the rock-paper-scissors game and can move to neighbor sites. If one properly constructs a kagome-lattice network where the agents move, the agent system exhibits effective vector potentials similar to that in a quantum anomalous Hall system and thus shows a robust boundary density wave. We note that the governing equations are the coupled Lotka-Volterra equation \cite{Kingsland1995}, a nonlinear equation of a particle-number distribution as the rate equation of a biochemical network (\ref{chemical_rate_equation}), while the linear analysis around the steady state bridges such nonlinear equations, and one can employ the notion of topology as is the case in active hydrodynamics. 

Recently, chiral edge modes have been experimentally observed in the electrical potential of a cell network \cite{Ori2023}. Such a chiral current  can be realized by genetically engineered human embryonic kidney cells that express an inward-rectifier potassium channel (K$_{\rm ir}$2.1) and a voltage-gated sodium channel (Na$_{\rm V}$1.5). Each channel of a cell can show a different active potential depending on the concentration of ions. Then, the boundary between the K$_{\rm ir}$2.1-expressing cells and the Na$_{\rm V}$1.5-expressing ones exhibits a chiral current observed as a wave-like propagation of the active potentials. While this chiral current has been analyzed by using the Hodgkin-Huxley-like model \cite{Hodgkin1952} and coupled FitzHugh-Nagumo model \cite{Fitzhugh1961,Nagumo1962}, the linear couplings used in these models are not reminiscent of those of the models of topological insulators. Instead, they may utilize the nonlinearity of models to realize nontrivial topology.

At the end of this section, we discuss a possible application of the active-matter analogs of the quantum (anomalous) Hall effect. One prominent feature of a topological edge mode is its unidirectional propagation (cf.~Fig.~\ref{QHE_fig}(a)). Utilizing the unidirectionality, one may construct phononic counterparts of diodes (heat diodes) \cite{Yang2015}. As is discussed above, the advantage of active metamaterials compared to passive ones is that they require no external forces to realize effective magnetic fields that lead to topologically nontrivial band structures.

\subsection{Non-Hermitian topology in active systems}\label{sec:5c}
\subsubsection{Skin effect}\label{sec:5c1}

\begin{figure}[]
\centering
\includegraphics[width=8cm]{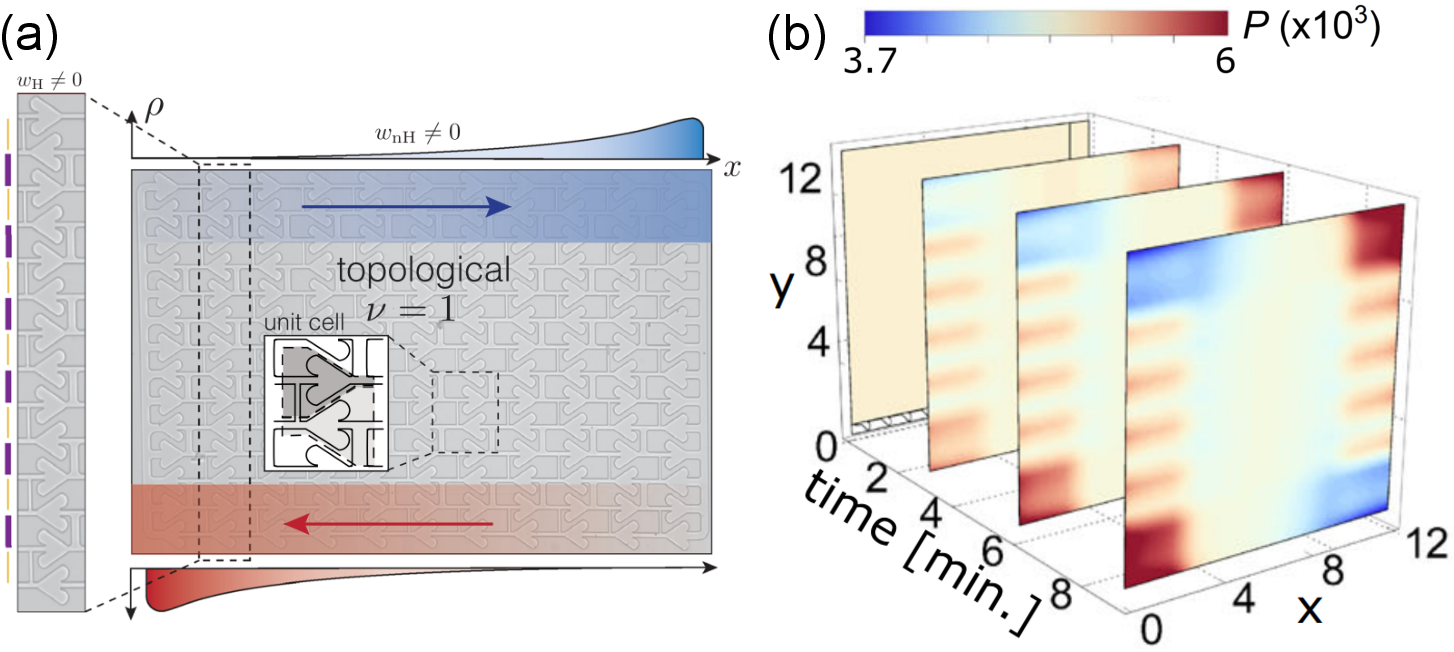}
\caption{\label{skineffect_actmat_fig} Second-order skin effect in active matter. (a) Heart-shaped channels induce unidirectional motion of active matter and realize the second-order non-Hermitian skin effect. (b) Dynamics of the proposed system. The color shows the density distribution of the active particles, which exhibits the localization at the corner of the system. These figures are adapted from L. S. Palacios et al., ``Guided accumulation of active particles by topological design of a second-order skin effect'' Nat. Commun. 12, 4691 (2021) \cite{Palacios2021}, licensed under a Creative Commons Attribution 4.0 International License (http://creativecommons.org/licenses/by/4.0/).}
\end{figure}

In the previous section, we have focused on the Hermitian topology of active matter, while the effective Hamiltonian of active matter can naturally become non-Hermitian as can be seen from Eq.~(\ref{linearized-chiral-active-hydrodynamics2}). Such non-Hermiticity is fairly common because active matter constantly consumes energy stored inside (e.g., animals get energy from their food and use it to move), which leads to dissipation. Utilizing such non-Hermiticity, active matter can exhibit various topological phenomena, such as skin effects (cf.~Sec.~\ref{sec:4c}) and exceptional edge modes (cf.~Sec.~\ref{sec:4d}).

An experimental study shows that active matter in engineered microchannels [Fig.~\ref{skineffect_actmat_fig}(a)] can exhibit the second-order non-Hermitian skin effect \cite{Palacios2021}. Here, it imitates a ladder-like model of the second-order skin effect discussed in Sec.~\ref{sec:4c2} \cite{Okugawa2020}. The nonreciprocal coupling is realized by using heart-shaped ratchets that are located on the way of the horizontal microchannels. Such ratchets convert the direction of motion of active particles; active particles coming from the front of the heart-shaped channel get into the channel and move along the channel's wall. When they come out of the channel, the direction of motion will be changed. Inversely, active particles from the other side seldom get inside the channel due to the shape of the entrance. Therefore, there exist more active particles moving to the front of the heart-shaped channel than ones moving in the opposite direction, which indicates the nonreciprocal transport of active particles. The probability of moving to the neighbor lattice point depends on the width of the channels. Finally, one obtains the microchannels whose hopping amplitudes imitate those of the tight-binding model of the second-order skin effect. In real experiments, they observed the distribution of the active colloids (cf.~Sec.~\ref{sec:2a4}) in the micro channels and confirmed its localization to the corner of the system shown in Fig.~\ref{skineffect_actmat_fig}(b).

A very recent theoretical work \cite{Cheng2025} has also discussed further localization of chiral edge modes by the non-Hermitian skin effect in a nonreciprocal kagome network. They have analyzed such active fluid in a nonreciprocal network by using a transfer matrix and Floquet (i.e., time-periodic-system) techniques \cite{Rudner2013} and found that chiral edge states characterized by the Chern number or a Floquet topological invariant can have a larger amplitude at one corner than they do at the others.

The non-Hermitian skin effect has been also observed in chiral active matter without using complicated periodic structures. Cells mentioned in the previous section \cite{Yamauchi2020} also show the localization of the bulk modes and the modification of the dispersion relation, i.e., the non-Hermitian skin effect. One intriguing aspect of the skin effect observed here is the direction of the localization depends on the wavenumber in the other direction. This wavenumber-dependent localization can be analyzed by one directional winding number (cf.~Eq.~(\ref{energywinding})) defined in each wavenumber sector. Similar wavenumber dependence is also discussed in electrical circuits \cite{Hofmann2020}.

From a broad perspective, the non-Hermitian skin effect is ubiquitous in various active systems. For example, agents playing the rock-paper-scissors game on a one-dimensional lattice can exhibit the non-Hermitian skin effect \cite{Knebel2020,Yoshida2022}, due to the imbalance of the rate of moving toward the left and right directions. Similar localization can be observed in a stochastic system with a periodic configuration space and nonreciprocal hopping rates \cite{Murugan2017}. In such a stochastic model, a localized steady state is obtained and the direction of the localization depends on which direction of hoppings is dominated. One can also observe the non-Hermitian skin effect as the difference in the system-size scaling of the relaxation time of stochastic processes \cite{Sawada2024} as discussed in Sec.~\ref{sec:4c4}.  We note that some biological systems, such as kinesins on a biofilament \cite{Leduc2012}, magnetotactic bacteria \cite{Klumpp2016,Rupprecht2016}, and cell adhesions \cite{dAlessandro2021,Parsons2010}, can be described by a stochastic process with nonreciprocal hopping, and thus non-Hermitian topology in stochastic dynamics may be realized by active matter.
This effect can be also seen in the quantum version of active matter proposed as a non-Hermitian spin model \cite{Adachi2022,Takasan2024}.

In practice, the non-Hermitian skin effect may play an important role in metamaterial designing and biophysics. Active metamaterials can realize nonreciprocal coupling by using active devices driven by external energy injection. In particular, a time-reversal odd response known as odd elasticity \cite{Scheibner2020b} has attracted interest because it can realize non-Hermitian properties in mechanical metamaterials. Localization of all the bulk modes in the skin effect implies the amplification of the response to the perturbation to one side. Thereby, this amplification would be utilized in lasing devices \cite{Schomerus2020} or sensors \cite{Mcdonald2020}. As an example, in a nonreciprocal mechanical metamaterial~\cite{Ghatak2020}, Ref.~\cite{Schomerus2020} has shown a highly large response to perturbations due to nonorthogonality of the topological edge states. Since the nonorthogonality also emerges at singularity points on the generalized Brillouin zone~\cite{Yokomizo2023v2}, it would be intriguing to investigate a non-Hermitian response from the viewpoint of the non-Bloch band theory (cf. Sec.~\ref{sec:4c3}). The localization and nonreciprocal coupling are also ubiquitous in biological processes \cite{Leduc2012,Mayer2010}, which we have discussed in Sec.~\ref{sec:2b}. We may understand such localization from a topological point of view, which might bring us a universal theory of the robustness of biological systems.

Although recent intensive studies have shed light on the point-gap topology leading to the non-Hermitian skin effect in active systems, a comprehensive understanding of the line-gap topology is still an open problem. Pioneering works have theoretically proposed that metamaterials with odd elasticity can realize topological edge states under the non-Hermitian skin effect~\cite{Zhou2020,Scheibner2020}. Remarkably, Ref.~\cite{Zhou2020} has shown that, in a rotor chain emulating the SSH model, the winding number defined from the generalized Brillouin zone (cf. Sec.~\ref{sec:4c3}) can predict the existence or absence of topological edge states. The previous work has also applied a similar non-Bloch framework to a 2D rotor lattice emulating the Haldane model to predict the topological edge states. However, the relation between the framework and the amoeba formulation (cf. Sec.~\ref{sec:4c4}) remains unclear.

\begin{figure}[]
\centering
\includegraphics[width=8cm]{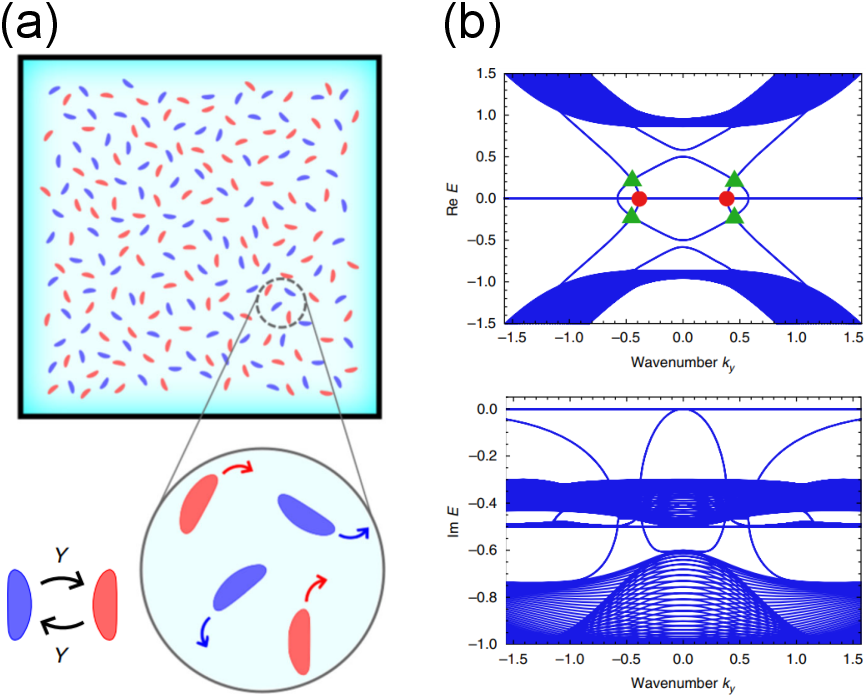}
\caption{\label{exceptional_edgemode_actmat_fig} Exceptional edge modes in chiral active matter. (a) Mixture of clockwise and counterclockwise rotating particles can exhibit exceptional edge modes. Non-Hermitian coupling can arise from the aligning interaction between the particles with different chirality and the flip of the direction of the rotation. (b) Dispersion relation of the exceptional edge modes in chiral active matter. The red circles represent the exceptional points that protect the exceptional edge modes from the gap opening. These figures are adapted from K. Sone, Y. Ashida, and T. Sagawa, ``Exceptional non-Hermitian topological edge mode and its application to active matter'' Nat. Commun. 11, 5745 (2020) \cite{Sone2020}, licensed under a Creative Commons Attribution 4.0 International License (http://creativecommons.org/licenses/by/4.0/).}
\end{figure}

\subsubsection{Exceptional point and exceptional edge modes}\label{sec:5c2}
Exceptional points, yet another non-Hermitian topological feature, can play an important role in active-matter dynamics. One example can be found in the exceptional edge modes of chiral active matter. A study of exceptional edge modes \cite{Sone2020} has proposed that a mixture of right- and left-rotating active particles in Fig.~\ref{exceptional_edgemode_actmat_fig}(a) can show exceptional edge modes. As discussed above, right- or left-rotating particles consist of counterparts of a Chern insulator for each.

To discuss the existence of exceptional edge modes in such a mixture of chiral active matter, Ref.~\cite{Sone2020} has first considered a particle-based model of chiral active particles. It introduces polar or antipolar interaction and chirality flips and assumes that the particles are dilute and thus three or more particle collisions do not occur. Then, the stochastic dynamics of the model is described as
\begin{eqnarray}
 \mathbf{r}_j(t+dt) &=& \mathbf{r}_j(t) + v_0 dt\mathbf{n}(\theta_j),\label{CAM-particle1}\\
 \theta_j(t+dt) &=& \left\{
  \begin{array}{c}
  \theta_j(t) + \omega^{c(j)} dt + \xi_j \\ \mbox{(no confliction)} \\ {} \\
  \Phi^{c(j),c(l)}(\theta_j(t),\theta_l(t)) + \omega^{c(j)} dt + \xi_j \\ \mbox{(when $j$th and $l$th particles conflict)}
  \end{array}
  \right.
\label{CAM-particle2}
\end{eqnarray}
where $\omega^{c(j)} = \pm \omega_0$ represents the circular motion of chiral active matter and $c(j)=\pm 1$ represents the direction of rotation. $\Phi^{i,k}(\theta_j(t),\theta_l(t))$ determines the direction of an active particle after the collision, which depends on the chiralities of two particles; in the case of the same chirality $c(j)=c(l)$ (different chiralities $c(j)\neq c(l)$), it represents the polar (anti-polar) interaction. The model also includes the chirality flip by flipping the sign of $c(j)$ at the rate of $\gamma$.

To combine the particle-based model with the Hamiltonian of exceptional edge modes, the previous paper has further derived the hydrodynamics of chiral active matter by using the Boltzmann equations similar to Eq.~(\ref{Boltzmann_eq}) in Sec.~\ref{sec:2c2}. Compared to the Boltzmann equation of achiral active matter in Eq.~(\ref{Boltzmann_eq}), that of the present model has a linear chirality term $\omega^i\partial_\theta f^i$ with $i=\pm1$ representing the chirality and $f^i$ being the distribution function of chiral active matter with the chirality $i$. The chirality flip also introduces a new linear term $\gamma (f^{-i} - f^i)$, and the other differences are included in the collision integral (cf. Eq.~(\ref{collision_integral})). Remaining the leading order of these newly emerging terms, one obtains linear terms in its hydrodynamics that are unique to the present model. Unfortunately, the odd viscosity does not appear as a leading-order term, while it ubiquitously appears in chiral active matter (cf. Sec.~\ref{sec:2c3}) and plays an essential role to define the topology of active fluid as discussed in the previous section. Thus, adding an odd-viscosity term by hand and linearizing the hydrodynamic equations, one finally obtains the following effective Hamiltonian:
\begin{equation}
H =   \left(
  \begin{array}{cccc}
   H_0 +A & C \\
   -C^{\ast} & H_0^{\ast} +A
  \end{array}
  \right) \label{CAM-Hamiltonian},
\end{equation}
where $H_0$ is an effective Hamiltonian of active matter counterpart of the quantum Hall effect
\begin{equation}
H_0 =   \left(
  \begin{array}{ccc}
   0 & -i \partial_x & -i \partial_y \\
   -i\partial_x & 0 & -i(\omega_0+\nu^o \Delta) \\
   -i \partial_y & i(\omega_0+\nu^o \Delta) & 0
  \end{array}
  \right) \label{CAM-H0},
\end{equation}
and $A$ and $C$ are 
\begin{equation}
A =   \left(
  \begin{array}{ccc}
   -i\gamma & 0 & 0 \\
   0 & -i\beta & 0 \\
   0 & 0 & -i\beta
  \end{array}
  \right) \label{CAM-diagonal},
\end{equation}
\begin{equation}
C =   \left(
  \begin{array}{ccc}
   i\gamma & 0 & 0 \\
   0 & i\beta & 0 \\
   0 & 0 & i\beta
  \end{array}
  \right) \label{CAM-coupling},
\end{equation}
where $\nu_o$ and $\beta$ represent the strength of the odd viscosity and the anti-poplar interaction between different chirality, respectively. The alignment interaction and chirality flips lead to the non-Hermitian coupling terms like dissipative terms discussed in Sec.~\ref{sec:4d}. Consequently, the topological gapless modes form a pair of exceptional points as shown in Fig.~\ref{exceptional_edgemode_actmat_fig}(b). Since chirality is broadly relevant to biological active matter, such as bacteria \cite{DiLuzio2005} and motility assays \cite{Sumino2012} (cf.~Secs.~\ref{sec:2a2} and \ref{sec:2a3}), exceptional edge modes should appear in such chiral active matter if one can realize the coexistence of left- and right-moving particles and their interactions.

Exceptional points can also affect the asymptotic behavior of the (dynamical) phase transition of active matter. If one considers chasers and preys on a circle, the eternally chasing (i.e., oscillatory) phase and the caught (i.e., static) phase can emerge depending on the parameters. A similar active system with nonreciprocal interactions can be realized in a collection of robots \cite{Fruchart2021} (cf.~Sec.~\ref{sec:2a4}). By analyzing such robotic active matter, the transition to the oscillatory phase accompanies exceptional points in the parameter space. The difference from the conventional phase transition emerges in, e.g., their upper critical dimension $d_c=8$ \cite{Hanai2020}, which is different from passive systems such as the Ising model ($d_c=4$). Similar transitions and appearance of the exceptional points are also discussed in active Cahn-Hilliard models \cite{Saha2020,You2020,Suchanek2023}, which is originally a continuum model of the phase separation and has an additional nonreciprocal term that expresses activity.

\subsection{Other activity-induced topological phenomena}\label{sec:5d}
Non-Hermiticity in active systems can further enrich topological phenomena associated with localized edge modes. A study on biological network systems \cite{Tang2021} reveals that nonreciprocal hoppings reproduce a chiral edge current in a steady state of a stochastic process. Specifically, it considers the reaction network in Fig.~\ref{chemical_EvelynTang_fig} and numerically shows that the steady state supports the unidirectional edge current. We note that in a stochastic process with a Hermitian rate matrix (in other words, a purely dissipative stochastic process), a steady-state current is prohibited, while the non-Hermiticity due to the nonreciprocal hoppings induces the edge current. The emergence of the edge current can be schematically understood; once the particle arrives at the edge, the hopping to the edge site is much larger than that to the bulk site. Therefore, many particles stay around the edge and are transported along the direction of the hopping, which induces the chiral edge current.

Activity can also drive the system into a nonequilibrium state, which enables a spontaneous excitation of the edge modes in topological metamaterials. Such excitation of the topological edge modes has been discussed in a mass-spring model with a honeycomb-lattice structure \cite{Liao2020}. If one considers a mass-spring system influenced by active matter moving around it, each mass is affected by a colored noise, which is often utilized to model the stochastic dynamics of active particles \cite{Fodor2016}. Such a colored noise enables the excitation of the edge modes, which cannot be achieved by using white noise. Thus, edge oscillations can spontaneously occur under such an active circumstance.

\section{Perspectives}\label{sec:6}
\subsection{Topological active matter in one and three dimensions}\label{sec:6a}
While topological active matter has been well explored in two-dimensional systems, topological phenomena should be also abundant in the other dimensions as we have discussed in Secs.~\ref{sec:3} and \ref{sec:4}. One possibility is a one-dimensional case, which can be realized by utilizing, e.g., molecular motors \cite{Leduc2012}, microswimmers in a channel \cite{Katuri2018}, and cell adhesions \cite{dAlessandro2021,Parsons2010}. In particular, while the skin effect is realized by using microchannels (cf.~Sec.~\ref{sec:5c1}), the associated breakdown of the bulk-boundary correspondence (cf.~Sec.~\ref{sec:4c3}) has so far not been confirmed. Due to the nonreciprocity inherent to active matter, one-dimensional active systems should offer an ideal platform to study non-Hermitian topology.

Active matter also exists in three-dimensional systems, including a school of fish, a flock of birds, and bacterial turbulence. Nevertheless, topological features of such three-dimensional active matter are largely unexplored. In three-dimensional systems, topological materials can exhibit unique properties without two-dimensional counterparts, such as Fermi arcs in topological semimetals \cite{Armitage2018} and exceptional topological insulators \cite{Denner2021}. Meanwhile, active hydrodynamics in three-dimensional systems is less established than in two-dimensional ones. In particular, chirality of active matter may behave differently from a uniform effective magnetic field, while it still breaks the time-reversal symmetry and thus may introduce a synthetic gauge field. Thus, advancing our understanding of three-dimensional active hydrodynamics can stimulate further studies on topological active matter.

\subsection{Nonlinear topology}\label{sec:6b}
To derive the effective Hamiltonian of active matter, we have linearized the hydrodynamic equations of active matter for the sake of simplicity (cf.~Eqs.~(\ref{linear-toner-tu-eq1},\ref{linear-toner-tu-eq2})). While the effective Hamiltonian can describe the dynamics of small fluctuations of the density and velocity fields around the steady state, the whole dynamics is ultimately governed by the nonlinear hydrodynamic equations. Recently, studies on topology of nonlinear systems have been initiated given the fact that nonlinearity is fairly common in classical systems, including photonics \cite{Lumer2013} and mechanical lattices \cite{Chen2014}. So far, studies of nonlinear topology have revealed interplay between topology and well-known nonlinear phenomena such as topological edge solitons \cite{Leykam2016} and topological synchronization \cite{Kotwal2021,Sone2022b}.

Recent studies \cite{Tuloup2020,Zhou2022,Sone2024} have also discussed a possible extension of the topological invariants to nonlinear systems on the basis of the nonlinear eigenvalue problems. Specifically, when we consider a nonlinear dynamical equation $i\dot{\mathbf{u}} = f(\mathbf{u})$, a periodically oscillating state $\mathbf{u}(t) = e^{-iEt} \mathbf{v}$ satisfies 
\begin{equation}
f(\mathbf{u}) = E\mathbf{u}, \label{nonlinear_eigen_eq}
\end{equation}
with $E$ being a real number. Then, one can employ Eq.~(\ref{nonlinear_eigen_eq}) as a nonlinear eigenequation, where $\mathbf{u}$ is a nonlinear eigenvector, and $E$ is a corresponding nonlinear eigenvalue. Furthermore, if one assumes an ansatz state of a plane wave such as $\mathbf{u}(\mathbf{x}) = e^{i\mathbf{k}\mathbf{x}} \mathbf{u}$, one can rewrite Eq.~(\ref{nonlinear_eigen_eq}) into the wavenumber-space description as in conventional condensed matter physics. By using the nonlinear eigenvectors in the wavenumber-space description, one can define nonlinear topological invariants such as the nonlinear Chern number 
\begin{equation}
C = \frac{1}{2\pi i} \int _{\rm BZ} \sum_{a,b,j} \epsilon_{ab} \partial_{a} \{ \left[u_j/||\mathbf{u}||\right] \partial_{b} \left[u_j/||\mathbf{u}||\right] \} dS,
\end{equation}
with $u_j$ being the $j$th component of the nonlinear eigenvector, $||\mathbf{u}||$ being the norm of the nonlinear eigenvector, $\epsilon_{ab}$ satisfying $\epsilon_{xy}=1$, $\epsilon_{yx}=-1$, and $\epsilon_{xx}=\epsilon_{yy}=0$, and $\partial_{a,b}$ being the wavenumber derivative in the $a$, $b$ direction. 

The crucial difference from linear systems is that a multiple of a nonlinear eigenvector is not necessarily a nonlinear eigenvector because of the absence of the superposition principle. Said differently, a nonlinear eigenvector also depends on its norm $||\mathbf{u}||$, and one should impose $||\mathbf{u}||=w$ independently of the wavenumber $\mathbf{k}$ by introducing an additional parameter $w$ of the nonlinear eigenequation. The norm dependence of the nonlinear eigenvectors and associated nonlinear topological invariants reveals the existence of the nonlinearity-induced topological phase transition, where the existence or absence of topological edge modes is tuned by the amplitude of nonlinear waves \cite{Hadad2016}. Such amplitude dependence of topological edge modes is a genuinely nonlinear effect on topological insulators and can lead to autonomous control of topological edge modes in nonlinear systems.

Since active hydrodynamics includes nonlinear terms derived from both convection and active forces, active matter can be an ideal platform to study nonlinear topological phenomena. In particular, large nonlinearity can be found in energy dissipation by active force described by $(\alpha-\beta|\mathbf{v}|^2)\mathbf{v}$ in Eq.~(\ref{toner-tu-eq2}). Thus, the stability of topological phenomena in active matter should be analyzed from the non-Hermitian and nonlinear point of view. 
Another possibility to observe nonlinear topological phenomena is using mechanically connected active particles. A recent paper \cite{Kotwal2021} has studied a nonlinear topological electrical circuit and revealed its possible connection to active Brownian particles. One can effectively describe the active Brownian particles by using a Mexican-hat-like potential with an active driving term. Therefore, their dynamics are effectively described by fourth-order nonlinear equations and might realize nonlinear topological waves.

\subsection{Topology induced by active forces}\label{sec:6c}
Active elements in metamaterials can induce nonreciprocal forces that are unique to nonequilibrium systems. In particular, if the metamaterial breaks the time-reversal symmetry, one can observe an active force perpendicular to the direction of the tension, which is known as odd elasticity \cite{Scheibner2020b}. Another time-reversal odd response is seen in odd viscosity \cite{Banerjee2017}, which we have already discussed in topological chiral active matter (cf.~Secs.~\ref{sec:2c3} and \ref{sec:5b1}).

Recent studies \cite{Zhou2020,Scheibner2020} have proposed topological edge modes under the non-Hermitian skin effect using odd elasticity. Since the breakdown of the time-reversal symmetry is ubiquitous in chiral active matter, one may utilize the active force to realize topological active matter. In fact, a recent experimental study \cite{Tan2022} has observed signatures of odd elasticity and chiral edge current in collections of starfish embryos, while the topological origin of the chiral edge current is not clarified. Furthermore, since odd elasticity realizes symmetry-breaking and nonreciprocal interactions that are difficult to implement in conventional materials, such active forces can be useful for investigating non-Hermitian topological phenomena. It is also noteworthy that time-reversal odd responses are analogous to the Hall conductivity in condensed matter, and thus their direct interplay with nontrivial topology (e.g., quantization) should also be investigated.

\subsection{Real-space topology}\label{sec:6d}
The properties of topological defects in biological experiments have been gaining interest owing to accumulating examples including those found in real animals \cite{Maroudas2021} as well as in the collective dynamics of bacteria \cite{Copenhagen2021,Meacock2021,Shimaya2022,Prasad2023,Doostmohammadi2016} and cultured cells \cite{Saw2017,Kemkemer1998,Duclos2017,Kawaguchi2017}. 3D experiments have also been conducted using systems involving microtubules and fast-scanning microscopy \cite{Duclos2020}.

More theoretically, the dynamics of defects and their relation to the conventional condensed matter phenomena, such as the Berezinskii-Kosterlitz-Thouless transition have been studied~\cite{Shankar2018,Shankar2019}. 
Topological defects at higher order have been largely unexplored, while the effect of anisotropy in the methods to introduce activity has started to be investigated \cite{Adachi2022,Solon2022,Nakano2024}.

In high-energy physics, topological defects can be regarded as quasiparticles and the emergence of the defects is considered as a potential key to understand the early universe. The emergence of topological defects by spontaneous symmetry breaking can be described by the Kibble-Zurek mechanism \cite{Kibble1976,Zurek1985}. Recent studies have also discussed the application of the Kibble-Zurek mechanism to condensed matters, such as cold atoms \cite{Lamporesi2013} and colloids \cite{Deutschlander2015}.

Despite the abundance of studies of topological defects in various fields of physics, those found in active matter have been mainly restricted to defects in two dimensions \cite{Saw2017,Kawaguchi2017,Mueller2019,Roshal2020}. Thus, exploring topological defects in, e.g., three-dimensional active matter \cite{Duclos2020} can enrich the interplay between the real-space topology and active matter dynamics. In addition, the Kibble-Zurek mechanism may bring new insight into the analysis of the dynamics of topological defects. Since topological defects can induce effective boundaries and emergent gauge fields, the nontrivial interplay between the band topology and topological defects may enrich the topological phenomena in active matter.

\subsection{Biological functions}\label{sec:6e}
So far, there has been no example of a topological edge mode at play in real biological systems (i.e., in multicellular tissues, inside cells, or even in purified biomolecular systems \textit{in vitro}). In cultured cells, bioengineering of edge dynamics at the interface between tissues has been implemented in Refs.~\cite{Ori2023,Scheibner2024}. While the topological origin of these edge dynamics is still under investigation, their existence implies the possibility that topological edge modes can work in real biological systems. In a broader perspective, some characteristics of topological edge modes, such as localization \cite{Leduc2012}, chiral motion \cite{Wioland2013}, and robustness \cite{Hopfield1974} might be found throughout biology. For example, robustness should be important for biological cells to adjust to changing environments. It will be interesting if the studies of topological active matter uncover the hidden topological structures in biology and elucidate the mechanism of, e.g., the robustness in biological functions.

\ack
We are grateful to Zongping Gong, Daiki Nishiguchi, Takahiro Sagawa, Taro Sawada, and Tsuneya Yoshida for fruitful discussions. 
K.S. acknowledges support from JSPS KAKENHI Grant No. JP21J20199, JP24K22848, and World-leading Innovative Graduate Study Program for Materials Research, Information, and Technology (MERIT-WINGS) of the University of Tokyo. 
K.Y. acknowledges support from JSPS KAKENHI through Grant No. JP21J01409. 
K.K. acknowledges support from JSPS KAKENHI Grants No. JP19H05795, No. JP19H05275, No. JP21H01007, and No. JP23H00095. 
Y.A. acknowledges support from the Japan Society for the Promotion of Science (JSPS) through Grant No. JP19K23424 and from JST FOREST Program (Grant Number JPMJFR222U, Japan) and JST CREST (Grant Number JPMJCR23I2, Japan). 

\section*{References}
\bibliography{reference}

\end{document}